\documentclass[11pt,a4paper]{article}
\usepackage{jcappub}
\usepackage{multirow}
\usepackage{portland}
\usepackage{rotating}
\usepackage{longtable}
\usepackage{threeparttable}
\def\Journal#1#2#3#4{{#1} {\bf #2}, (#3) #4}
\def\cir#1{{\GCN} #1}
\def\rep#1{{\GCR} #1}

\def\etal{{\it et al.}}
\def\AA{\em A.\& A.}

\def\AIP{\em AIP Conf.Proc.}
\def\ANA{\em Astro. Nachri.}

\def\APJ{\em ApJ.}
\def\APJL{\em ApJ.Lett.}
\def\APJS{\em ApJ.Suppl.}
\def\APP{\em Astropart. Phys.}

\def\ASS{\em Astrophys.Space.Sci.}
\def\AST{\em Astron. J.}

\def\GCN{\em GCN Circ.}
\def\GCR{\em GCN Rep.}
\def\GRG{\em Gen. Rel. Grav}

\def\JPg{\em J. Phys. G}
\def\JPJ{\em J. Plasma Sci. Japan}

\def\MRA{\em MNRAS}

\def\NAT{\em Nature}

\def\NPA{{\em Nucl. Phys.} A}

\def\PAS{\em Pub. Astro. Soc. Japan}
\def\PFB{{\em Phys. Fluids} B}

\def\PPL{\em Phys. Plasma}
\def\PRA{{\em Phys. Rev.} A}

\def\PRL{\em Phys. Rev. Lett.}

\def\PRE{\em Phys. Rep.}

\def\SCI{\em Science}

\def\SSR{\em Space Sci. Rev.}

\def\be{\begin{equation}}
\def\ee{\end{equation}}
\def\bea{\begin{eqnarray}}
\def\eea{\end{eqnarray}}
\def\bes{\begin{equation*}}
\def\ees{\end{equation*}}
\def\beas{\begin{eqnarray*}}
\def\eeas{\end{eqnarray*}}

\def\hubbleunit{km~sec$^{-1}$~Mpc$^{-1}$}
\def\plotdir{./}

\onecolumn

\title{Broad band simulation of Gamma Ray Bursts (GRB) prompt emission in presence of an external 
magnetic field}

\author[a]{Houri~Ziaeepour} 
\author[b,c]{and Brian Gardner}
\affiliation [a] {Max Planck Institut f\"ur Extraterrestrische Physik (MPE), \\
Giessenbachstra$\mathbf{\beta}$e 1, 85748 Garching, Germany}
\affiliation [b] {Mullard Space Science Laboratory,\\ Holmbury St Mary, Dorking, Surury, UK}
\affiliation [c] {Department of physics,\\ Exeter University, Exeter, UK}
\emailAdd{houriziaeepour@gmail.com}
\emailAdd{brgardner@hotmail.co.uk}

\abstract {The origin of prompt emission in GRBs is not yet well understood. The simplest and most 
popular model is Synchrotron Self-Compton (SSC) emission produced by internal shocks inside an 
ultra-relativistic jet. However, recent observations of a delayed high energy component by the 
Fermi-LAT instrument have encouraged alternative models. Here we use a recently developed 
formulation of relativistic shocks for GRBs to simulate light curves and spectra of synchrotron and 
self-Compton emissions in the framework of internal shock model. This model takes into account the 
evolution of quantities such as densities of colliding shells, and fraction of kinetic energy 
transferred to electrons and to induced magnetic field. We also extend this formulation by 
considering the presence of a precessing external magnetic field. These simulations are very 
realistic and present significant improvement with respect to previous phenomenological GRB 
simulations. They reproduce light curves of 
separated peaks of real GRBs and variety of spectral slopes at $E > E_{peak}$ observed by the 
Fermi-LAT instrument. The high energy emission can be explained by synchrotron emission and a 
subdominant contribution from inverse Compton. We also suggest an explanation for extended tail 
emission and relate it to the screening of the magnetic field and/or trapping of accelerated 
electrons in the electromagnetic energy structure of the plasma in the shock front. Spectral slopes 
of simulated bursts at $E \ll E_{peak}$ are consistent with theoretical prediction and at 
$E < E_{peak}$ can be flatter if the spectrum of electrons is roughly flat or has a shallow slope.  
The observed flat spectra at soft gamma-ray and hard x-ray bands is the evidence that there is a 
significant contribution at $E < E_{peak}$ from lower Lorentz factor wing of electron distribution 
which have a roughly random acceleration rather than being thermal. This means that the state of 
matter in the jet at the time of ejection is most probably nonthermal.
As for the effect of a precessing external magnetic field, we show that due fast variation of other 
quantities, its signature in the Power Distribution Spectrum (PDS) is significantly suppressed and 
only when the duration of the burst is few times longer than the oscillation period it can be 
detected, otherwise either it is confused with the Poisson noise or with intrinsic variations of 
the emission. Therefore, low significant oscillations observed in the PDS of GRB 090709a are most 
probably due to a precessing magnetic field.}

\arxivnumber{1101.3909}

\begin{document}
\maketitle

\section{Introduction} \label{sec:intro}
Progenitors of long GRBs are believed to be massive stars. At the end of their short lifetime 
of few hundred million years their cores collapse to neutron stars or black holes and their 
envelops are ejected making supernovae (SN) of type Ib or Ic. Under special conditions that we 
ignore they make also gamma-ray bursts. However, the absence of a SN signature in 
some nearby GRBs such as GRB 060614~\cite{swiftgrb060614} is probably the evidence 
that sometimes the ejected envelop - if any - is very thin and significant amount of energy is 
ejected as an ultra-relativistic jet which produces a GRB. We ignore details of necessary 
conditions for the occurrence of each case. 

The simplest model for the origin of GRBs is the synchrotron emission produced by collision between 
plasma shells inside an ultra relativistic jet~\cite{intext,intext1}. This model is a variant of 
{\it fireball} model which considers that a flash of gamma-ray is emitted when a fireball 
of $e^\pm$ becomes optically thin~\cite{fireball1,fireball2,epplasma}. However, non-thermal spectra 
of gamma-ray emission of GRBs as well as broad band observations of afterglows prefer a baryonic 
fireball or a jet.\footnote{In leptonic fireballs a long-lasting afterglow can exist only if 
leptons or gamma-ray photons deposit part of their energy in the surrounding baryonic material. 
Thus a special setup of surroundings is necessary. No signature of absorption of gamma-ray in GRBs 
is observed. Therefore, observations disfavor this type of models.} In the latter case, 
dissipation of the jet's kinetic energy by collisions and shocks between 
matter shells inside the jet and their synchrotron self-Compton emission is the origin of the prompt 
emission, and the collision of the jet with the ISM produces the afterglow. In this framework a 
pre-burst tight interaction between constituent i.e. a fireball is not necessary and the main 
mechanism for transfer of energy to photons is the shock, see e.g. ~\cite{piranrev} for a review. 
An alternative model for the prompt emission is a Poynting flow produced by energy extraction from 
a rotating black-hole through a Blandford-Znajek process~\cite{bhpoyting} or from the strong 
magnetic field of a neutron star/magnetar~\cite{poytingflow}. In this case, the dissipation of 
energy by field lines reconnection and acceleration of electrons during this process is the 
origin of GRB prompt emission. We should also remind that although in modern GRB models the 
nature of the ejecta diverges significantly from a fireball as a dense and initially opaque 
concentration of energy in a small region, many GRB literature use the word {\it fireball} as 
synonym to jet/ejecta/outflow. This leads to confusion, thus in this work the expression 
{\it fireball} is used only in the context of what is called {\it standard fireball} model.

None of the models mentioned above are flawless. In particular, since the launch of the Fermi 
satellite which is able to detect GRBs at high energies inaccessible to previous space missions e.g. 
BATSE and Swift, a number of issues and doubts have been raised against the most popular GRB models 
including internal shock-SSC, Poynting flow, and standard fireball model. In~\cite{critisism} 
authors have reviewed these problems. For instance, the absence of expected thermal emission from 
the fireball photosphere with a temperature $\sim 10^6$ eV~\cite{rphoto,thermal0,thermal1,thermal2,
thermal3,thermal4,thermal5,thermal6} is the main drawback of both leptonic and baryonic fireball 
models. 
The broad band observed spectra of GRBs up to $\mathcal{O}(100)$ GeV is very different from a 
thermal 
spectrum. The main problem of the Poynting flow model is its difficulty to explain the observed fast 
variations in prompt emission of all GRBs up to shortest time resolution of available instruments. 
The flow of the electromagnetic energy in this model is quite smooth and probability of field line 
reconnection per unit time is also small. Moreover, simulations show that a high magnetic to matter 
energy at the base of the flow becomes only a moderate value when the flow arrives at a distance of 
$\sim 10^{16}$ cm where the formation of instabilities dissipates the electromagnetic 
energy~\cite{poytingsimul0,poytingsimul1}. Therefore, such a Poynting flow does not transfer the 
initial energy to a GRB as efficiently as the original suggestions expected. Internal shocks-SSC 
model is not without shortcomings either. For instance, the expected slope of the spectrum below 
the peak energy is mush steeper than what is observed~\cite{batcat}, and doubt about this models 
was raised~\cite{deathline,deathline1}. Another 
issue for which none of these models has a simple solution is the observation of a tail 
emission in high energies $E \gtrsim 100$ MeV. It fades much slower than the flux of low energy 
$E \lesssim 100$ MeV photons\cite{grb090510obs,grb090510,grb090902b,grbcompare0, grbcompare1,
poytingprobl1}, and 
looks like to be a separate component because in some bursts e.g. GRB 090902B it has asynchronous 
variations with respect to low energy bands.

Various explanations or solutions have been suggested for each of these problems. For instance, 
the absence of a thermal component in baryonic fireballs may be explained if the ejecta is neutron 
rich~\cite{neutronrich}. The decay of neutrons when the fireball becomes optically thin produces 
high energy electrons and neutrinos, and a significant fraction of energy is dissipated as GeV 
neutrinos. Moreover, Compton scattering of photons with electrons reduces or smears the expected 
thermal spectrum~\cite{thermalmod0,thermalmod1}. A number of variations of Poynting model are 
also proposed~\cite{poytingmodel,critisism, jitter} to overcome the shortages of the original 
suggestion. However, it does not seem that either of these models can reproduce the featureless 
spectrum of GRBs~\cite{poytingprobl0,poytingprobl1} without fine-tuning. 

By contrast, it seems that internal shock-synchrotron self-Compton model can more easily explain 
all the issues raised here~\cite{simulother0,simulother1}. For instance, in the framework of this 
model, the ejecta does need to be a fireball with a large internal energy density because, as 
mentioned above, the main source of energy to be transferred to radiation is kinematic 
(see ~\cite{piranrev} and references therein). Therefore, matter inside the jet can be 
cool~\cite{simulother0}.

In the context of original $e^\pm$ or tightly coupled baryonic fireball models, the role of the 
photosphere is crucial because GRB gamma-ray emission is considered to be due to the simultaneous 
release of trapped photons when the fireball becomes optically thin. Even if the ejecta/jet is a 
fireball, the absence of a thermal emission can be explained if internal shocks occur inside the 
photosphere. In this case, chaotic collision, acceleration of particles, expansion, and finally 
emission of synchrotron photons can completely smear the initial thermal state of the ejecta. 
The radius of the photosphere of a standard fireball is estimated to be $R_{ph} \sim 6 \times 
10^{14} (L/10^{52}$ erg sec$^{-1})\zeta^{-1}(\Gamma/100)^{-2}$ cm~\cite{thermal0,rphoto,thermal2,
poytingprobl0}, where $L$ is the total luminosity of the baryonic outflow of the 
fireball~\cite{poytingprobl0}, $\zeta \equiv L_{in}/c^2\dot{M}$ is its initial dimensionless entropy, 
and $\Gamma$ is the Lorentz factor of fireball~\cite{rphoto}. From the Fermi~\cite{fermi} and 
Swift~\cite{swift} observations a Lorentz factor of $\Gamma \gtrsim 1000$. If total outflow luminosity 
is $L\sim 10^{53-55}$ erg and $\zeta \sim 1-100$, photospheric radius $R_{ph}$ is in range $10^{10-14}$ 
cm~\cite{rphoto}. If thermal radiation is the only or the dominant contributor to 
observed radiation, it must have a temperature $\sim 1$ MeV. Under some conditions 
(see next paragraph) internal shocks can occur inside this radius. Note that in internal shock+SSC 
models the ejecta is usually considered to be cold and there is no issue with the absence of a 
thermal component. In these models shocks usually occur outside the photospheric radius and the 
photospheric emission is weak and at low energies.

The distance of internal shocks from the central engine is estimated by the time necessary for a 
fast shell with Lorentz factor $\Gamma_f$ to catch and collide with a slower shell with Lorentz 
factor $\Gamma_s$ ejected $\Delta t$ sec earlier:
\be
R_{coll} = \frac{2c \Delta t \Gamma_f^2 \Gamma_s^2}{\Gamma_f^2-\Gamma_s^2} 
\approx 2c \Delta t \Gamma_s^2 \quad\quad \Gamma_f \gg \Gamma_s \label{rcoll}
\ee
All quantities in (\ref{rcoll}) are in the rest frame of the central source. GRB literature 
usually assume $\Delta t \sim 0.1$ sec as the typical time separation between shells - density 
anisotropies in the flow. This leads to $R_{coll} \sim 10^{14}$ cm i.e. the same order as the 
photosphere radius of a standard fireball for some choice of parameters - specially when a low 
$\Gamma$ is assumed. However, the choice of $\Delta t \sim 0.1$ sec is arbitrary because in both 
long and short bursts we see variations at much smaller time scales. In fact, the whole duration of 
some short bursts is less or comparable to this value. On the other hand, simulations show that 
collimated jets come from the most inner part of the accretion disk. The size of neutron stars and 
stellar-size black holes is few kilometers. Thus, the innermost radius of the accretion disk must 
be of the same order. Assuming that spatial separation between density anisotropies of the disk is 
comparable with this inner radii, the natural variation time scale for relativistic accretion disk is 
$\sim (R_{disk}/3$ km$) \times 10^{-5}$ sec. This leads to $R_{coll} \sim 10^{10-12}$ cm for progenitors 
of mass $\sim 2-200$ solar mass. Even if we consider the orbital period of spinning material around 
black holes with masses a few times $M_\odot$, the Keplerian approximate period for the innermost 
quasi-stable orbit $\tau = (4\pi GM/c^3)^{1/2}$ is of order $\sim 10^{-3}$ sec. This leads to 
$R_{coll} \sim 10^{12}$ consistent with the upper limit of the range given above. Besides, simulations 
of magneto-rotational instabilities in accretion disks show the presence of a "channel" type radial 
flow both inward and outward~\cite{radialvelo}. This radial motion is necessary for 
reducing the viscosity of the accreted material which suppresses the conductivity of the flow and 
thereby the formation of a strong magnetic field and a jet. Therefore, $R_{coll}$ range obtained by 
assuming the radius as the distance scale between anisotropies is reliable. Detail modeling of some 
bursts with good data (see e.g.~\cite{grb060607a,grb100316b}) also confirms these 
estimations\footnote{We should remind that in GRBs variability is observed in all scales. This means 
that the accretion disk and thereby the jet don't have a special variation distance scale and are 
roughly scaleless. This ensure that the process of internal shocks is not limited to small scales 
and GRB emission as a whole is dominantly due to this process.}.

If the ejecta is a standard fireball, short distances suggested here for internal shocks mean that 
their initiation inside the photosphere is not ruled out if the input entropy $\zeta$ and the 
Lorentz factor are small or if $\Delta t$ is very small. Nonetheless, following the internal shocks, 
acceleration of electrons, formation of a random magnetic field, synchrotron/self-Compton scattering, 
and sudden blow up of the ejecta, the initial strong coupling and thermal distribution of particles 
will be interrupted and their memory in the distribution of high energy electrons will be in large 
extend lost~\cite{fermiaccspec}, see sections \ref{sec:simulatedspec} and \ref{sec:lowener} 
for more details. Therefore in general during the whole or large fraction of burst time no 
photosphere is present and photons are free to propagate. We should also mention that according to 
PIC simulations despite metamorphoses mentioned above, a dominant low energy thermal population of 
electrons seems to persist, at least in $e^\pm$ plasma. On the other hand, internal shock models do 
not need a fireball, thus issues related to shocks inside or outside the photosphere do not arise.
This scenario is consistent with non-thermal spectrum of prompt emission in high energy bands. 
Other issues of internal shock-SSC model such as small efficiency, steep slope of the spectrum at 
low energies which is not observed, etc. are related to the micro-physics of the shock and can be 
realistically addressed only when more realistic PIC simulations become possible. If internal 
shocks occur at short distances from the central source, the densities of the shells and surrounding 
are expected to be large and optical depth small. We address this issue in Sec. \ref{sec:proc}

In order to better understand internal shock-SSC model in a previous work~\cite{hourigrb} we 
have developed a simple and consistent formulation of ultra-relativistic shocks. This formalism 
and corresponding simulations are essentially similar to work by other authors 
specially~\cite{simulother0,simulother1}. The latter authors have emphasized on the phenomenological 
aspects of the formulation and given a qualitative view on the role of various quantities on 
observable properties of GRBs. For this reason these topics have not been discussed 
in~\cite{hourigrb} or here. On the other hand, we have tried to discuss in details mathematical 
aspects of the problem as much as our aim to keep the formulation simple allowed. Moreover, this 
formulation includes time variations of micro-physics parameters such as densities and fraction of 
energy transferred to accelerated electrons and to the induced magnetic field in the shock which in 
the context of fast varying prompt emission are crucial. The advantage of our method is that it can 
be relatively easily extended and time variation of other quantities can be included. In addition, 
giving the fact that at least 3 moving reference frames are involved in the modeling of relativistic 
shocks, a detailed formulation reduces the risk of confusion for the reader. For each simulation we 
present light curves and spectra such that they can be directly compared with observations. Due to 
much larger parameter space in these simulations with respect to the modeling of~\cite{simulother0}, 
we leave its exploration to a future work. Nonetheless, we have included results for bursts with 
different durations, brightness, and hardness to show that practically all the features seen in 
real GRBs can be reproduced. Notably, the variety of spectral slopes which has been recently claimed 
as to be unexplainable with the simple internal shocks-SSC model are obtained.

Preliminary simulations of both prompt and afterglow according to this formulation~\cite{hourigrb1} 
have shown that they well reproduce the behaviour of GRBs in various intervals of their evolution. 
Here, we improve those results by simulating the whole prompt emission of GRBs. In addition, we 
extend the formalism of~\cite{hourigrb} to include an external magnetic field, and we study its 
effect on the prompt emission of simulated GRBs. Here by {\it external field} we mean {\bf not 
induced by shocks.} It can be of one or a combination of the following origins: 
\begin{enumerate}
\item A Poynting flow i.e. a slowly varying electromagnetic wave;
\item A magnetic field frozen in the jet's plasma i.e. the magnetic field in a 
magnetohydrodynamical flow;
\item Magnetic field of the central source in case of a proto-neutron star, magnetar or accretion 
disk around a black hole;
\item Remnant of the dynamo field in the star's envelop. 
\end{enumerate}
Considering the distance to central source obtained earlier, shocks can be surrounded by this 
envelop if it is not yet ejected. In the case 1 precession of the field seems unlikely. In case 2 
the field precesses only if the plasma has a helical 
twister-like movement. In cases 3 and 4 the magnetic field can have a precession. In case 3 the 
precession period is expected to be of millisecond scale. If the outer layers of the star continue 
their differential rotation before being completely disrupted and ejected, their dynamo magnetic 
field (case 4) should have a rotation period of few seconds. We do not specify the type of the 
field in the simulations and consider the general case of a precessing field. A non-precessing 
field can be assumed to have a precession period much longer than prompt emission of simulated GRBs. 
The simple phenomenological model used for simulations here is not sensitive to these details. 
Thus we simply assume that the magnetic field influencing accelerated electrons has 2 components: 
one is induced by the shock and is parametrized as usual by fraction of energy of falling ions 
transferred to electrons, see equation (\ref{magener}) in the next section, and the other 
component is assumed to be independent of the shock and a function of the distance from central 
source only.

We have also calculated the distribution of Compton scattered photons by electrons for these 
simulations. We show that its contribution is very small. This is consistent with observations 
and is another evidence of the validity of internal shock-synchrotron model. Compton scattering 
between protons and photons, photon-photon annihilation, and Compton self-absorption are not 
considered. It is known that Compton emission extends longer in time than synchrotron, but we show 
that it cannot explain observations by the Fermi-LAT~\cite{grb090510obs,grb090902b}. On the 
other hand, based on the results of the state-of-art Particle-In-Cell (PIC) simulations we 
related the high energy tail to the structure of induced electromagnetic field in the shocks. 
It can play an important role in a delayed dissipation of most energetic electrons. The screening 
of the external magnetic field by the shock can be also involved. This probably explains why the 
delayed tails have been observed only in bright hard bursts in which, as we will show here, an 
external field can have significant contribution. However, the phenomenological model of 
ultra-relativistic shocks used here cannot address these complexities.

We begin by briefly discussing evidence for the presence of precessing magnetic fields in 
astronomical sources and the case of GRB 090709A in Sec. \ref{sec:magprecess} as an observational 
evidence 
in favour of an external magnetic field in GRBs. The formulation of dynamics in presence of an 
external field is discussed in Sec. \ref{sec:shock}. In Sec. \ref{sec:simul} we present a number of 
simulated bursts, discuss their properties and compare them with properties of observed bursts. 
This section also includes results of the simulation of Inverse Compton contribution 
in light curve and spectrum of GRBS, as well as an explanation for the high energy tail emission. 
Finally Sec. \ref{sec:conclude} includes outlines of this work. In Appendix \ref{app:a} an 
analytical expression is given for the contribution of the external magnetic field in the 
evolution of bursts kinematic. In Appendix \ref{app:b} the projection of a precessing magnetic 
field on the shock front surface is obtained. Finally, Appendix \ref{app:c} gives a qualitative 
description of how a magnetic field influences the formation of instabilities in charged plasmas.

\section{Precessing magnetic fields in astronomical sources} \label{sec:magprecess}
There are a number of motivations for presence of other sources of magnetic field than shocks 
in the environment of relativistic jets. For instance, due to turbulent rotation of the accretion 
disk that surrounds the central source the disk is highly magnetized. Along with dynamical and 
gravitational forces, the pressure created by this field plays an important role in the 
formation and collimation of ultra-relativistic jets (see e.g. ~\cite{jetcollim} for a review). 
In fact observations of astronomical sources in which long lasting jets are formed such as 
Active Galactic Nuclei (AGNs)~\cite{agnprecess} and accreting X-ray binary 
stars~\cite{binaryprecess,binaryprecess1}, tell us that both mechanical and magnetic pressures must 
be involved (see e.g.~\cite{precessionaccre} for a review). In all these cases jets are associated 
with a rotating and precessing accretion disk~\cite{warpeddisk}. The mechanical pressure at the 
inner side of the disk 
compresses and pushes part of the falling material up and ejects it out to free the remnant 
momentum and angular momentum of in-falling matter. Therefore, it seems that the presence of an 
external magnetic field in GRBs environment is inevitable. 

In addition to providing acceleration and collimation, a strong magnetic field explains the 
puzzling and apparently inconsistent fluence of high energy component of the bursts observed by 
the Fermi-LAT instrument. Simulations presented in the next sections show that when an additional 
magnetic field is included, bursts are few times brighter and harder. This soften the efficiency 
issue raised by Particle In Cell (PIC) simulations which find that only $\sim 10\%$ 
of kinetic energy of the jet is transferred to electrons~\cite{fermiaccspec}. Although a larger 
magnetic field does not solve the problem of insufficient energy transfer from bulk of material to 
electrons, it makes synchrotron emission more efficient. The low efficiency of energy transfer 
means that the total energy released by the engine must be even larger. However, simulations 
by~\cite{fermiaccspec} are performed for a $e^\pm$ plasma. Due to the large mass difference between 
nucleons and electrons, the behaviour of a baryonic plasma can be very different from a $e^\pm$ one. 
Indeed PIC simulations with larger mass difference between particles confirm this point. In 
particular they show that the presence of a parallel magnetic field facilitates the acceleration 
of electrons~\cite{fermiaccspec1}.

Due to the fast precession of accretion disk and probably the central source itself, it is expected 
that during a collapsar phase the external field at the position of internal shocks precesses. This 
can be the case even if the jet is aligned with the magnetic field at the time of its ejection. 
Moreover, if the argument given in the Introduction is correct and GRBs occur in a distance of 
$\sim 10^{10-12}$ cm, at the time of their formation internal shocks are yet inside the star's 
external envelop, thus the flux of the field can be significant. 

\subsection{A special case: GRB 090709A} \label{sec:grb}
The absence of a clear evidence of a precessing field in prompt emission of GRBs was a puzzle, but 
the situation has been changed with the observation of GRB090709a by Swift-BAT, ~\cite{grb090709a0}, 
Konus-Wind~\cite{grb090709a1}, INTEGRAL-SPI-ACS~\cite{grb090709a2}, and 
Suzaku-WAM~\cite{grb090709a3} instruments. Initial doubts on the identification of this transient 
as a GRB has been removed by further observations~\cite{grb090709axrt} including the detection of 
a candidate host galaxy for this burst~\cite{grb090709ahost}. The signature of an oscillating 
component with a period of $\sim 8$~sec superposed on two FRED-like spikes is detected by all the 
instruments that observed the prompt emission in gamma-ray. Although from further analysis of 
the data~\cite{grb090709a4} it has been claimed that the detected oscillations in the PDS have only 
a significance of $\lesssim 3\sigma$, the fact that they are detected by multiple independent 
instruments and in different energy bands is the proof that they are genuine. Oscillations are 
particularly significant in the power spectrum of light curves rather than in the differential 
spectrum obtained after removing a smoothed curve. The latter is necessarily depends on the way 
the smoothed component is determined and can include artifacts to power spectrum. 

\begin{figure*}
\begin{center}
\begin{tabular}{ll}
\includegraphics[width=7cm]{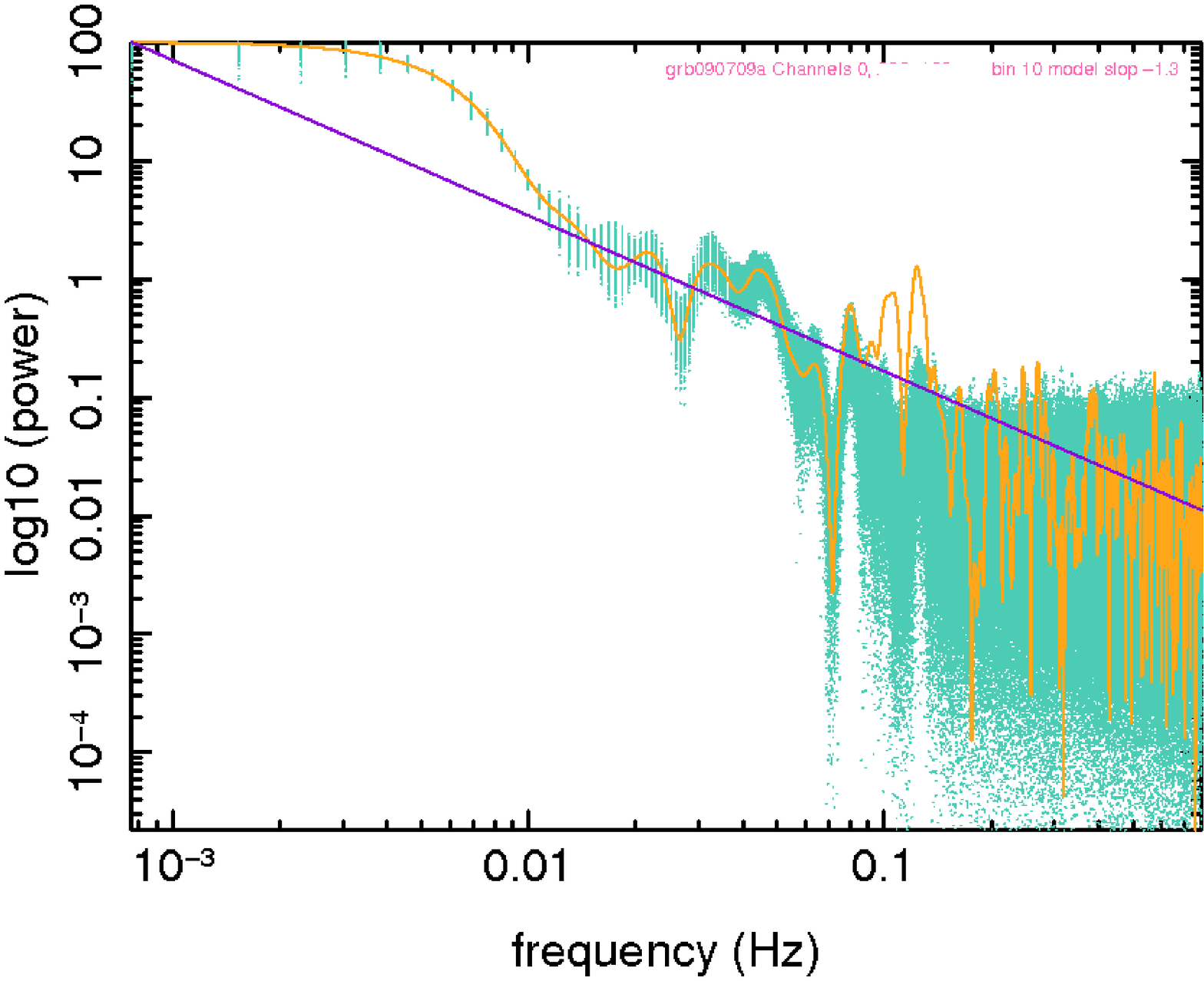} & 
\includegraphics[width=7cm]{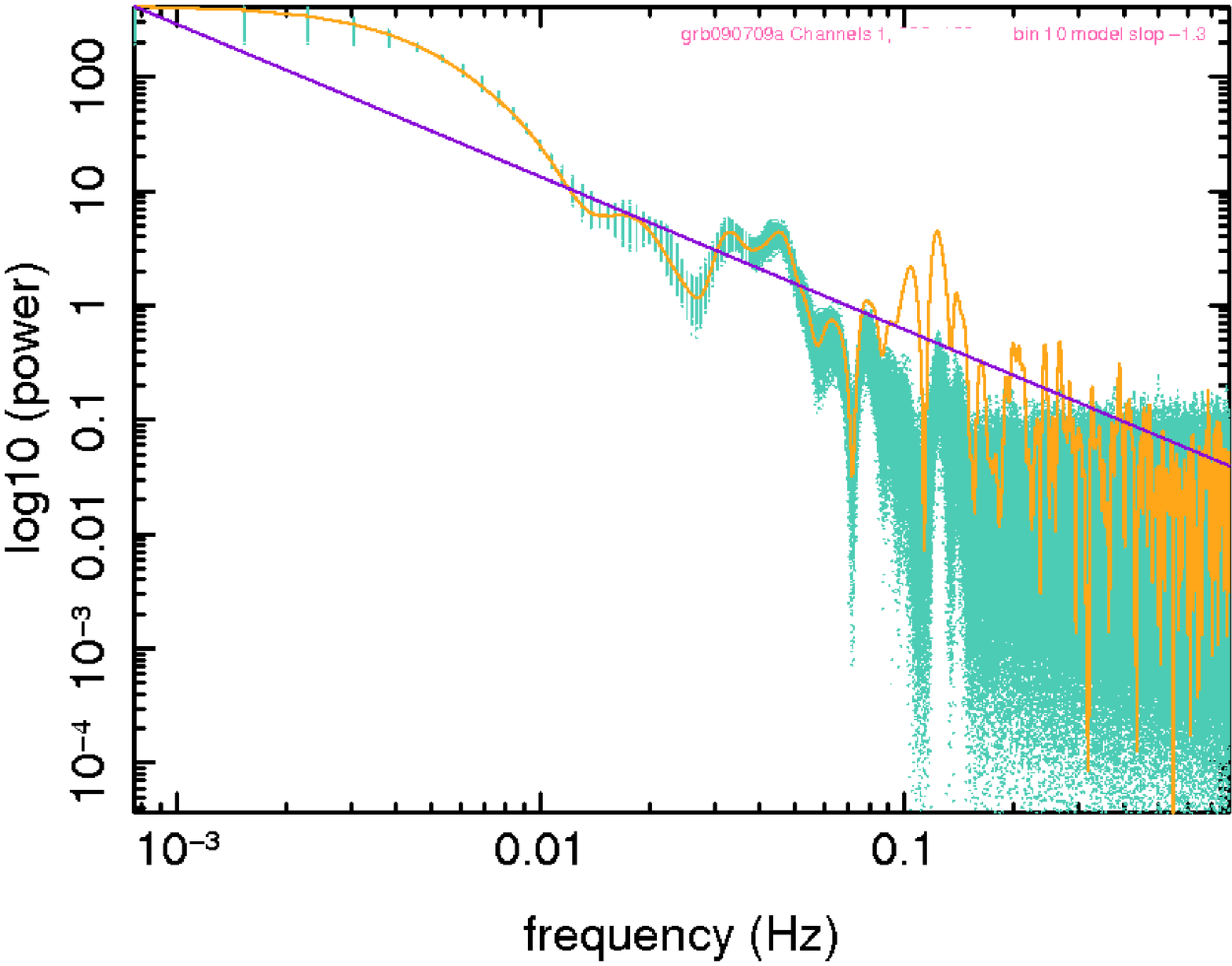} \\
\includegraphics[width=7cm]{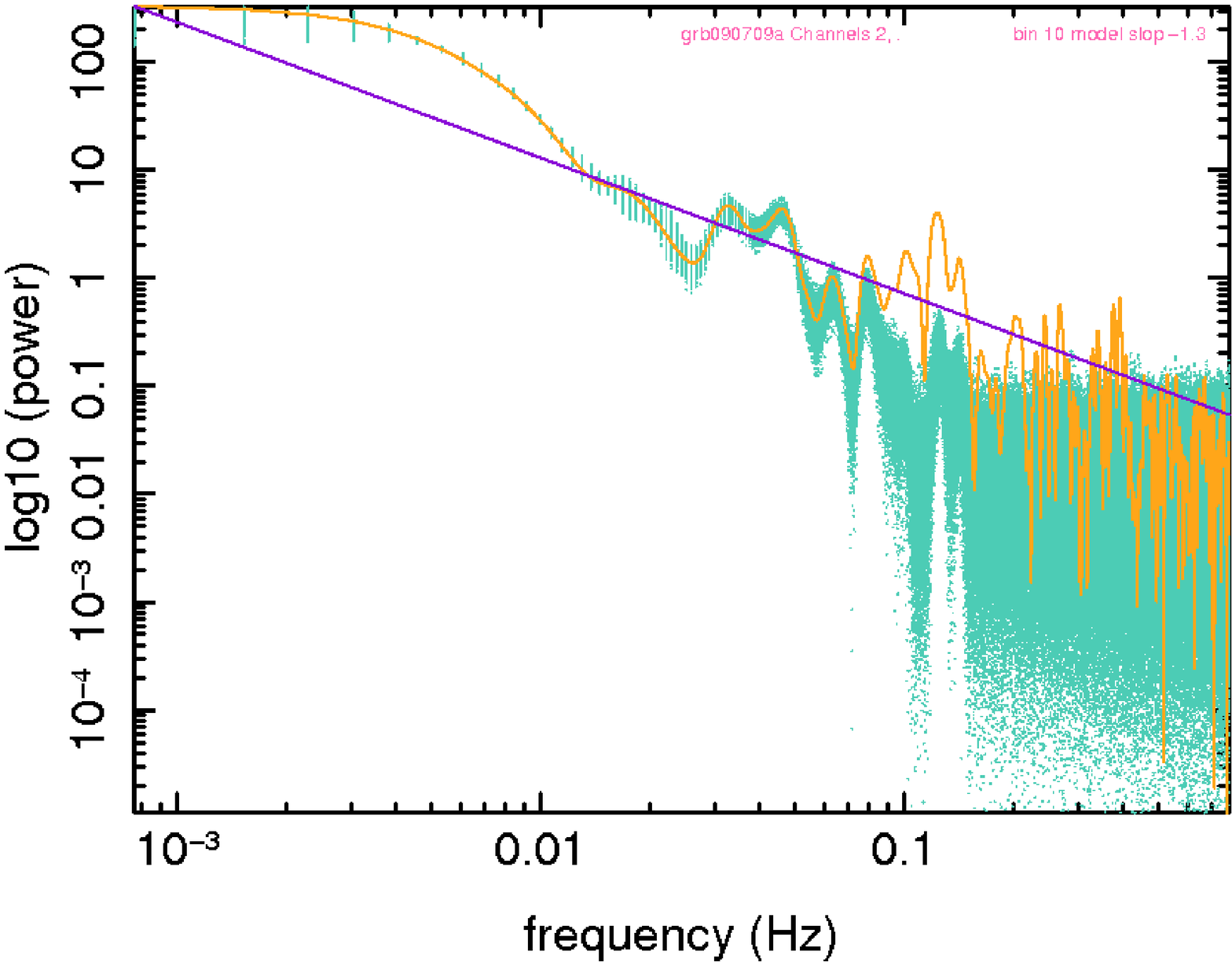} & 
\includegraphics[width=7cm]{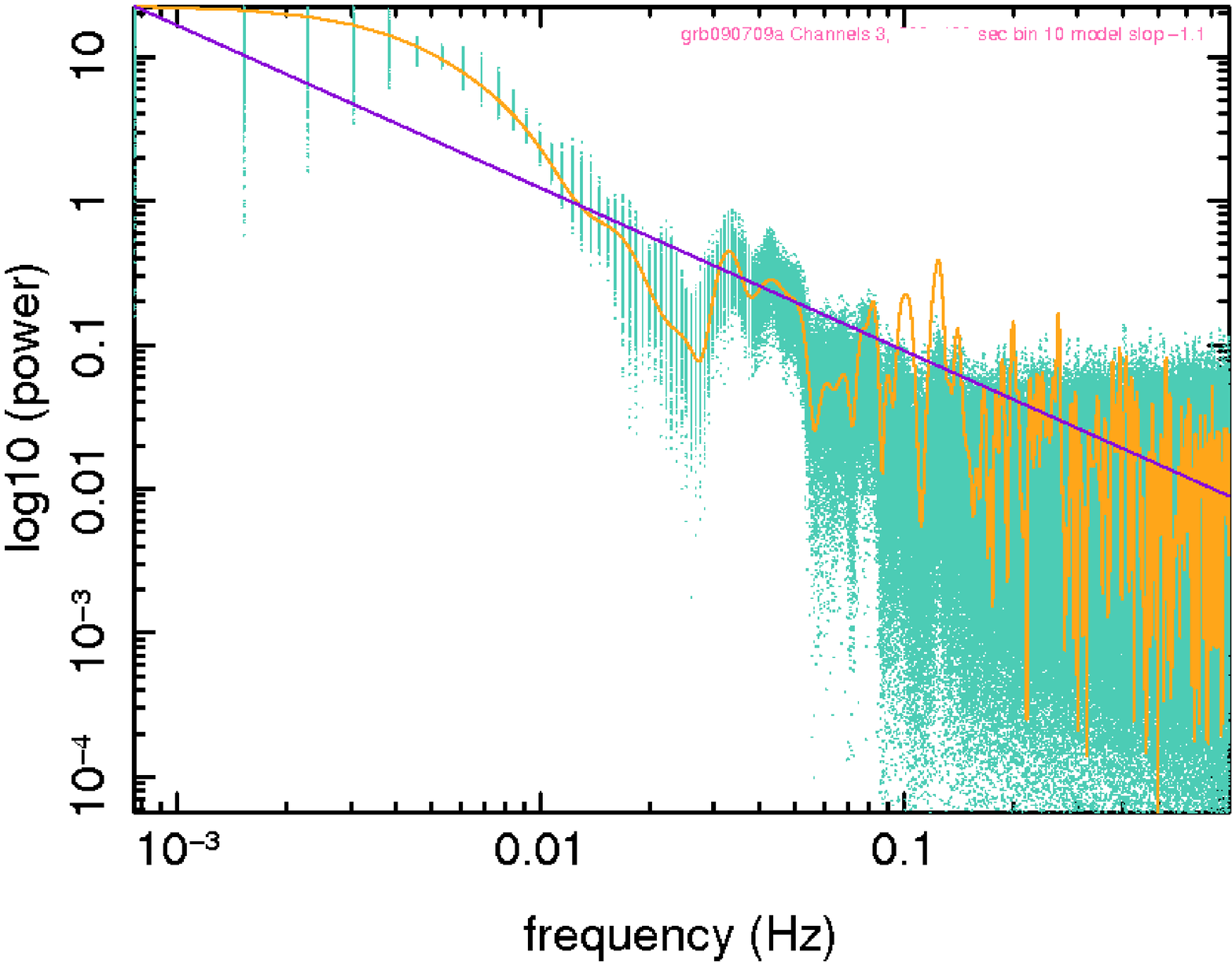}
\end{tabular}
\caption{Orange (light grey): Power spectrum of GRB 090709A in 4 non-overlapping Swift-BAT bands, 
from top left to bottom right: 15-25 keV, 25-50 keV, 50-100 keV, 100-350 keV. We have used BAT 
light curves with $64$ msec resolution and binned them (bin-size = 10). Turquoise (mid-gray) points 
present 500 spectra of modified light curves by adding to each data point a random value with 
Gaussian distribution and a standard deviation equal to the estimated $1\sigma$ error of the 
Swift-BAT data. Purple (dark grey) line presents a fit to obtain an average slope index of 
$\alpha = -1.3$ of the spectrum (for an ideal bilateral-triangle shape peak $\alpha = -2$). The 
peak around $f=0.12$ Hz is clearly distinguishable from the noise and is consistent with the 
analysis of other groups. There are also some other peaks at higher frequencies but with much less 
significance. Inspection of plots with linear time scale shows that most probably they are not 
noise but the harmonics of $f=0.12$ Hz oscillation. The two broad peaks at low frequency limits 
are due to the main peaks in the light curves and their first harmonic.
\label{fig:grboscil}}
\end{center}
\end{figure*}

Fig. \ref{fig:grboscil} shows the power spectrum of GRB 090709A in 4 non-overlapping Swift-BAT 
bands. In place of modeling the noise, we have calculated the random variation of the average 
spectrum by adding a Gaussian noise with a standard deviation equal to the estimated error for each 
data point. We have repeated this procedure 500 times. In Fig. \ref{fig:grboscil} the plateau 
formed by the simulated noise shows precisely the level of white noise in the data. Features 
covered by points are not distinguishable from noise. The peak around $f=0.12$ Hz is about 
one order of magnitude higher than the noise in all 4 non-overlapping energy channels. Considering 
the number of iterations, we obtain a significance of $3.5 \sigma > 99\%)$ per channel for this 
feature. Therefore, assuming that the measurement errors in different channels are independent, the 
probability of coincidence in all 4 channels is $< 10^{-4}$. We have performed the same analysis 
for a number of other GRBs detected by the Swift-BAT instrument and didn't find any other burst 
showing an oscillatory component in its PDS with such a high significance. The result of this 
study will be reported elsewhere.
No significant evidence of oscillation was found in the Swift-XRT Window-Timing (WT) data beyond 
$150$ sec after trigger~\cite{grb090709axrt}. This is consistent with our simulations which show 
that a significant oscillation exists only in the very early prompt emission because when the jet 
leaves the envelop of the star, the magnetic field decreases very rapidly.

To interpret these oscillations we briefly review their plausible sources. An oscillatory 
modulation of emission from an astronomical source can arise from a number of processes: 
\begin{description}
\item {\bf {$\bullet$} Precession of a hot spot:} In the framework of internal shocks and 
synchrotron/inverse-Compton emission model such explanation does not apply to GRBs. After 
the ejection of the relativistic jet, gravitational and kinematic influence of the central 
engine on the ejecta are negligible. The exception could be the emission very close to a 
rotating black hole. In this case the relativistic ejecta could be dragged by the hole for 
sometimes after its ejection. For instance, the ejecta can have a helical 
orbit~\cite{bhhelicalorb} before the 
gravitational influence of the hole or precession of a jet around the rotation axis of a Kerr 
black hole is stopped. In this case the effect should be energy dependent because less energetic 
photons should be less affected by the black hole drag.
\item {\bf {$\bullet$} Regular and oscillation-like density anisotropies in the ejecta} 
Density perturbations in the ejecta are expected and are one of the 
main ingredients of the internal shock model. However, it does not seem plausible that anisotropies 
be able to keep their coherence for a long time to induce a regular modulation in the emission. 
Expansion of the ejecta, turbulence, and minor internal shocks should very quickly destroy the 
coherence of different bunches leaving only random variations - substructures - in the emission. 
Nonetheless, if the source of oscillations is external, for instance if they are produce under the 
influence of an oscillating magnetic field, distribution of the particles in the plasma can be 
modulated (see also Appendix {app:c}).
\item {\bf {$\bullet$} External magnetic field:} The presence of a pre-collapse large scale 
strong magnetic field, in the range of few hundred to few kG and as large as few tens of kG in 
O-stars and Ap/Bp stars respectively, has been confirmed by many observations 
(see ~\cite{massstarmag,massstarmag1} for a review). Moreover, the iron core of collapsars 
progenitors are expected to have a fast 
rotation with a period of $\sim 1-8$~sec~\cite{snrot}. The combination of an external magnetic 
field and rotation not only helps but perhaps is indispensable~\cite{snbipolar,snbipolar1} for 
the formation of an ultra-relativistic jet~\cite{snjet,snjet1} and a GRB.
\end{description}
Therefore, it seems that a rotating magnetic field is the most plausible explanation for the 
cyclic modulation of peaks in GRB 090709a. In fact if our understanding of the structure of 
massive stars and supernova explosion is correct, the effect of magnetic field of the central 
engine should have been seen in most GRBs. In Sec. \ref{sec:simul} we show that due to fast 
variation of quantities that control the luminosity of a burst, coherent oscillation of quantities 
such as magnetic field is smeared or destroyed and even when they are present in the light curves, 
only under special conditions they can be distinguished from from intrinsic fluctuations of the 
emission.

\section {Synchrotron emission by relativistic shocks in presence of an 
external magnetic field}
\label{sec:shock}
In this section after a quick review of the simplified relativistic shock and synchrotron model 
discussed in~\cite{hourigrb}, we reformulate the model in presence of an external magnetic 
field\footnote{As we have mentioned in the introduction what we call an 'external magnetic field' 
can have various origins, and is not necessarily a Poynting flow. Moreover, relatively small fields we 
consider do not make the jet a 'magnetized outflow' i.e. Poynting energy dominant (see 
Fig. {fig:kinetic}), and a dust approximation is adequate in what concerns the kinematic of the flow. 
We also remind that this work is not meant to be a magnetohydrodynamics modeling of GRBs which is 
definitely a much more complex problem. Apart from PIC simulations some authors have used simplified 
magnetohydrodynamics algorithms to study dynamics of relativistic shocks between magnetized plasma 
shells without considering any radiation emission~\cite{maghydoshock}. In their model synchrotron 
emission is added analytically~\cite{maghydoshocksyn}. In fact, even in absence of an external magnetic 
field, a correct formulation of the jet needs a magnetohydrodynamics modeling because most probably the 
jet is ionized, includes current sheets, and thereby a magnetic field - although it can be small 
and has a small coherent length. Our analytical model and simulations presented here are evidently 
much simpler, nonetheless, they allow to investigate microphysics parameters and can be used to model 
real bursts.}.

A shock is formed when two shells of material with different densities collide. The 
discontinuity in physical properties of the shells exist only at the beginning of the collision, 
thus is considered as an initial condition (see e.g.~\cite{relfluid}). During collision one 
can distinguish two {\it shocked zone} in the opposite side of the initial discontinuity. For 
an observer at rest with respect to the shock front - which is no longer a discontinuity but 
the place of a sharp variation of density and other physical properties - some of the 
particles are accelerated and ejected to downstream. On the upstream side by contrast particles 
are decelerated. This process induces turbulence and creates coherent electric and transversal 
magnetic fields which accelerate electrons by Fermi 
processes~\cite{fermiacc,fermiaccspec,fermiaccspec1}. These electrons move back and forth across 
the shocked zone. Their energy is dissipated mainly by synchrotron emission in places where the 
induced magnetic field is strong and transversal. If the velocity of massive particles - presumably 
baryons - are reduced to relativistic sound speed in the upstream, a secondary {\it reverse shock} 
front which moves in the opposite direction - with respect to the main {\it forward shock} would 
be formed~\cite{revshockform}.

Although it was expected that the difference between the dominant synchrotron frequency and 
time evolution of light curves from forward and reverse shocks make their separation possible, 
multi-band and early observations of GRBs showed the contrary. In fact since the first GRBs detected 
by Swift there was a significant effort by observers to detect what is called an {\it optical flash} 
with steep slope characteristic of the emission from reverse shock according to some theoretical 
models~\cite{revshockoptflash,reverseshock}. Despite interpretation of some bright early optical 
afterglows as being the prompt reverse shock, none of them have the necessary properties - the 
rising slope - to be indisputably interpreted as a reverse shock peak. In all such cases either 
the observed peak in optical coincides with flares in X-ray and/or gamma-ray, or can be 
interpreted as the passage of the synchrotron characteristic frequency of the continuum to optical 
bands followed by a fast decline~\cite{sampaper}. A good example is GRB 061121~\cite{grb061121} 
in which the burst was detected during its precursor and optical data are available during its 
main peak. Simulations of shocks (including ours in this work) reproduce this continuum-like peak 
in low-energy bands few tens of seconds after the prompt peak in gamma-ray~\cite{maghydoshocksyn}. 
It is produced by forward shocks. Simulations in~\cite{maghydoshocksyn} also show the reverse 
shock peak occurs before forward shock peak. However, such a peak has not been observed in any 
bursts with optical observations before a major peak such as GRB 050820a~\cite{grb050820a,
grb050820a1}, GRB 061121~\cite{grb061121}, and GRB 070721B~\cite{grb070721b}. Therefore, based on 
these observations one can conclude that in the shocks which make the prompt emission of GRBs 
despite energy dissipation in the shock, the velocity of particles in the jet is not reduced 
down to relativistic sound speed such that a reverse shock be produced~\cite{revshockform} 
or the reverse shock is too weak to be detectable, or the delay between arrival time of forward and 
reverseshocks emission to observer is too short to make them distinguishable. As for PIC 
simulations which are more precise, shocks in $e^\pm$ shows very little reverse shock, see 
Fig 1-e,h,i in~\cite{fermiaccspec}. No realistic simulation of baryonic relativistic shocks is 
yet available, see e.g.~\cite{fermiaccspec1}, but one can argue that if light particles such as 
electrons with large $e/m$ can be barely reflected by a relativistic shock, it would be even more 
difficult for massive particles like protons.

In this case emissions from the prompt shock is better modeled by assuming a single emitting 
region that we call the {\it active region}. It is dominantly in this region that transfer of 
kinetic energy from fast shell to slow shell and radiation occurs. The active region is expected 
to be smaller than shocked region - the zone in which two shells are mixed. Notably, in contrast 
to the shocked region which increases continuously until two shells are coalesced or pass through 
each other, the width of active region first increases, arrives to a peak value, then decreases 
when the discontinuity between colliding shells changes to a continuous transfer region. To 
simplify the model further we also assume that 
the thickness of this emitting region is small, thus the propagation time of photons in this 
region is smaller than time resolution of this model. In fact for objects moving with ultra 
relativistic speeds with respect to an observer, time and distance are approximately proportional: 
$r' (t') = \beta' (t)ct' \approx ct'$.\footnote{Through this work quantities with a prime are 
measured with respect to the rest frame of the slow shell and without prime with respect to a 
far observer at the redshift of the central engine. Parameters used for parametrization do not 
have a prime even when parametrization is in the slow shell frame.} Under these approximations 
evolving quantities only depend on the average distance of the active region from central engine 
(or a far observer). Mathematically, this approximation is equivalent to assuming a wave-like 
behaviour for dynamical quantities i.e they depend on $r'- c\beta't'$ rather than $r'$ and $t'$ 
separately. When $\beta' = const$, i.e. when there is no collision or dissipation, this is an 
exact solution. In this case the solution at every point can be obtained from the solution of 
one point. 

Conservation of energy and momentum determines the evolution of this system. The velocity $\beta'$ 
of the fast shell/active region decreases due to absorption of particles from slow shell and 
dissipation of kinetic energy as radiation due to synchrotron and self-Compton interactions. 
After a variable change the dynamic equations - energy-momentum conservation equations - for the 
active region can be written:
\bea
&& \hspace{-0.7cm}\frac {d(r'^2 n' \Delta r' \gamma')}{dr'} = \gamma' \biggl (r'^2 
\frac{d(n'\Delta r')}{dr'} + 2r' (n'\Delta r')\biggr ) + 
r'^2 (n'\Delta r') \frac{d\gamma'}{dr'} = n'_0(r) r'^2 - \frac{dE'_{sy}}{4\pi m c^2dr'} 
\label {enercons} \\
&& \hspace{-0.7cm}\frac {d(r'^2 n' \Delta r' \gamma' \beta')}{dr'} = \beta' \gamma' (r'^2 
\frac{d(n'\Delta r')}{dr'} + 2r' (n'\Delta r')) + r'^2 (n'\Delta r') 
\frac {d(\beta' \gamma')}{dr'} = - \frac{dE'_{sy}}{4\pi m c^2dr'} \label {momcons}
\eea
where $r'$ is the average distance of the active region from central engine, $n'$ is the baryon 
number density of the fast shell measured in the slow shell frame, $n'_0$ is the baryon number 
density of the slow shell in its rest frame and in general it depends on $r'$. Here we assume 
that $n'_0(r') = N'_0 (r'/r'_0)^{-\kappa}$. For the ISM or thin shells where density difference 
across them is negligible $\kappa = 0$, i.e. no radial dependence. For a wind surrounding the 
central engine usually $\kappa = 2$~\cite{windexpo} is assumed. For a thin shell/jet expanding 
adiabatically also $\kappa = 2$ if we neglect the transverse expansion in the case of a jet 
(collimated ejecta). If the lifetime of the collision is short we can neglect the density change 
due to expansion during the collision and assume $\kappa = 0$. $\Delta r'$ is the thickness of 
the active region, $\gamma'$ is the Lorentz factor of the fast shell with respect to the slow 
shell, $\beta' = \sqrt {\gamma'^2 - 1} / \gamma'$, $m = m_p+m_e \approx m_p$, $E'_{sy}$ is the 
total emitted energy, and $c$ is the speed of light. The evolution of the average radius of the 
shell is:
\be
r' (t') - r' (t'_0) = c \int_{t'_0}^{t'} \beta' (t'') dt'' \label {revol}
\ee
where the initial time $t'_0$ is considered to be the beginning of the collision. 

If the thickness of the active region is not negligible, as a first approximation, we can assume 
that it can be divided to layers with different characteristics which are in a quasi-steady state 
with respect to each others. In this case equations (\ref{enercons}) and (\ref{momcons}) can 
be written separately for each layer and the emission can be integrated. Note also that the 
thickness of the shell in these equations always appears as $n'_c \equiv n' \Delta r'$, i.e. the 
column density of the shell. Thus in the limit of shells with infinitesimal thicknesses we can 
simply replace $n' \Delta r'$ with $n'_c$ which is in addition a Lorentz invariant 
quantity.\footnote{During a collision the active region is essentially part of the overlap of two 
shells. In a radiative collision or even when the exchange of energy between shells is 
purely elastic, this region has necessarily a smaller Lorentz factor than yet uncollided part of 
the jet.}

After integration of equation (\ref{enercons}) we obtain the evolution equation of the column 
density of the active region:
\be
n'_c = \frac {N'_0 {r'}_0^3 ((\frac {r'}{{r'}_0})^{3-\kappa} - 1) + 
(3-\kappa) {r'}_0^2 n'_c({r'}_0) \gamma'_0 (1-\beta'_0)}{(3-\kappa) r'^2 \gamma' (1-\beta')} 
\label{columndenssol}
\ee
where $\gamma'_0 \equiv \gamma'(r'_0)$ and $\beta'_0$ is the corresponding $\beta'$. By solving 
this equation along with (\ref{momcons}) we obtain the evolution equations for $\gamma'$ and $n'_c$ 
with respect to radius or equivalently time.

The power of synchrotron radiation emitted by the active region is:
\bea
P'= \frac {dE'_{sy}}{dt'} = c\beta' \frac {dE'_{sy}}{dr'} & = & \frac {16\pi}{3} r'^2 
\Delta r' \sigma_T c \gamma'^2 \frac{{B'_\bot}^2}{8\pi} \int n'_e (\gamma_e) \gamma_e^2 d\gamma_e 
\label {synchpower}
\eea
where $n'_e$ is the number density of accelerated charged leptons - electrons and possibly 
positrons - with a Lorentz factor $\gamma_e \gg \gamma'$ with respect to slow shell frame. 
We assume that the thermal motion of charged particles is negligible with respect to their boost,  
thus, $B'_\bot$ is defined as the total magnetic field component perpendicular to the direction of 
accelerated electrons. ${B'_\bot}^2/8\pi$ is the magnetic energy density and $\sigma_T$ is the 
Thompson cross-section.

The normalization of the electrons distribution is defined as:
\bea
\int_{\gamma_m}^{\infty} n'_e (\gamma_e) d\gamma_e & = & n'_a \label {enum} \\
\int_{\gamma_m}^{\infty} \gamma_e n'_e (\gamma_e) d\gamma_e & = & 
\frac {\gamma'^2 m_p n'_0 \epsilon_e}{m_e} \label {eener}
\eea
where $n'_a$ is the number density of accelerated charged leptons and $\epsilon_e$ is the 
fraction of the kinetic energy of the falling baryons transferred to the accelerated leptons 
in the slow shell frame. Here we assume that it is equal to the number density of the slow 
shell accelerated to fast shell i.e. $n'_a = \gamma'n'_0$. Note also that (\ref{eener}) assumes 
an abrupt break in the electrons Lorentz factor distribution. This simplifying assumption has 
crucial consequences for the spectrum of synchrotron emission that we will describe in 
Sec. \ref{sec:lowener}.

From constraints (\ref{enum}) and (\ref{eener}) the minimum Lorentz factor of electrons 
${\gamma_m}$ and the normalization of $n'_e (\gamma_e)$ can be obtained. For a power-law 
distribution:
\bea
n'_e (\gamma_e) & = & N_e \biggl (\frac {\gamma_e}{\gamma_m}\biggr)^{-(p+1)} 
\mbox{for } \gamma_e \geqslant \gamma_m \label{nedistpow} \\
N_e & = & \frac {n'_a p}{\gamma_m} = \frac{p^2 m_e {n'}_a^2}{(p-1) \epsilon_e 
\gamma'^2 m_p n'_0} \label{neamp} \\
\gamma_m & = & \frac {(p-1)\epsilon_e \gamma'^2 m_p n'_0}{p m_e n'_a} \label{gammam}
\eea
In the simulations presented here we also consider the case of a distribution with an exponential 
cutoff at high energies:
\bea
n'_e (\gamma_e) & = & N_e \biggl (\frac {\gamma_e}{\gamma_m}\biggr)^{-(p+1)}\exp (-\frac{\gamma_e}
{\gamma_{cut}}) \quad , \quad \gamma_e \geqslant \gamma_m \label{nedistpowc} \\
\gamma_m & = & \frac {\epsilon_e \gamma'^2 m_p n'_0}{m_e n'_a} \times 
\frac {\frac {\gamma_m}{\gamma_{cut}}\Gamma (-p, \frac{\gamma_m}{\gamma_{cut}})}
{\Gamma (-p+1, \frac{\gamma_m}{\gamma_{cut}})} \label{gammamcut} \\
N_e & = & \frac{n'_a}{\gamma_m \biggl (\frac{\gamma_m}{\gamma_{cut}} \biggr )^p 
\Gamma (-p, \frac{\gamma_m}{\gamma_{cut}})} \quad , \quad \Gamma (a,x) \equiv 
\int_x^\infty dy~y^{a-1} e^{-y} \label{neampcut}
\eea
In simulations present in next sections we assume that $\gamma_m/\gamma_{cut}$ factor is constant 
during each regime. Therefore in equation (\ref{gammamcut}) this factor is a parameter rather 
than a variable. From properties of incomplete Gamma function we find that 
when $\gamma_m/\gamma_{cut} \rightarrow 0$, the solutions for $\gamma_m$ and $N_e$ are 
approximately equations (\ref{gammamcut}) and (\ref{neamp}). We also use a broken power-low 
distribution for electrons in the simulations:
\bea
&& n'_e = 
\begin {cases} N_e \biggl (\frac {\gamma_e}{\gamma_m}\biggr)^{-(p_1+1)} & 
\gamma_m \leqslant \gamma_e \leqslant \gamma_{cut}\\ 
N'_e \biggl (\frac {\gamma_e}{\gamma_m}\biggr)^{-(p_2+1)} & \gamma_e > \gamma_{cut}, \quad 
N'_e = N_e \biggl (\frac {\gamma_m}{\gamma_{cut}}\biggr)^{p_1 - p_2}
\end{cases} \label{nedistpowbr} \\
&& N_e = \frac{n'_a p_1}{\gamma_m \biggl (1 - (1-\frac{p_1}{p_2})
(\frac{\gamma_m}{\gamma_{cut}})^{p_1}\biggr )} \label{nepowbr} \\
&& \gamma_m = \frac {(p_1-1)\epsilon_e \gamma'^2 m_p n'_0}{p_1 m_e n'_a} \times 
\frac{1 - (1-\frac{p_1}{p_2})(\frac{\gamma_m}{\gamma_{cut}})^{p_1}}{1 - (1-\frac{p_1 - 1}{p_2 - 1})
(\frac{\gamma_m}{\gamma_{cut}})^{p_1 - 1}} \label{gammampowbr}
\eea

In absence of an external magnetic field the total field $\mathbf {\vec {B}'}$ is equal to 
the magnetic field induced by turbulence and Weibel instabilities~\cite{weibel} in the shocked 
material (the active region) $\mathbf {{\vec {B}'}_{in}}$. Its direction is almost perpendicular 
to the direction of motion of charged particles i.e. to the surface created by the shock 
front~\cite{fermiaccspec1}.\footnote{In reality electrons have a helical movement in the direction 
parallel to the jet axis~\cite{montecarlo0}, but we neglect these details in the simple model 
presented here.} This means that $\mathbf {{\vec {B}'}_{in\bot}} 
\approx \mathbf {{\vec {B}'}_{in}}$. The amplitude of $\mathbf {{\vec {B}'}_{in}}$ is expected 
to be proportional to the energy density of particles falling into the shock front and can be 
parametrized as:
\be
\frac {B_{in}'^2}{8\pi} = \epsilon_B c^2 {\gamma'}^2 m_p n'_0 \label {magener}
\ee
where $\epsilon_B$ is the fraction of protons energy transferred to the induced magnetic field. 

In presence of an external magnetic field the total field is $\mathbf {\vec {B}'} = 
\mathbf {{\vec {B}'}_{in}} + \mathbf {{\vec {B}'}_{ex}}$. In this case the amplitude of the 
magnetic field in the shock front surface is:
\be 
{B'}^2 = {B'}_{in}^2 + {B'}_{ex}^2 + 2 \mathbf {{\vec {B}'}_{in}} \cdot 
\mathbf {{\vec {B}'}_{ex}} \label{magmag}
\ee
If the external magnetic field precesses around a fixed axis, both its amplitude and its angle 
with the shock front surface $\Sigma$ are time dependent and periodic. On the other hand, 
$\mathbf {{\vec {B}'}_{in}}$ is a circularly polarized random field across the shocked zone 
(see e.g. both 1D and 3D simulations in~\cite{fermiaccspec1}), and ${B'}_{in}$ presents the mean 
value of its amplitude. Its projection on $\Sigma$ is also random 
with a uniform distribution if the ejecta has a cylindrical symmetry - the case we consider here. 
Therefore, the angle between $\mathbf {{\vec {B}'}_{in}}$ and $\mathbf {{\vec {B}'}_{ex}}$ is 
random. If its random variation takes place much faster than the precession of the external field, 
the average scalar product of 
$\langle \mathbf {{\vec {B}'}_{in}} \cdot \mathbf {{\vec {B}'}_{ex}}\rangle = 0$. 
Therefore, for determining the total power of synchrotron emission and the spectrum we only need 
to determine the projection of the precessing field on the shock front surface. The details of 
the calculation can be found in Appendix \ref{app:b}.

From dynamic equations (\ref{enercons}) and (\ref{momcons}), the solution obtained for the column 
density $n'_c$ (\ref{columndenssol}), and the relation between synchrotron power and the 
amplitude of the magnetic field (\ref{synchpower}), we obtain the following equation for the 
evolution of $\beta'$:
\bea
&& \hspace{-2cm}\frac {d}{d\biggl(\frac{r'}{r'_0} \biggr )}\biggl [\frac {(\frac {r'}
{{r'}_0})^{3-\kappa} - 1 + \frac {(3-\kappa) n'(r'_0)\Delta r' 
(r'_0)}{n'_0 r'_0}~\gamma'_0 (1-\beta'_0)) \beta'}{(3-\kappa) (1-\beta')} 
\biggr ] = -\frac {{\mathcal A} \gamma'^7 \Delta r'}{\beta' \Delta r' 
(r'_0)} \biggl (\frac{r'}{r'_0}\biggr )^{2 -\eta} - \frac {{\mathcal A}_1 \gamma'^5 \Delta r'}
{\beta' \Delta r' (r'_0)} \biggl (\frac{r'}{r'_0}\biggr )^{2 -\eta_1} \label {synchcoupling} \\
&& {\mathcal A} \equiv \frac{4\alpha m_p^2 \sigma_T n'_0 \Delta r' (r'_0) \epsilon_e^2 (r'_0) 
\epsilon_B (r'_0)}{3 m_e^2} \label{biga} \\
&& {\mathcal A}_1 \equiv \frac{\alpha m_p \sigma_T \Delta r' (r'_0) \epsilon_e^2 (r'_0) 
{B'}_{ex\bot}^2 (r'_0)}{6\pi c^2 m_e^2} \label{bigap}
\eea
where $\eta \equiv 2\alpha_e + \alpha_B + 2\kappa$ and $\eta_1 \equiv 2\alpha_e + \kappa + 
\alpha_x$, where $\alpha_e$, $\alpha_B$, and $\alpha_x$ are respectively power-law indices of 
$\epsilon_e$, $\epsilon_B$, and external magnetic $B_{ex}$ defined as $x \propto (r/r_0)^{-\alpha}$ 
for each of them.

Note that the two terms in the right hand side of 
(\ref{synchcoupling}) have very similar forms. Although this differential equation is first order 
and can be solved easily, it is also highly nonlinear. Indeed, the solution includes a 
polynomial of order 6 of $\beta'$ and a term proportional to $(1-\beta')^{1/2}$. Such an 
algebraic equation must be solved numerically at each point and is not suitable for numerical 
simulations. Another approach - which is explained in 
details in~\cite{hourigrb} - is a perturbative/iterative approximation. It consists of the 
integration of right hand side of (\ref{synchcoupling}) by replacing $\beta'$ with its order 
$n-1$ approximate solution to obtain its order $n$. The solution of order zero corresponds to 
absence of synchrotron emission, i.e. the right hand side of (\ref{synchcoupling}) is set to 
zero. The general solution at order $n$ has the following form:
\bea
&& \beta'_{(n)} = \frac {{\mathcal D} - \frac {{\mathcal A}{\mathcal M}_{(n-1)}(r') + 
{\mathcal A}_1{\mathcal N}_{(n-1)}(r')}{\Delta r' (r'_0)}}{\frac {1}{3-\kappa}
((\frac {r'}{{r'}_0})^{3-\kappa} - 1) + \frac {{\mathcal D}}{\beta'_0} - 
\frac {{\mathcal A}{\mathcal M}_{(n-1)}(r') + {\mathcal A}_1{\mathcal N}_{(n-1)}(r')}
{\Delta r' (r'_0)}} \label {betan} \\
&& {\mathcal M}_{(n-1)}(r') = \frac {1}{\Delta r' (r'_0)} \int_1^{\frac{r'}{r'_0}} 
\frac {{\gamma'}_{(n-1)}^7 \Delta r'}{\beta'_{(n-1)}} x^{2 -\eta}dx \label{mmath} \\
&& {\mathcal N}_{(n-1)}(r') = \frac {1}{\Delta r' (r'_0)} \int_1^{\frac{r'}
{r'_0}}\frac {{\gamma'}_{(n-1)}^5 \Delta r'}{\beta'_{(n-1)}} x^{2 -\eta_1}dx \label{nmath} \\
&& {\mathcal D} \equiv \frac {n' (r'_0) \Delta r' (r'_0) \beta'_0 \gamma'_0}
{n'_0 r'_0}  \label{dconst}
\eea
The advantage of this approach is that one can use the approximate analytical solution to study the 
effect of various parameters and quantities on the evolution of dynamics of the ejecta and its 
synchrotron emission. To proceed, we have to consider a model for the evolution of $\Delta r'(r')$ 
which cannot be obtained from the simple formulation of relativistic shocks presented here. We 
consider the following phenomenological models:
\bea
&& \Delta r' = \Delta r'_{\infty} \bigg [1-\biggl (\frac{r'}{r'_0} \biggr )^
{-\delta}\biggr ] \Theta (r'-r'_0) \quad \text {Steady state model} \label {drquasi} \\
&& \Delta r' = \Delta r_{\infty} \bigg [1- \exp (-\delta' \frac{r'}{r'_0}) \biggr ] 
\Theta (r-r'_0) \quad \text {Exponential model} \label {expon} \\
&& \Delta r' = \Delta r'_0 \biggl (\frac {\gamma'_0 \beta'}{\beta'_0 \gamma'} 
\biggr )^{\tau}\Theta (r'-r'_0) \quad \text {dynamical model} \label {drdyn} \\
&& \Delta r' = \Delta r'_0 \biggl (\frac{r'}{r'_0} \biggr )^{-\delta} 
\Theta (r'-r'_0) \quad \text {Power-law model} \label {drquasiend} \\
&& \Delta r' = \Delta r'_0 \exp \biggl (-\delta'\frac{r'}{r'_0} \biggr )
\Theta (r'-r'_0) \quad \text {Exponential decay model} \label {expodecay}
\eea
The initial width $\Delta r'(r'_0)$ in the first two models is zero, therefore they are suitable 
for description of initial formation of an active region in internal or external shocks. The last 
three models are suitable for describing more moderate growth or decline of the active region. 
The integration of (\ref{mmath}) and (\ref{nmath}) for models containing power-law terms can be 
obtained from the solution for dynamical model. The results for ${\mathcal M}_0(r')$ is presented 
in~\cite{hourigrb} and for (\ref{nmath}) can be found in Appendix \ref{app:a}. For models 
containing an exponential term analytical calculation is too complex and only numerical integration 
is used in our studies.  

For the simulations described in the next section we have tried both approximated analytical 
solutions for ${\mathcal M}_0(r')$ and ${\mathcal N}_0(r')$ - in the cases for which an analytical 
solution is available, and numerical integration of (\ref{mmath}) and (\ref{nmath}). We have also 
used the exact solution of differential equation (\ref{synchcoupling}) and its linearization to 
find an approximate solution for $\beta'$. The results of these 3 methods differs by a few 
percents. Therefore, considering other simplifications and uncertainties in this model, it is not 
important which one of them is used. However, in practice it is easier to use the approximate 
solution (\ref{betan}) with numerical integration of integrals because this procedure can be 
applied to all models for $\Delta r'(r')$ and new models can be added to simulation code without 
additional analytical calculations.

The next step is the determination of the synchrotron emission spectrum and 
flux. In~\cite{hourigrb} (and references therein) it is shown that energy (intensity) angular 
spectrum of synchrotron depends on $B'_\bot$ the transverse component of the field with respect 
to the direction of motion of electrons only through the dependence on the synchrotron 
characteristic frequency:
\be
\omega'_c\equiv \frac {3e \gamma_e^2 B'_\bot}{2c m_e} \label{syncchar}
\ee
Therefore, when an external magnetic field is present, the formulation developed in~\cite{hourigrb} 
can be used without any modification. We only need to consider the total field in the determination 
of $\omega'_c$. For the sake of completeness we repeat here the final expression obtained for the 
synchrotron emission in~\cite{hourigrb}:
\be
\frac {dP}{\omega d\omega} = \frac{4\sqrt {3} e^2}{3\pi} r^2 \frac{\Delta r}{\Gamma (r)} 
\int_{\gamma_m}^\infty d\gamma_e n'_e (\gamma_e)\gamma_e^{-2} K_{2/3} (\frac{\omega'}
{\omega'_c}) + {\mathcal F}(\omega, r) \label{powerdopcorr}
\ee
where ${\mathcal F}(\omega, r)$ includes subdominant terms and terms depending on the curvature 
of emission surface. In~\cite{hourigrb} it is argued that these terms are much smaller 
than the dominant term, thus we neglect them in the simulations. The slow shell Lorentz factor 
$\Gamma (r)$ can be expressed as:
\be
\Gamma (r) = \Gamma_f \gamma' (r)(1 + \beta_f \beta' (r)) \label {gammabulk}
\ee
where $\Gamma_f$ is the Lorentz factor of coalesced shells with respect to far observer at the 
same redshift. Note that synchrotron emission spectrum in (\ref{powerdopcorr}) is integrated 
with respect to to angular distribution. The description of angular dependent spectrum can be 
found in~\cite{hourigrb}, equations (56) and (60) for slow shell and observer rest frame 
respectively. We use (\ref{powerdopcorr}) in our numerical simulation rather than angular 
dependent spectrum because it makes the comparison with observations easier. However one should 
remember that the total emission energy obtained from this formula is $E_{iso}$ and one has to 
take into account the effect of relativistic beaming of the emission in the rest frame of the 
observer if one wants to compare the energy output with the initial kinetic energy of the ejecta, 
see Sec. \ref{sec:efficiency}

\section {Simulation of light curves and spectra of GRBs in presence of 
an external magnetic field}\label{sec:simul}
In this section we present light curves and spectrum of simulated GRBs according to the simplified 
relativistic shock model introduced in Sec. \ref{sec:shock}. They are more complete and realistic 
than first simulations of this model in~\cite{hourigrb1}. Not only they include the effect of an 
external magnetic field, but more importantly they present the whole prompt emission during the 
time that a typical burst is observable. 

Each simulated burst consists of few {\it regimes} (at least 3): Beginning of the formation of 
an active region, the main emission, and a tail during which the emission becomes softer and 
fades off. Each regime corresponds to one of the models (\ref{drquasi})-(\ref{drquasiend}) for 
the evolution of width of the active region. All simulated bursts begin with steady state or 
exponential models because their initial width is zero. Some of the other parameters, specially 
indices parametrizing the evolution of parameters also can change with regime, but continuity 
conditions should be respected, see Sec \ref{sec:continu}.

An important issue to be considered in numerical calculation of the synchrotron flux is the 
integration over Lorentz factor distribution of accelerated electrons in (\ref{powerdopcorr}). 
The upper limit of this integral is $\infty$, but in numerical calculations we must stop 
integration at a finite upper limit. It is easy to see that after a variable change the argument of 
the Bessel function in (\ref{powerdopcorr}) becomes: 
\be
y \equiv \omega'/x^2 \omega'_m \quad ,\quad \omega'_m \equiv \omega'_c|_{\gamma_e = \gamma_m} 
\quad ,\quad x \equiv \gamma_e/\gamma_m \geqslant 1 \label{intparam}
\ee
For $\gamma_e\rightarrow \infty$, $x\rightarrow \infty$ and $K_{2/3}(y) \rightarrow \infty$. 
Therefore, we must choose the upper limit of numerical integration such that at high energies 
where $\omega'/\omega'_m \gg 1$, the variable $x$ take much higher values such that 
$y \rightarrow 0$, otherwise the flux at high energies is under-estimated. For this 
reason we choose an energy dependent upper limit, $x_{max} = 200 \omega'/\omega'_m$. Evidently, when 
electrons distribution has an exponential cutoff, it automatically reduces the contribution of 
high $\gamma_e$ in the integral. Nonetheless, the above argument stays valid because the 
decreasing exponential and increasing Bessel function at $y \rightarrow 0$ cancel each others 
effect.

\subsection {Continuity conditions} \label{sec:continu}
To have a smooth transition from one regime to another - the necessary condition for having 
continuous light curves - we have to respect some continuity conditions. Table \ref{tab:param} 
summarizes parameters of the simulated models. In each 
simulation all the relevant parameters should be given as initial values for the first regime. 
In the following regimes some of the parameters can be changed. They include the exponents/indices 
that determine the evolution of quantities such as $\alpha_B$ and $\alpha_e$ that respectively 
parametrize the evolution of the fractions of kinetic energy transferred to electric and magnetic 
fields $\epsilon_e$ and $\epsilon_B$, $\alpha_x$ the index determining variation with distance of 
the external magnetic field, $\alpha_p$ index of power-law factor in electrons energy distribution, 
$\kappa$ index parametrizing the distance dependence of slow shell density, $\gamma_{cut}$ the 
cutoff Lorentz factor of electrons when a power-law with exponential cutoff or broken power-law 
distributions is considered, and finally $\tau$, $\delta$, and $\delta'$ that parametrize 
the evolution of active region width. As we explained before, the latter cannot be determined from 
first principles in this simplified model of relativistic shocks and needs input from detailed 
simulation of microphysics such as PIC simulations. 

Initial values of other quantities including 
the distance to the central engine $r_0$, quantities that determine the kinematic of the shells 
such as $\gamma' (r'_0)$, $\Gamma (r_0)$ and their corresponding $beta$, the column density of 
the fast shell, the density of the slow shell, and the amplitude of the external magnetic field 
must be continuously transferred from one regime to the next. This means that their initial 
values in a new regime must correspond to their values at the end of the previous regime. Other 
quantities including the final $\Gamma$ when shells are coalesced, the phase, and frequency of 
the external magnetic field stay the same in all regimes by definition. Note that although for 
the sake of simplicity we assume a constant precession period for the central engine and the 
external magnetic field, in reality they can change due to the slow-down of the engine and 
deformation of disk or envelop material that create the field. Nonetheless, if the slow-down 
time scale is much longer than burst duration this should be a valid assumption. If the engine 
collapses to a long life or a temporary magnetar before collapsing to a black hole, it is expected 
that its rotation rate and ejection of material continues roughly steadily for hundreds 
of seconds~\cite{grbmagform}.

In addition to the matching of initial values, we must be also careful and use correct initial 
values for $r'_0$, $\beta'_0$, etc. which are important for the continuity of fluxes. We should 
remind that evolution of physical quantities with distance/time in the model discussed here 
depend always on $r'/r'_0$ where $r'_0$ is the distance from the engine at the beginning of the 
regime - not at the beginning of the simulation. Therefore, in each regime we must use the 
initial value of $r'$, $\beta'$, etc. of that regime in (\ref{betan})-(\ref{nmath}) and 
(\ref{drquasi})-(\ref{drquasiend}). 

\subsection {Processes and parameters}\label{sec:proc}
These simulations assume that the main energy dissipation process is synchrotron. According to 
previous analysis and simulations~\cite{shocksynch,shocksynch1} this is 
a reasonable assumption specially at middle range of energies. Compton scattering between high 
energy electrons and photons can be significant at very high or low energies, see next section. 
$p-\gamma$ scattering has been already shown to be insignificant and therefore its effect on the 
dynamics of the shock can be neglected~\cite{protonscatter,highener080916c0,highener080916c1,highener080916c2}. $\gamma-\gamma \longrightarrow e^-e^+$ can 
only affect photons of $E \gtrsim 5 \times 10^5 \Gamma \sim 5 \times 10^8$ eV. Moreover, most 
synchrotron photons are emitted in the direction perpendicular to the shock front and therefore 
are very close to collinear with each other. In this case the Center of Mass (CM) energy of photon 
pairs is very small, thus they cannot annihilate each other to $e^\pm$. The situation is very 
different for a standard fireball model in which released photons have a high energy thermal 
distribution and their direction is random. As for the synchrotron self-absorption, it is only 
important at very low energies which are not crucial for the prompt emission.

Table \ref{tab:param} summarizes the parameters and initial values of physical quantities for the 
simulations presented in this section. Based on physical expectations not all of the parameters 
used in this model 
are independent. For instance, density, relative Lorentz factor of shells, and the fractions of 
kinetic energy transferred to electric and magnetic fields must be somehow related. However, the 
complexity and precision of plasma physics simulations do not yet allow a detailed exploration 
of parameter space to understand the underlying correlation between physical properties. 
Nonetheless, some of the relations between parameters are more evident. For instance, there is a 
degeneracy between the distance of the shock from the central engine and densities or column 
densities of shells that produce a given synchrotron flux, because due to adiabatic 
expansion the density of a shell decreases as $r^2$ even in absence of any energy dissipation. 
Therefore, larger the distance to the central engine, smaller the density of shells to generate 
the same flux. Other simple and obvious relations are: Emission hardness defined by $\omega'_m$ is 
proportional to the magnetic field; Strength of the shock depends inversely on the quantity 
${\mathcal D}$ defined in (\ref{dconst}); Flux of synchrotron emission is proportional to the 
fraction of kinetic energy transferred to electrons; Duration of a burst depends on how quickly 
the induced magnetic field and accelerated electrons - equivalently electric field - approach to 
their maximum value and then decrease. We use these relations to {\it design} simulated bursts i.e. 
select initial value of parameters. Another guide in the selection of plausible values are 
observations. For instance, densities of shells are chosen based on the density of massive stars 
envelops, indices based on observations of GRBs, etc. Despite these relations, due to the large 
number of parameters in Table \ref{tab:param}, a systematic exploration of the parameter space is 
not simple. Nonetheless, the consistency of simulations with observations significantly reduce 
the range of parameters. 

In the Introduction we raised the issue of radiation absorption inside the shocks. Here we address 
it in more details. High energy electrons are mostly ejected in front of the shock in slow 
shell~\cite{fermiaccspec,fermiaccspec1}. 
Assuming that the width of slow shell is roughly equal to the width of active region, the column 
density traversed by photons for the densest shells considered in these simulations is 
$N'_0 \sim 10^{15}$ cm$^{-2}$ is $\sim 10^{24}$ cm$^{-2} / \gamma'$ where $\gamma'$ is the Lorentz 
factor of the fast shell with respect to slow shell. The characteristic attenuation column density 
$\lambda$ for photons with energies in the range of $10-100$ keV is $\lambda \sim 3-4$ gr cm$^{-2}$ 
or equivalently $\lambda \sim 2.5 \times 10^{24}$ cm$^{-2}$ and larger at higher 
energies~\cite{particledata}. Therefore even for  
these high density shells, photons absorption inside the emission region is small $\lesssim 25\%$. 
In the majority of our simulations the density of slow shell is much smaller and absorption is 
completely negligible. Evidently, the possibility of absorption by other material surrounding the 
central source cannot be ruled out. This issue is true irrespective of the distance to the central 
source. But, no strong absorption was detected in the GRB prompt emission. This is the evidence that 
surroundings of the source is cleared before GRB-making shocks occur - most probably by earlier 
ejecta. In fact the detection of precursors before the main peak and flares - softer 
emissions for thousands of seconds after gamma-ray spikes - is the evidence that activities of the 
central source last for much longer time than its peak activity observed as a GRB. Previous ejecta 
which were too soft to emit a gamma-ray flash most probably clear the surrounding of the source 
such that when the most energetic shells are ejected, there is no intercepting material around to 
attenuate them. Finally we remind that in equation (\ref{powerdopcorr}) which describes the emission 
power, the value of $r,~\Delta r,~\Gamma (r)$, and $n'_e \propto N'_0$ are degenerate. Therefore, 
a larger distance to central source can be compensated by a smaller active region and lower 
densities for electrons. Thus, simulations presented here can be considered as representatives of a 
set of bursts with roughly similar $\chi \equiv r_0^2 \Delta r_0 n_0 \Gamma_f$. The degeneracy is 
not exact because at longer distances the external magnetic field and thereby $\omega_m$ are smaller. 
This leads to a softer emission. Nonetheless, it is always possible to adjust parameters to obtain a 
burst with the same characteristics occurring at larger distances. However, for longer distances from 
central engine, it would be more difficult - if not impossible - to explain the observed very short 
variabilities of emission. See also Sec. \ref{sec:efficien} for the issue of efficiency.

Although at ejection jets are expected to be along the magnetic field lines, if the central source 
precesses at later times their direction can deviate from each other. The precession is due to 
internal forces and independent of the properties and behaviour of the jet after its release. 
Inside Alfv\'en radius $R_A \sim R_{disk}$ acceleration and collimation is due to magneto-centrifugal 
forces that align the jet with the magnetic field~\cite{alfvenradius}. 
Outside this zone kinetic energy of the jet dominates and collimation is most probably due to 
a helical induced magnetic field~\cite{helicalmag,helicalmag1}. In this case the change of magnetic 
field direction does not significantly change the direction of ejected material, and in fact it 
helps its collimation by providing a transversal field component~\cite{helicalmag,helicalmag1}. 
Simulations show that progenitors of supernovae (collapsars) and their accretion disk indeed 
precess rapidly and irregularly~\cite{snbipolar}. Therefore it is expected that at the time of GRB 
formation a precessing magnetic field be present. Fig. \ref{fig:proj} shows precession axis and 
its projection on the shock front surface. $\Psi$ is the angle between shock front and precession 
axis and $\rho$ is the precession angle. They depend on the following quantities: direction and 
speed of the precession of the central source; precession of its magnetic field which can have 
an axis and period deviated from source ones; presence of a frozen field or a Poynting 
flow~\cite{diskmag1}; distance of the shock front from the central source; etc.. The nature and 
value as well as relation between these quantities are mostly unknowns. For this reason we have 
considered arbitrary values for $\Psi$ and $\rho$. Also in all simulations their values are the 
same such that we can compare light curves without the influence of these poorly understood 
quantities.

Apriori we do not have any rule for the duration of each regime. However, giving the fact that 
all evolving quantities depend on $r/r_0$ where $r_0$ is the initial radius of the current regime 
and the fact that $\beta \lesssim 1$, a given $r/r_0$ presents longer times for longer $r_0$. This 
gives a rule of thumb for choosing the duration of a regime e.g. the initial rise of $\epsilon_e$ 
and $\epsilon_B$, etc. Another rule of thumb is the fact that if quantities become unphysically 
large, the solution of equation (\ref{betan}) leads to $\beta \lesssim 0$ which is unphysical. 
This is an indicator that the value of evolving parameters are not anymore realistic. This seems 
very arbitrary, but sudden change of exponents in nonlinear systems most probably arise due to 
similar relations and internal feedback between quantities and conservation/dynamical constraints 
that control their behaviours. Evidently, the ultimate constraint for us is obtaining simulated 
bursts which look like real ones, thus some try and error is necessary at this level.

In~\cite{hourigrb1} we have presented plots of various physical quantities involved in the model. 
For this reason we do not repeat them here and only present light curves and spectra of some 
examples of long and short simulated bursts which can be directly compared with observed GRBs.

\subsection {Simulated bursts: Light curves}\label{sec:simulatedlc}
In this section we describe properties of light curves and spectra of simulated bursts. To 
simplify their comparison with each others and with real bursts we have transferred all long 
bursts to redshift $z=2$ and all short bursts to $z=0.5$ in a flat FLRW cosmology with 
$H_0 = 73$~\hubbleunit and $\Omega_{\Lambda} = 0.73$. 

\subsubsection{Oscillations} \label{sec:simuloscil}
\begin{figure*}
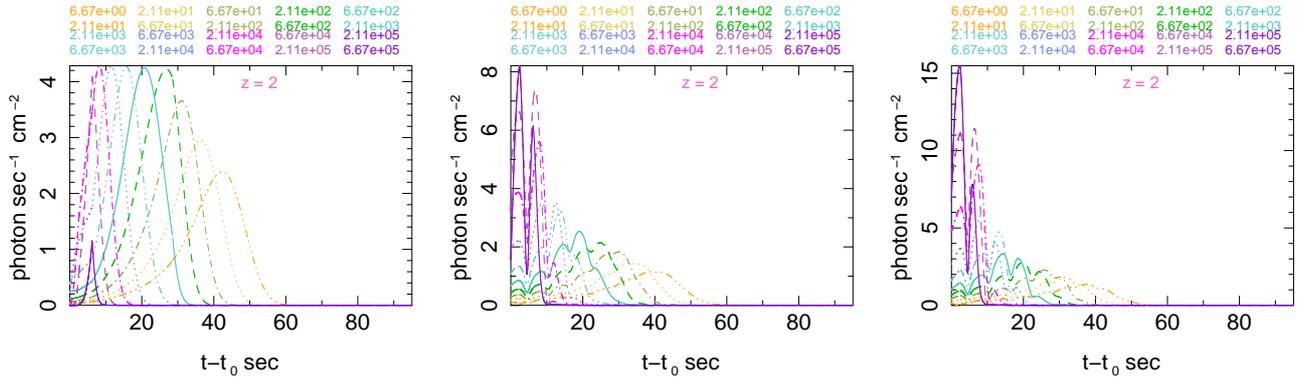

\begin{center}
\begin{tabular}{lll}
\includegraphics[width=5cm,angle=-90]{\plotdir/spec-m6p-3222-lin-lc-mag0-0.ps} &
\includegraphics[width=5cm,angle=-90]{\plotdir/spec-m6p-3222-lin-lc-mag200-05.ps} &
\includegraphics[width=5cm,angle=-90]{\plotdir/spec-m6p-3222-lin-lc-mag400-05.ps}
\end{tabular}
\caption{Simulation No. 1 with no external field (left)and with $|B| = 35$ kG (centre), $|B| = 70$ kG 
(right). In all plots of light curves energy bands are written on the top of the plot and the color 
of their fonts correspond to the color of light curves of the band.\label{fig:spikes}}
\end{center}
\end{figure*}

Between many simulations we have performed only a subset which are representative of the variety of 
observed bursts are discussed here. We remind that these simulations include only a single 
spike/peak of a burst. Well separated or even superimposed peaks in observed GRBs most 
probably originate from collision of shells which are separated in spacetime, or collision of 
successive shells with each others. In this case the accumulation of material in the slow-downed 
front shell leads to a stronger shock when lately emitted shells catch up and collide with it. 
Some of the brightest GRBs detected by the Swift Satellite such as GRB 060105~\cite{grb060105}, 
GRB 061007~\cite{grb061007,grb061007-1} and GRB 061121~\cite{grb061121} could be produced by 
such a process that we call {\it highway collision} because of its similarity to pile up 
collisions in highways. An analogue explanation is suggested for precursors that are seen in  
majority of bursts~\cite{shelldecel,precursdecel}. On the other hand, even single or well 
separated peaks include substructures. The simplest explanation for their presence is 
inhomogeneities in the ejecta density. Nonetheless, our simulations show that substructures or even 
separate spikes can be produced in presence of a precessing magnetic field. In fact nonlinearity 
of the dynamics, dissipation processes, and fast variation of all physical quantities including 
the magnetic field can produce resonance emission, suppression, and/or create rapid variations 
in the emission i.e. spiky light curves with an acyclic appearance. A good example of such cases 
is simulation No. 1. in Table \ref{tab:param}. Fig. \ref{fig:spikes} shows its simulated light curves 
without and with an external magnetic field. Without the presence of an external field as expected 
there is only one smooth peak in all energy channel.
Moreover, lags between soft gamma-ray channels are somehow larger than what is usually observed 
in long bursts, i.e. up to few hundreds of milliseconds. When a precessing magnetic field with 
precession period of $T = 2$ sec (or equivalently $f = 0.5$ Hz) in the rest frame of the engine 
is present, for the observer this burst includes 2 successive hard peaks and a third separated 
softer peak.\footnote{For the sake of simplicity in all of the simulations presented here the 
precession frequency is the same in all the regimes. In reality we expect a variation due to 
acceleration of the central source after accretion of material or its slow-down due to energy 
loss by gravitational wave radiation or other forms of dissipation.} There are also other peaks 
observable only in X-ray and softer channels. These peaks have different durations, and during 
about 12 sec that such a burst would be observable by a detector with sensitivity of the 
Swift-BAT, no periodic signature would be detectable in its PDS. Long 
observations in lower frequencies usually performed tens of seconds later i.e. at larger distances 
from central engine where the strength of the source field is significantly reduced. In fact, in 
this simulated burst there is practically no signature of oscillations in UV and optical because 
of the lag between their peaks. In addition, observations in these bands need at least few seconds 
of exposure and all data must be binned to remove/reduce the shot noise. This additionally smears 
the close to periodic behaviour of light curves. Note also that in presence of an external 
magnetic field the burst is harder, stronger and peaks have smaller lags. This burst looks like 
some of bright GRBs detected by Swift such as GRB 050525A and GRB 080409. In this simulation we 
have chosen relatively high densities to make a brightness comparable with very powerful GRBs 
such as GRB 080319B and GRB 080607. Evidently weaker bursts are easier to simulate and a number 
of examples will be shown later.

\begin{figure*}
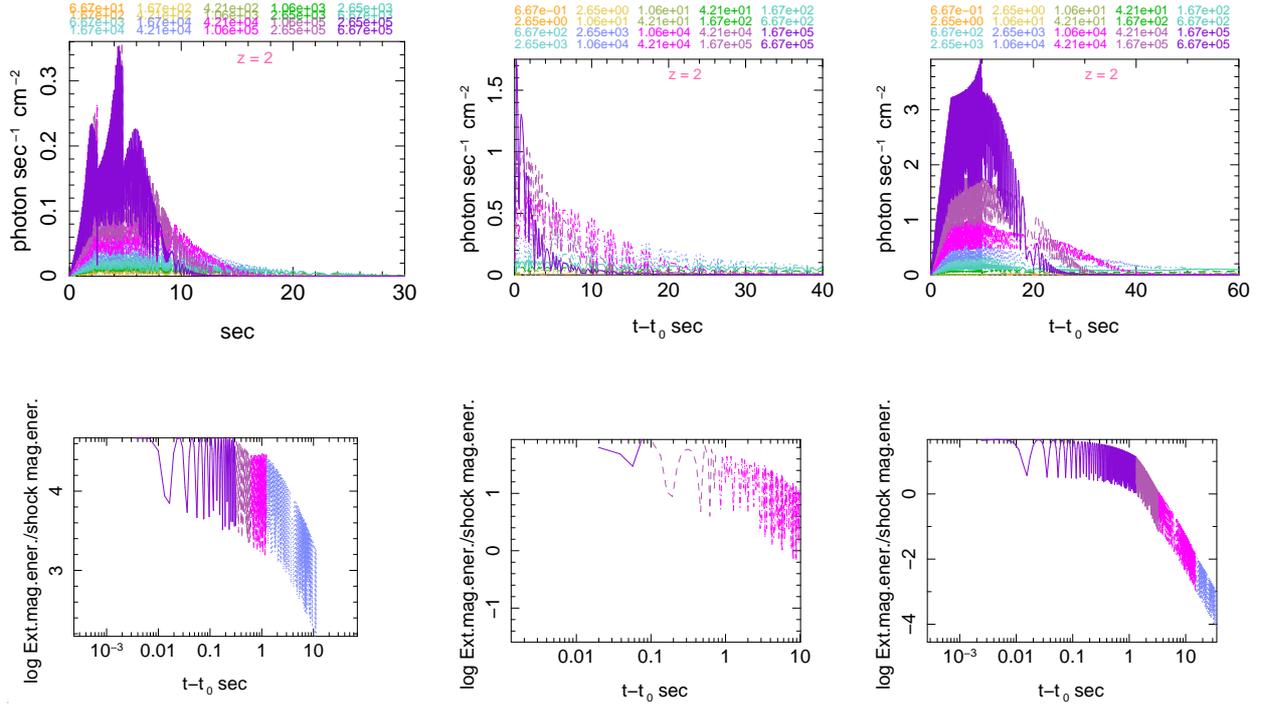

\begin{center}
\begin{tabular}{lll}
\includegraphics[width=4.5cm,angle=-90]{\plotdir/spec-m5-10222-lin-lc-600-50-gamma400.eps} &
\includegraphics[width=4.5cm,angle=-90]{\plotdir/spec-m2-1222-lin-lc-mag200-400.eps} &
\includegraphics[width=4.5cm,angle=-90]{\plotdir/spec-m6-1222-lin-lc-mag600-50.eps} \\
\includegraphics[width=4.5cm,angle=-90]{\plotdir/spec-m5-10222-param-600-50-gamma400.ps} &
\includegraphics[width=4.5cm,angle=-90]{\plotdir/spec-m2-1222-param-mag200-400.ps} &
\includegraphics[width=4.5cm,angle=-90]{\plotdir/spec-m6-1222-param-mag600-50.ps} 
\end{tabular}
\caption{Top: Three examples of simulated burst with millisecond precession period at observe 
redshift. Simulation No. 2, $f=50$ Hz, $\Gamma_f = 100$ (left), Simulation No. 3, $f=400$ Hz (centre), 
Simulation No. 3, $f=50$ Hz (right). Bottom: Ratio of external magnetic field amplitude to shock 
induced field for these simulated bursts. In these plots time is at the redshift of the source.
\label{fig:highf}}. 
\end{center}
\end{figure*}

\begin{figure*}
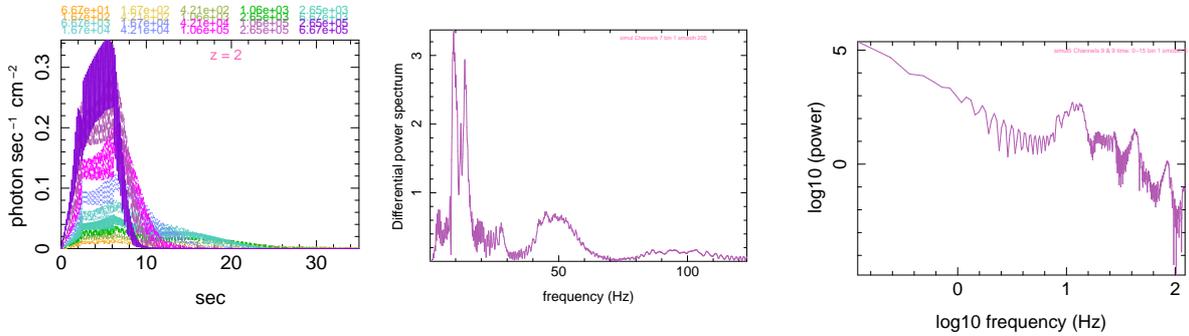

\begin{center}
\begin{tabular}{lll}
\includegraphics[width=4cm,angle=-90]{\plotdir/spec-m5p-10222-lin-lc-mag600-02-rot100.eps} &
\includegraphics[width=4cm,angle=-90]{\plotdir/spec-m5p-10222-lin-lc-mag600-02-rot100_1_10_1simul.ps} &
\includegraphics[width=4.5cm,angle=-90]{\plotdir/spec-m5p-10222-lin-lc-mag600-02-rot100_100_10_1simul-ch9.ps} 
\end{tabular}
\caption{Light curves (left), differential power spectrum (center), and power spectrum of simulation 
No. 5 with a two component external magnetic field with rest frame frequencies of $f_1 = 0.2$ Hz and 
$f_2 = 20$ Hz. Plots are at observer's redshift. Note that there is practically no signature of the 
long period magnetic field oscillations in PDS.\label{fig:twocomp}} 
\end{center}
\end{figure*}

GRBs seem to have fast varying prompt gamma-ray radiation - a jitter - up to shortest revolvable 
time scales. It is very difficult to know if this high frequency component is intrinsic or due 
to the Poisson noise in the arrival time of photons. Nonetheless, various origins have been proposed 
for this phenomenon~\cite{jitter}. Here we argue that the origin of high frequency variations can be 
the precession of magnetic field of a forming magnetar or the magnetic field of the accretion disk 
of the engine during GRB emission. Fig. \ref{fig:highf} shows the light curves of simulations 
with millisecond precession periods. 
It is evident that the precession of the field induces a fast varying component similar to the 
observations. Note that these light curves include substructures and overlapped peaks generated by 
variation of other quantities. Apriori both fast and slow varying magnetic fields can be 
simultaneously present. In fact it is well known that the differential rotation, 
turbulence, and instabilities in the accretion disk amplifies the seed magnetic field which in 
its turn produces a jet (see e.g. ~\cite{diskmag1,diskmag}). This field can be the source for 
a slow varying component, coexisting with a fast varying component from e.g. the central source. 
There can be also a magnetic field frozen in the outer layers - convective layers - of the 
progenitor. In this case the total magnetic field that affects accelerated 
electrons in the shock is the sum of all these components. Fig. \ref{fig:twocomp} shows an example 
of such situation. The external magnetic field energy is equally distributed in a slow component  
$f_1 = 0.2$ Hz and a fast component of $f_2 = 20$ Hz. The signature of the fast component and its 
harmonics is clearly detectable in the PDS. Note that due to the dependence of $\omega_m$ on the 
amplitude of the magnetic field, oscillation frequency is two times precession frequency. 
Only a non-significant peak at expected frequency of slow oscillation is detectable in the 
differential PDS. Fig. \ref{fig:twocomp} shows that with available time resolution of detectors, 
even for bright bursts the high frequency part of PDS is dominated by Poisson noise and any 
signature of oscillation is smeared.
Fig. \ref{fig:highf} includes also the evolution of ratio of external field to shock induced 
field for each simulated burst. From these plots it is evident that weaker the shock, more important 
the role of an external field in synchrotron emission. They also show how the field geometry affects 
its contribution at large distances and reduces the probability of detecting oscillations in the 
prompt low energy bands emission. We must also remind that in all the simulations presented here 
the strength of the external magnetic energy is orders of magnitude smaller than the kinetic energy 
of matter, thus bursts are not Poynting energy dominated.

\begin{figure*}
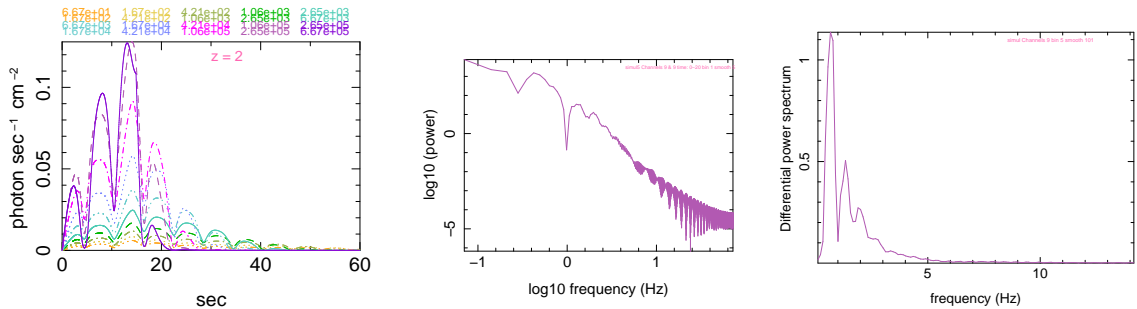

\begin{center}
\begin{tabular}{lll}
\includegraphics[width=4cm,angle=-90]{\plotdir/spec-m5-1022-log-lc-mag200-05.eps} &
\includegraphics[width=4cm,angle=-90]{\plotdir/spec-m5-1022-log-lc-mag200-05_100_10_1simul-ch9.ps} &
\includegraphics[width=4cm,angle=-90]{\plotdir/spec-m5-1022-log-lc-mag200-05_1_10_1simul.ps}
\end{tabular}
\caption{Light curves (left), differential power spectrum (center), and power spectrum of simulation 
No. 2 with rest frame frequency $f_1 = 0.5$ Hz. \label{fig:oscil}} 
\end{center}
\end{figure*}

To be able to detect periodic features, the length of the interval during which the prompt 
emission is observable must be long enough such that it covers a few cycles. An example of such 
bursts and their PDS is shown in Fig. \ref{fig:oscil}. 
The precession period of the external magnetic field in this simulation is not very different from 
previous examples but the burst is longer - about 34 sec in gamma-ray channels. The light curves 
in all channels contain more features and a close to periodic behaviour is evident even by eye. 
But, it is noticeable that even in this case peaks are not completely periodic. Therefore, if the 
detection noise is high, a wide peak in the PDS of the burst can be interpreted as the effect of a 
white noise. In fact even in the noiseless PDS shown in  Fig. \ref{fig:oscil} the peak due to the 
periodic component is not very significant. This explains the low significance of the peak in the 
PDS of GRB 090709A, if its origin is an external magnetic field.

In conclusion, only under special conditions a periodic variation in the light curve of a GRB is 
observable. Notably, when the precession of the external magnetic field is very fast - a situation 
which is expected if the burst occurs after the formation of a temporary or long-life millisecond 
neutron star/magnetar - the signature of fast oscillations even for very bright bursts is strongly 
smeared by the Poisson noise and binning. Nonetheless, the clear detection of a periodic component 
in the light curve of GRB 090709a proves that maybe in this source the burst was formed before the 
rotation of the central source accelerates by accretion of material from the disk. Alternatively 
this relatively slow component can be due to the magnetic field of outer part of the disk or the 
envelop of the star, or formation of a long range density oscillation in the disk due to the 
presence of a magnetic field, see Appendix \ref{app:c} for a brief review. As mentioned in 
Sec. \ref{sec:grb}, PDSs of many bursts seem to have weak periodic components, but due to the 
weakness of the signal and complexities described in the previous paragraph, these observations 
are not conclusive. In particular, as simulation examples here demonstrate, when burst duration 
is only few times the precession period, the effect of external magnetic field variation appears 
as substructures or simple peaks rather than coherent oscillations. This makes the detection of 
imprint of the precession very difficult.

We single out two reasons for the absence of periodic features. The first one is intrinsic to the 
source. If oscillations have high frequencies, they would be smeared by the Poisson noise. 
Moreover, the geometry of the field plays a very important role both in the strength of the emission 
and in presence of oscillations specially in low energy bands. For instance, if the field is 
confined in a torus around the source, it affects the emission only at very early times when 
the jet traverses the torus. The second reason is the nonlinear dissipation and evolution of 
microphysics. Variation of indices that model the evolution of quantities such as $\epsilon_e$, 
$\epsilon_B$, and densities along with nonlinearity of equations can produce sudden and resonant 
variations in the emission which breaks the coherence of slow precession of the magnetic field. 
Evidently, random variations of quantities due to turbulence in the jet environment also help to 
smear coherent oscillations.

As we explained above, in the simulations presented here indices vary discretely. Although the 
variation of these quantities may be more gradual in the real bursts - equivalent to having a large 
number of regimes each lasting for a very short time - it is well known that in nonlinear systems 
phase transitions are usually first order i.e. discontinuous. Therefore, discontinuous variation of 
quantities implemented here can be an enough close imitation of real bursts. Moreover, a close 
inspection of simulated light curves show that they do not produce the features explained above. 
In the next section we present more light curves in some of which the moment of regime change is 
visible and demonstrate that sudden changes in indices plays a minor role in creation of features.

\subsubsection{Phenomenological models} \label{sec:phenomen}
In contrast to~\cite{hourigrb1} simulations presented here cover the whole duration in which the 
prompt emission is observable by present instruments. Therefore, in this section we test the 
validity and differences of the phenomenological models (\ref{drquasi})-(\ref{expodecay}) for the 
evolution of active region in internal shocks.\footnote{We leave the study of the propagation 
of internal shocks remnant to large distances and formation of external shocks with 
ISM/circumburst material to a future work.} We also investigate the parameter space of the model, 
but due to its large volume we only present some examples and leave the systematic exploration and 
comparison with properties of candidate progenitors of gamma-ray burst to a future work.

\begin{figure*}
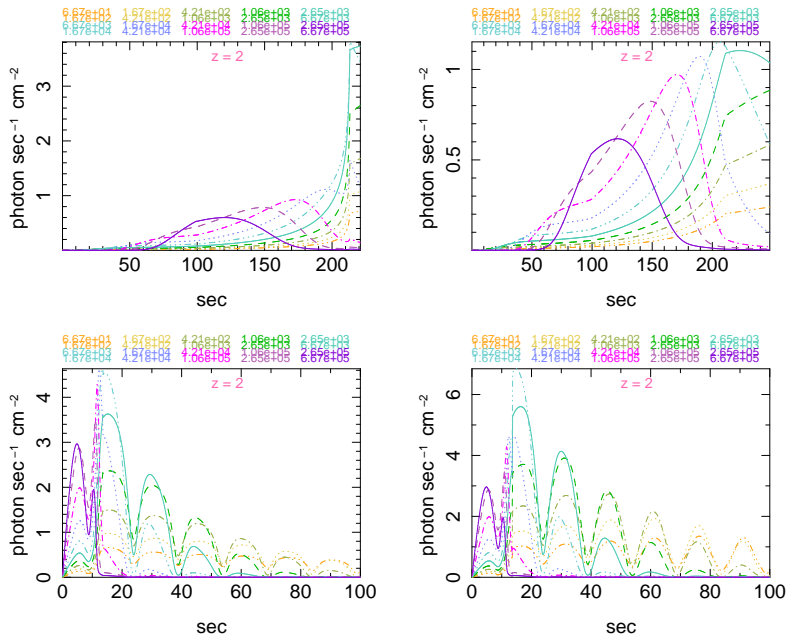

\begin{center}
\begin{tabular}{ll}
\includegraphics[width=4cm,angle=-90]{\plotdir/spec-m1p-102-lin-lc-mag0-0.eps} &
\includegraphics[width=4cm,angle=-90]{\plotdir/spec-m1p-122-lin-lc-mag0-0.eps} \\
\includegraphics[width=4cm,angle=-90]{\plotdir/spec-m1p-122-lin-lc-mag600-02.eps} &
\includegraphics[width=4cm,angle=-90]{\plotdir/spec-m1p-102-lin-lc-mag600-02.eps}
\end{tabular}
\caption{Simulation No. 7 without magnetic field (top row), with $|B| = 100$~kG, $f=0.2$ Hz 
(bottom row), middle regime dynamical (left column), middle regime steady state (right column). 
\label{fig:dynmodel}} 
\end{center}
\end{figure*}

\begin{figure*}
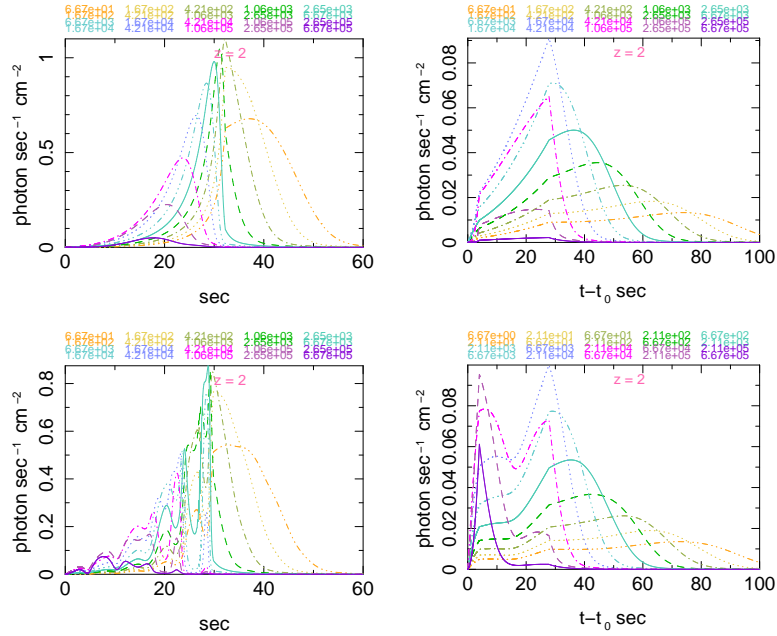

\begin{center}
\begin{tabular}{ll}
\includegraphics[width=4cm,angle=-90]{\plotdir/spec-m7-122-lin-lc-mag0-0.eps} &
\includegraphics[width=4cm,angle=-90]{\plotdir/spec-m2p-1222-lin-lc-mag0-0.eps} \\
\includegraphics[width=4cm,angle=-90]{\plotdir/spec-m7-122-lin-lc-mag100-05.eps} &
\includegraphics[width=4cm,angle=-90]{\plotdir/spec-m2p-1222-lin-lc-mag20-01.eps}
\end{tabular}
\caption{Simulation No. 8 without magnetic field (top left) and with $|B| = 12$~kG, $f=0.2$ Hz (top right).
 Simulation No. 9 without magnetic field (bottom left) and with $|B| = 2.5$~kG, $f=0.1$ Hz (bottom right) 
\label{fig:diverslc}} 
\end{center}
\end{figure*}

Fig. \ref{fig:dynmodel} shows light curves of 4 simulations. Each simulated burst includes 3 regimes. 
It is clear that when the middle regime is dynamical, the lags between gamma-ray bands are somehow  
large and peaks are softer. In absence of an external magnetic field irrespective of the 
middle regime the lags are too large to be consistent with observations. This proves that an 
external magnetic field can be crucial for generating bursts with observed properties. 
Nonetheless, considering the light curves in Fig. \ref{fig:spikes} this is not always the case and 
other physical properties of the shock are also important. This means that one cannot claim that the 
presence of an external magnetic field is a necessary condition for GRBs. In fact these simulations 
present very hard and bright bursts. Fig. \ref{fig:diverslc} shows examples of less bright and softer 
bursts with and without the presence of an external magnetic field. In this figure simulations 
without an external magnetic field are consistent with observations. The case of the simulation in 
bottom-left plot of this figure is interesting. It is soft and consistent with late soft peaks 
detected in many bursts. Delayed bright peaks in X-ray bands can be detected as a bright X-ray 
flare without significant emission in gamma-ray. Many of this type of flares have been detected 
by the Swift-XRT instrument. In the bottom-right simulation rapid slope changes due to regime 
change are clearly visible and show that they are not responsible for the formation of features. 
We find that although the lags between various energy bands in these simulations are consistent with 
observations, they are systematically larger. This means that at the beginning of the shock when it 
has its maximum strength, the synchrotron emission is hard and only when the shock is evolved and 
becomes softer, the emission in low energy bands becomes significant. This issue is directly related 
to the steeper spectral slope of simulated spectra in the framework of synchrotron model. In 
Sec. \ref{sec:lowener} we discuss this issue in details and relate it to the abrupt break in the 
distribution of Lorentz factor of electrons considered here.

\begin{figure*}
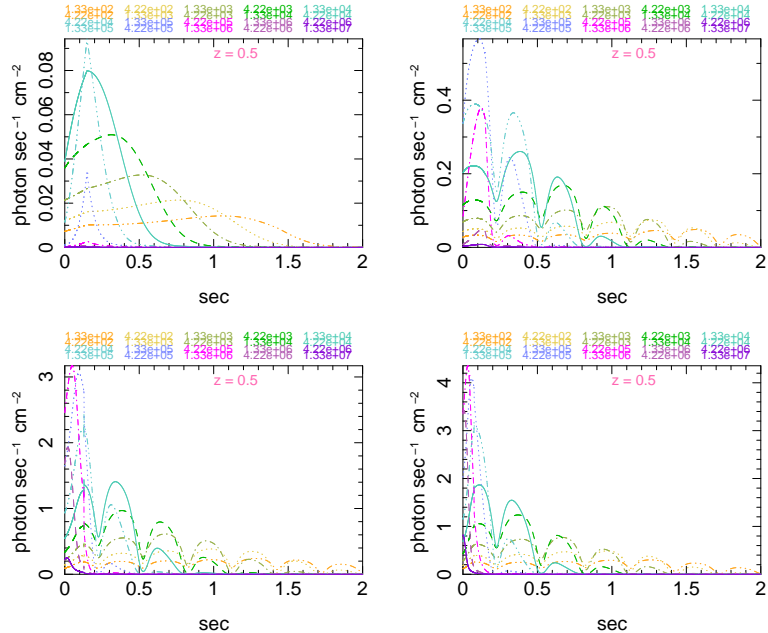

\begin{center}
\begin{tabular}{ll}
\includegraphics[width=4cm,angle=-90]{\plotdir/spec-m3-322-lin-lc-mag0-0.eps} &
\includegraphics[width=4cm,angle=-90]{\plotdir/spec-m3-322-lin-lc-mag60-5.eps} \\
\includegraphics[width=4cm,angle=-90]{\plotdir/spec-m3-322-lin-lc-mag400-5.eps} &
\includegraphics[width=4cm,angle=-90]{\plotdir/spec-m3-322-lin-lc-mag600-5.eps}
\end{tabular}
\caption{Simulation No. 10 without external magnetic field (top left), with $|B| = 10$~kG, $f=5$ Hz (top 
right), $|B| = 70$~kG (bottom left), $|B| = 100$~kG (bottom right). \label{fig:shortburst}} 
\end{center}
\end{figure*}

\begin{figure*}
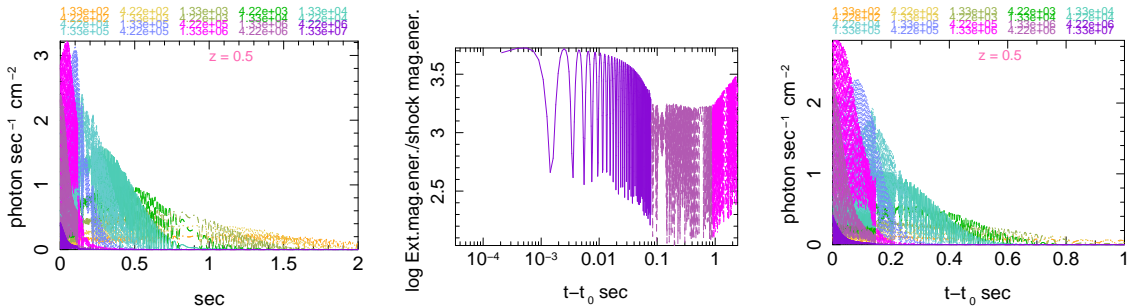

\begin{center}
\begin{tabular}{lll}
\includegraphics[width=4cm,angle=-90]{\plotdir/spec-m3-322-lin-lc-mag400-500.eps} &
\includegraphics[width=4cm,angle=-90]{\plotdir/spec-m3-322-param-mag400-500.ps} & 
\includegraphics[width=4cm,angle=-90]{\plotdir/spec-m3-322-lin-lc-mag400-500-index3.eps}
\end{tabular}
\caption{Light curves of simulation No. 10 with $|B| = 70$~kG, $f=500$ Hz (left); ratio of external 
to shock induced magnetic field (centre); light curves of the same model but with external magnetic 
evolving as $(r/r_0)^{-3}$ (right). \label{fig:shorthighf}} 
\end{center}
\end{figure*}

So far all the examples were long bursts. Short bursts are believed to be 
produced by the collision between two compact sources, e.g. 2 neutron stars, a neutron star and a 
black hole, etc. When one of the sources is a neutron star, one expects the presence of a strong 
magnetic field, specially at small distances from the central source. If the neutron star is young, 
the magnetic field on the surface of the star can be similar to magnetars $|B| \gtrsim 10^{15}$ Gauss. 
Even if the internal collisions occur at a distance of order $10^{10}$ cm and we assume that 
the magnetic field decreases as $r^{-3}$, the external magnetic field would be few kGauss 
at the shock site and can have significant effect on the synchrotron emission. 
Fig. \ref{fig:shortburst} shows the light curves of 4 short simulated bursts. All these bursts have 
the same parameters except for the strength of the external magnetic field. It is evident that 
stronger fields create both brighter and harder bursts. Indeed without an external magnetic, 
according to the model presented here, this burst would be very weak and too soft to be consistent 
with observations. Lags also are shorter for stronger and harder bursts, although in the case 
of this example lags are very close to zero even in absence of an external magnetic field. By 
contrast, the total duration of the burst does not significantly depends on the strength of 
the external field. Because the precession period of the external magnetic field in this burst 
is only few times smaller than its duration, cyclic behaviour can be observed in the softest 
channels if simultaneous observations of high and low energy channels could be performed. 
To be able to better understand the effect of the precession speed of the external field in the 
features of short bursts Fig. \ref {fig:shorthighf} shows a simulated burst with the same parameters 
as Fig. \ref{fig:shortburst} but a much faster precession $f = 500$ Hz. The comparison of light 
curves 
of these bursts show very little difference in the duration and strength of these burst, but when 
the precession is fast, the burst is slightly harder. We explain this observation by the fact that 
when the external magnetic field varies very quickly, the effect of having many field peaks at the 
initial part of the burst where it is harder leads to a slightly harder burst because a larger 
fraction of the available energy is emitted earlier, see the spectrum for each regime of this burst 
Fig. \ref{fig:diffp}. In Fig. \ref{fig:shortburst} we have also shown the ratio of the external to 
induced magnetic field. Given the fact that the distance of this burst from central engine as 
well as its duration are shorter than simulated bursts in Fig. \ref{fig:highf}, the external magnetic 
field is dominant in the whole duration of emission. Note also that in the middle of the burst 
when the shock is in its maximum strength, the contribution of external field decreases but it 
becomes dominant again close to the end when the shock becomes weaker. We also notice that the 
brightness and light curve of this simulation mimics closely the light curves of GRB 051221A the 
bright short GRB observed by the Swift-BAT.
As Table \ref{tab:param} shows, simulations of this burst and its variants are performed by 
assuming a relatively flat magnetic field which varies only as $(r/r_0)^{-1}$. To see how much 
the geometry of the field can affect a short burst we have also simulated the same burst with 
$(r/r_0)^{-3}$ during the whole burst. Fig. \ref{fig:shorthighf} shows also the light curves of this 
case. Their comparison shows that the burst becomes shorter and weaker. 

\begin{figure*}
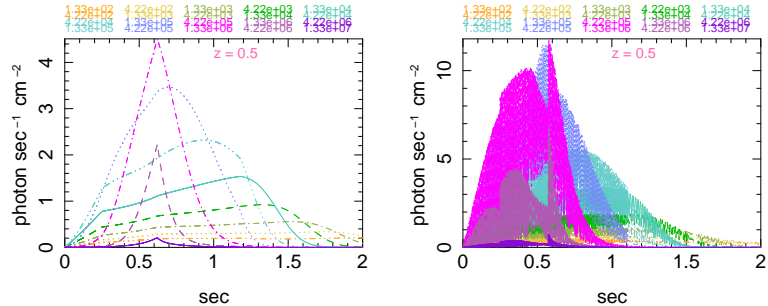

\begin{center}
\begin{tabular}{lll}
\includegraphics[width=4cm,angle=-90]{\plotdir/spec-m4-1022-lin-lc-mag0-0.eps} &
\includegraphics[width=4cm,angle=-90]{\plotdir/spec-m4-1022-lin-lc-mag200-500.eps} 
\end{tabular}
\caption{Light curves of simulation No. 4 without an external magnetic field (left) and with 
$|B| = 35$~kG, $f=500$ Hz (right). \label{fig:shorthf}} 
\end{center}
\end{figure*}

Fig. \ref{fig:shorthf} shows another example of a simulated short burst with and without an external 
magnetic field. In contrast to the previous example this burst is enough hard even in absence of an 
external magnetic field to be consistent with observations. However, without an external field 
lags between the softest channels in gamma-ray are large. As we mentioned above, this must be 
related to energy distribution of electrons. The main difference between this burst and the previous 
one is the number density of protons and electrons in both slow and fast shells. They are 10 times 
higher than in the previous example. Consequently, the emission without an external field is stronger 
and harder. The precession frequency considered here is very high and consistent with the precession 
of a millisecond pulsar. We note that there are substructures in the light curves that do not exist 
in absence of the external field. As we mentioned before they are formed by nonlinearities, 
resonance effect, and rapid change of the magnetic field and evolution of indices.

In these simulated short bursts the distance of shell collisions from central engine is at the 
shortest limit of the range given in the Introduction. Considering the large bulk Lorentz factor in 
these simulations, one obtains short variability time scales $\sim \Delta t \lesssim 10^{-5}$ sec 
for them. A simple calculation shows that in a few other simulations presented here such as 
simulations No. 2 
and 3 also $\Delta t$ is very short because although they have larger distances, they have also 
large Lorentz factors. Nonetheless, they both generate long bursts, see Fig. \ref{fig:highf}. In 
particular, if simulation No. 2 occurs at $z \sim 1$, its light curve looks like a number of 
bursts with a duration of $\sim 6-7$ seconds which, depending on their redshift, can be considered 
as long or short. The lag in this simulation is also consistent with observations of latter category 
i.e. longer than short bursts but shorter than typical long bursts. In simulation No. 3 lags are 
longer and consistent with normal long bursts. Therefore, variability is not a determining 
parameter for classifying the progenitor of bursts to collision of compact sources and collapsar. 
As we ignore any relation between the radius of accretion disk in these sources and Lorentz factor 
of shells, the only constraint which can be imposed is similarity of emission to observed GRBs. 
At present, time resolution and sensitivity of detectors do not allow to distinguish between shot 
noise and genuine variability, thus any constraint on the latter is purely hypothetical. Even 
the constraint discussed in the Introduction is just an upper limit because inhomogeneities in the 
accretion disk can be very close and next to each others. We should also 
remind that as we discussed in Sec. \ref{sec:proc}, there is a degeneracy between the distance of 
the shocks from central engine and their density/column density. Thus, one can obtain a similar burst 
at larger distances i.e. with longer time variability, but its efficiency would be smaller, 
see Sec. \ref{sec:efficien} for more details.

Considering the importance of an external magnetic field for synchrotron emission, the question 
arises whether a GRB can be formed without a shock. After all the internal shock model for GRB 
presumes shocks between shells that are accelerated to ultra-relativistic speeds by the engine 
through a mechanism which is not well understood but most probably is a combination of both 
mechanical and magnetohydrodynamical forces of an accretion disk. Due to the mass difference of 
electrons and protons it is expected that under the same force electrons are accelerated to much 
higher Lorentz factors. Therefore in presence of a transverse magnetic field they lose their 
energy by synchrotron emission. On the other hand, even in absence of a large scale precessing 
field a helical field is believed to be necessary for the collimation of jet outside Alfv\'en 
radius where it does not anymore follow the parallel magnetic field~\cite{helicalmag1}. 
This field dissipates electrons energy through synchrotron emission, and indeed 
recently this emission has been detected in numerical simulations of AGNs jets~\cite{helicalsynch}. 
The emission is expected to be low energy - in radio band for AGNs jets with $\Gamma \sim 10$ and 
in IR/optical band for ultra-relativistic jets of GRBs with $\Gamma \sim 1000$, assuming the same 
strength for the helical field. Such radiation from GRBs has never been observed. The reason is 
most probably the absorption of this emission be the jet itself because such emission occurs along 
the jet. Relatively low energy photons are absorbed or scattered during their propagation through 
the large column density of the jet. Moreover, electrons lose their energy very quickly and in a 
flash. In a collisionless flow they are not reaccelerated, and are simply dragged forward by 
protons flow to which they are electrically coupled. In this case if at later times the external 
magnetic field changes its direction no accelerated electron is available. Only when electrons are 
reaccelerated in the shock, the external transversal field can become active again and dissipate 
electrons energy by synchrotron emission. Shocks accelerate electrons are to higher Lorentz factors 
and generate additional magnetic field, thus the synchrotron emission is in much higher energy bands. 
Moreover, a shock can sustain a continuous acceleration of electrons for a relatively long time by 
gradually transferring the kinetic energy of colliding baryons to electrons. Therefore, even in 
the situation in which the magnetic field made by the shock is subdominant, the presence of a shock 
is a necessary condition for the formation of a GRB. In Poynting flow models reconnection of 
magnetic lines plays the role of shocks for accelerating electron. But, as mentioned before, 
the long distance of the process from the engine and rarity of line connection disfavour this model.

\begin{figure*}
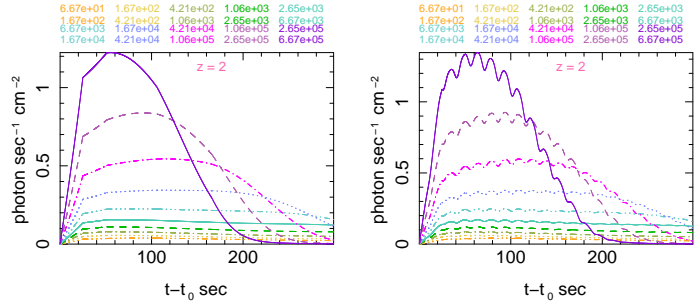

\begin{center}
\begin{tabular}{lll}
\includegraphics[width=4cm,angle=-90]{\plotdir/spec-m5pp-10222-lin-lc-mag0-0.ps} &
\includegraphics[width=4cm,angle=-90]{\plotdir/spec-m5pp-10222-lin-lc-mag06-02.ps} 
\end{tabular}
\caption{Light curves of simulation No. 11 without an external magnetic field (left) and with 
$|B| = 100$~G, $f=0.2$ Hz (right). \label{fig:largeepsilon}} 
\end{center}
\end{figure*}

Finally we notice that in all the simulations presented here the value of $\epsilon_e|_{max} 
\lesssim 0.15$. This is consistent with the values obtained in numerical simulations of electron 
acceleration in PIC~\cite{fermiaccspec} and Monte Carlo simulations~\cite{montecarlo0} 
(see~\cite{montecarlo1} for a review). On the other hand, this value is orders of magnitude 
larger than $\epsilon_e|_{max} \sim 10^{-5}$ found in some modification of basic SSC model to explain 
the behaviour of high energy spectrum of GRBs~\cite{upscatter}. In addition to a low efficiency, 
this type of models produce a significant bump at $E \sim 1$ GeV which has not been observed. 
We should also remind that $\epsilon_e$, density of shells, and distance to the central 
source are somehow degenerate i.e. a burst with roughly similar properties can be obtained for 
$\epsilon_e \sim 0.3$, $r_0 \sim 10^{12}$ cm and smaller densities. Fig. \ref{fig:largeepsilon} 
shows an example of such case for simulation No. 11. The reason for such degeneracies is clear. At large 
distances densities are smaller and therefore higher efficiency is necessary to produce the same 
amount of photons. At smaller distances a large $\epsilon_e$ leads to too efficient and hard 
emission and the duration of burst is usually very short, only consistent with very short hard 
bursts. Although apriori large $\epsilon_e$ can be consistent with observations, giving smaller 
densities at larger distances, they are unrealistic. Moreover, the peak in the light curves shown in 
Fig. \ref{fig:largeepsilon} is too long to be consistent with a usual GRB, although GRBs for which 
a supernova is observed e.g. GRB 060218~\cite{grb060218} are very faint and long. In contrast to 
other GRBs, they usually have a thermal spectrum with a temperature of few keV. The gamma-ray is 
in fact the high energy wing of this thermal emission which most probably originates from an 
uncollimated - or mildly collimated - ejecta. The lack of significant variability in such bursts 
and simulations here is another argument for occurrence of normal GRBs at much shorter distances.

\subsubsection{Tail emission} \label{sec:tail}
In real GRBs usually a long emission tail specially in X-ray is observed at the end of the main 
spikes with a very steep power-law or in some cases exponential decline after few hundreds of 
seconds. In these simulations we see the steep decline, see Fig. \ref{fig:synchcompton} where light 
curves are plotted in logarithmic scales. This figure also shows that the decline slope is flatter 
in lower energy bands which is consistent with the observations, and explains why X-ray and optical 
light curves of GRBs have different slopes and breaks are chromatic. However, in our simulations 
the steep decline occurs much earlier than in real GRBs. We have two explanations for this 
difference. In what concerns simulations, we consider only one shock with few regimes i.e. we end 
the shock quite quickly. In fact due to numerical complexities and validity range of approximations 
when the shock becomes very weak, simulations are usually stopped before $\beta' \approx 0$ i.e. 
before shells coalesce completely. In real bursts one expects that shocks fades more slowly, thus 
the soft emission continues for much longer time. The second reason is intrinsic to the activity 
of the engine. Usually late peaks are softer meaning that gradually the process responsible for the 
acceleration of the jet becomes weaker and less efficient. Therefore the tail of the prompt 
emission is due to soft shocks.

\begin{figure*}
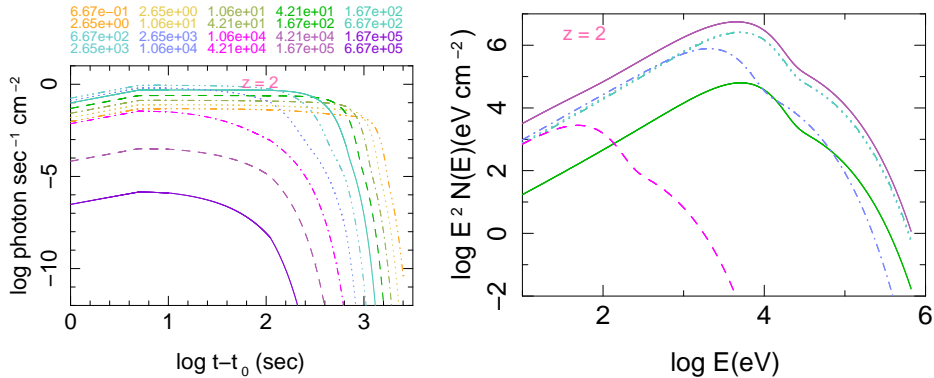

\begin{center}
\begin{tabular}{ll}
\includegraphics[width=5cm,angle=-90]{\plotdir/spec-m5ppp-10222-1.5-1-mag0-0-lc.eps} &
\includegraphics[width=5cm,angle=-90]{\plotdir/spec-m5ppp-10222-1.5-1-mag0-0-spect.eps}
\end{tabular}
\caption{Light curves (left) and spectra (right) of simulation No. 12 which simulates weak shocks or 
when the main shock is weakened due to close to complete coalescence of two shells. In right plot
the highest full line curve is the total spectrum and other curves are spectra in the time 
interval of 4 simulated regimes. They show how the peak of spectrum moves toward lower energies 
when the shock becomes weaker. \label{fig:tail}} 
\end{center}
\end{figure*}

Rather than following slow decay of simulated prompt emissions, we simulate an example of weak shocks 
using simulation No. 12 which is similar to simulation No. 11 but has a longer distance from 
central source 
and lower densities. The relative Lorentz factor is also much smaller to simulate weak shocks 
when shells are very close to complete their coalescence. Light curves and spectrum of this 
simulation are shown in Fig. \ref{fig:tail}. The first few seconds when shocks is formed must be 
ignored because we had to make the shock from beginning. The tail emission presents the declining 
regime of the light curves. We notice that similar to what is observed in many bursts, they are 
smooth and have a steep break after few hundreds of seconds. They confirm the conclusion taken from 
simulation No. 11 that longer the distance of the shock from the central source, smoother light 
curves and softer the burst. We also remind that these simulations do not take into account the 
curvature of the emission surface. It somehow flattens the break slope due to later arrival time of 
photons. In radiative shocks the steep break occurs when the two shells are coalesced. In the example 
presented here, the simulation is stopped when the Lorentz factor of the fast shell reduces to 
400.002 for an input value for the final Lorentz factor of 400, see Table \ref{tab:param}. 

This simulation is consistent with the results of~\cite{xrtafterglow} who relate the steep decline 
and the plateau regime of the X-ray flux respectively to the tail of prompt emission and its overlap 
with the rise of the afterglow (which is not simulated here). The peak of emission spectrum for this 
shock is at few hundreds of eV, much lower than in typical prompt emissions.

\subsection {Simulated bursts: Spectra}\label{sec:simulatedspec}
As we mentioned in the Introduction, since the launch of Fermi satellite and observation of prompt 
emission of GRBs in MeV and GeV energy bands, various models are suggested to explain unexpected 
results of these observations. Thus, an evident validity test for the model presented 
in~\cite{hourigrb} and extended here is its ability to reproduce the observed broad band spectra, 
specially at high energies. In the context of internal shock SSC model,~\cite{simulother0} 
and~\cite{simulother1} have already performed detailed simulations of 
GRBs spectra based on a formulation of shocks by~\cite{shocksynch}. They find a bump in the 
spectrum due to the inverse Compton scattering between high energy electrons and photons at 
$E \sim 10 GeV$, following by an exponential break. According to their predictions and depending 
on the parameters of their model, other features can also be present in the spectra at high energies. 
None of these features have been observed in the Fermi-LAT/GBM data. By contrast, in some bursts 
such as GRB 090510 it seems that the high energy part of the spectrum up to highest observed 
energies is best fitted by a power-law with positive index. 

In this section include only present the spectrum of synchrotron emission in simulated bursts. 
Compton emission is discussed separately in Sec. \ref{sec:compton} and we find that in most 
cases it has a negligible contribution to the total emission in GeV energy band. Because the 
spectra of simulated bursts discussed in the previous section have very similar properties we do 
not present the spectra of all the simulations and concentrate only on some of them which 
are representative.

\begin{figure*}
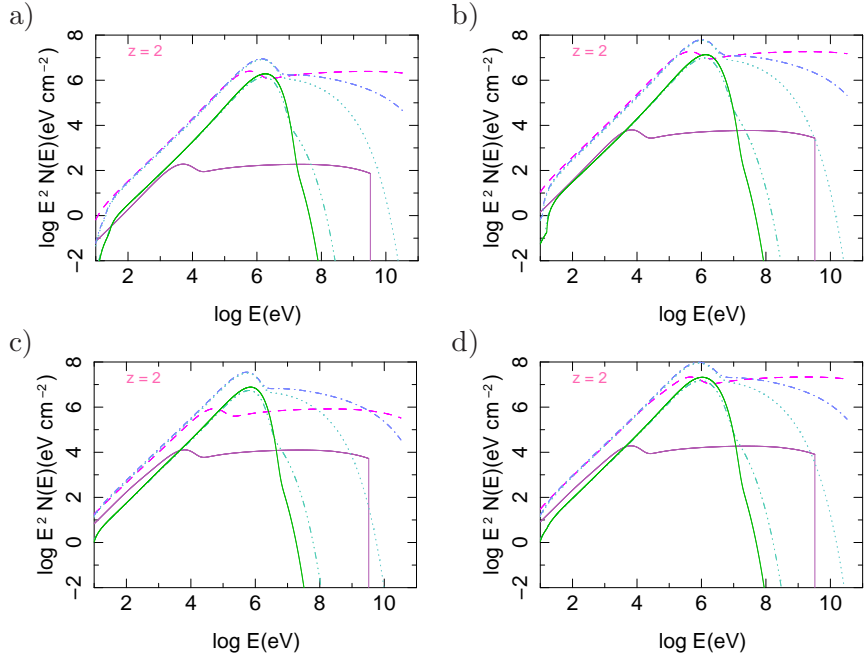

\begin{center}
\begin{tabular}{ll}
a)\includegraphics[width=4cm,angle=-90]{\plotdir/spec-m1p-122-mag60-02-expocut-highener-1.ps} &
b)\includegraphics[width=4cm,angle=-90]{\plotdir/spec-m1p-122-mag60-02-expocut-highener-2.ps} \\
c)\includegraphics[width=4cm,angle=-90]{\plotdir/spec-m1p-122-mag60-02-expocut-highener-3.ps} &
d)\includegraphics[width=4cm,angle=-90]{\plotdir/spec-m1p-122-mag60-02-expocut-highener-4.ps}
\end{tabular}
\caption{Simulation No. 7: a), b) and c) are respectively spectra of the first, second and third 
regimes. d) is the total spectrum. In each plot curves correspond to different values for $p$ and 
$\omega_{cut}/\omega_m$: $p = 2.5$ and $\omega_{cut}/\omega_m =$ 0.5 (full line), 1 (dash-3 dots), 
10 (dot), 100 (dash-dot); $p = 1.9$ and $\omega_{cut}/\omega_m = 1000$ (dash). The external 
magnetic field in these simulations is $10$ kG. The low amplitude full line has $p = 1.9$ and 
$\omega_{cut}/\omega_m = 1000$ but no external magnetic field. \label{fig:sepectmonep}}
\end{center}
\end{figure*}

Fig. \ref{fig:sepectmonep} shows the spectra of simulation No. 7 during separately for each regime 
as well as the total spectrum in observer frame. Each plot includes spectra for a set of index and 
exponential cutoff values. The peak of photon number per unit energy $N(E)$ appears as a 
bump with an amplitude roughly proportional to the index $p$ of the power-law factor of electrons 
Lorentz factor distribution.\footnote{This bump is not exactly at the same energy as the peak 
of $N(E)$ but they are close and from now on we call it the peak although it is not always the 
absolute peak of the fluence $E^2 N(E)$ which are plotted here. The reason for plotting the latter 
rather than $N(E)$ is that spectral features are enhanced in $E^2 N(E)$.} 
It is directly related to the average synchrotron characteristic frequency in 
the time interval in which the spectrum is determined~\cite{electrody} and its position determines 
the hardness of the burst. When $\omega_{cut}/\omega_m \sim 10-100$ the peak is followed by another 
bump at higher energies which is due to the exponential cutoff in electrons energy distribution. 
A good example of such bursts is GRB 090926A~\cite{grb090926a}. When $p < 2$ and 
$\omega_{cut}/\omega_m \gtrsim $ few 100, the bump becomes flatten and looks like a linear increase 
in the emission similar to behaviour of the spectrum of GRB 090510~\cite{grb090510obs,notforward}. 
These simulations show that this feature is due to the high value of cutoff in the electrons 
energy distribution and no additional process is needed to explain it. The spectrum of this burst 
also shows clearly the low energy peak around $\sim$ MeV energies. When the cutoff is low or $p > 2$, 
this peak becomes the absolute peak of the fluence. As Fig. \ref{fig:sepectmonep} clearly shows 
its position evolves with time. This is consistent with the observed evolution of spectral 
peak GRBs observed by the Fermi-LAT such as GRB 080916C~\cite{grb080916c,spect080916c0,spect080916c1}.

in these simulation at $E \ll E_{peak}$ the slope of the spectra approach the theoretical value of 
$4/3$. However, in the observed GRBs low energy slope of gamma-ray emission is usually smaller 
than this value and is on average $\sim 1$~\cite{batcat,fermispect}. Addition of inverse Compton emission slightly flatten the spectrum both at low and 
high wing of this simulated burst, see Fig. \ref{fig:synchcompton}. This is in agreement with 
simulations of~\cite{simulother1}. However, in contrast to some of previous simulations we find 
that in very hard bursts such as GRB 090510 Compton scattering does not have a significant 
contribution in the high energy emission. This explains the lack of any feature due to this process 
in the spectrum of GRB 090510 which was a short hard burst. See the next section for more discussion 
about low energy spectrum.
\begin{figure*}
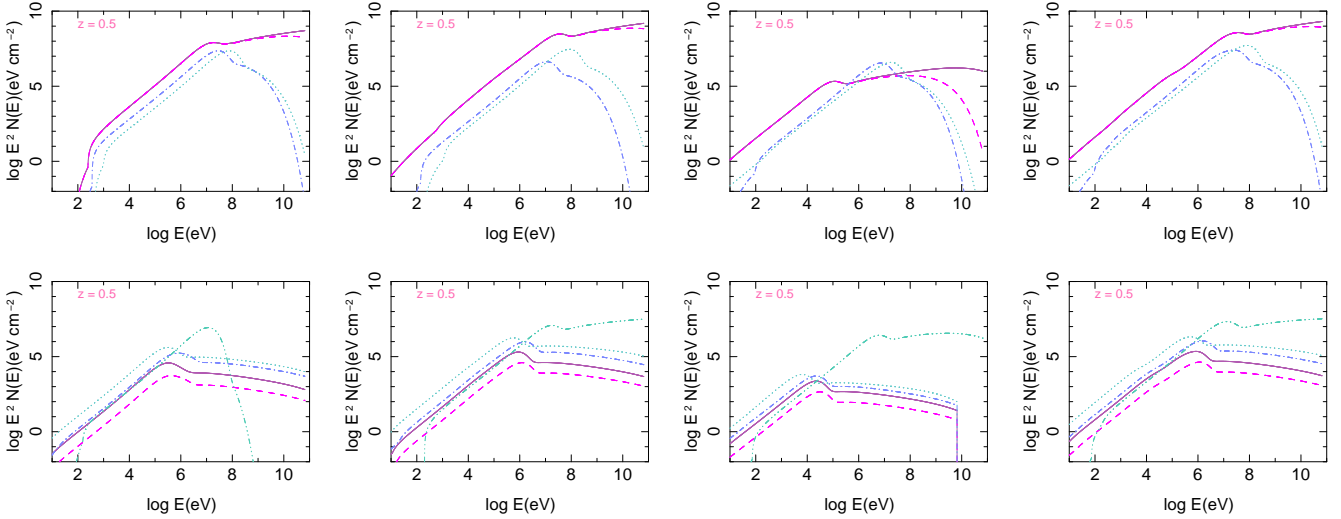

\begin{center}
\begin{tabular}{llll}
\hspace{-1cm}\includegraphics[width=3.25cm,angle=-90]{\plotdir/m3-322-variouscut-1.ps} &
\includegraphics[width=3.25cm,angle=-90]{\plotdir/m3-322-variouscut-2.ps} &
\includegraphics[width=3.25cm,angle=-90]{\plotdir/m3-322-variouscut-3.ps} & 
\includegraphics[width=3.25cm,angle=-90]{\plotdir/m3-322-variouscut-4.ps} \\
\hspace{-1cm}\includegraphics[width=3.25cm,angle=-90]{\plotdir/m3-322-variousepsilon-1.ps} &
\includegraphics[width=3.25cm,angle=-90]{\plotdir/m3-322-variousepsilon-2.ps} &
\includegraphics[width=3.25cm,angle=-90]{\plotdir/m3-322-variousepsilon-3.ps} & 
\includegraphics[width=3.25cm,angle=-90]{\plotdir/m3-322-variousepsilon-4.ps}
\end{tabular}
\caption{Simulation No. 10 with power-law + exponential cutoff for electrons, from left to right 
spectra of 3 regimes and total spectrum. Top row: Simulations with various values of power-law 
index and cutoff and $|B_{ext}| = 70$ KG: $\omega_{cut}/\omega_m = 1000$, $p = 1.5$ (full line); 
$\omega_{cut}/\omega_m = 100$, $p = 1.5$ (dash); $\omega_{cut}/\omega_m = 3$, $p = 2$ (dot-dash); 
$\omega_{cut}/\omega_m = 3$, $p = 2.5$ (dot). Bottom row: Simulations with various values of 
parameters and $|B_{ext}| = 100$: $\omega_{cut}/\omega_m = 1000$, $p = 2.5$, $\epsilon_e = 0.002$, 
$\Gamma = 500$ (full line); $\omega_{cut}/\omega_m$ and $p$ as previous case and $\epsilon_e = 
0.02$, $\Gamma = 50$ (dash); $n'_0 = 5 \times 10^{15}$ cm$^{-3}$ and other parameters as the 
previous case (dot-dash); varying $p$ with index $-0.2,~0,~0.5$, initial $p = 2.5$ and 
$\epsilon_e = 0.002$ (dot); initial $p = 1.8$, the same indices as previous, and 
$\omega_{cut}/\omega_m = 0.5,~1000,~100$ (dash-3 dots). Different spectral shape of the last 
simulation in different time intervals is similar to observed spectra of GRB 090510.
Parameters which are not specified here are equal to what is given in Table \ref{tab:param}.
\label{fig:diffp}}
\end{center}
\end{figure*}
Although the high energy spectra of some bursts are consistent with an exponential cutoff, many 
bursts show broader spectra. Such a behaviour can be reproduce be considering a broken power-law 
distribution for electrons at high energies rather than an exponential cutoff. Few examples such 
bursts are shown in Fig \ref{fig:powercut}-a \& b. It is clear that in this case the spectrum is 
more featureless and broader. 
\begin{figure*}
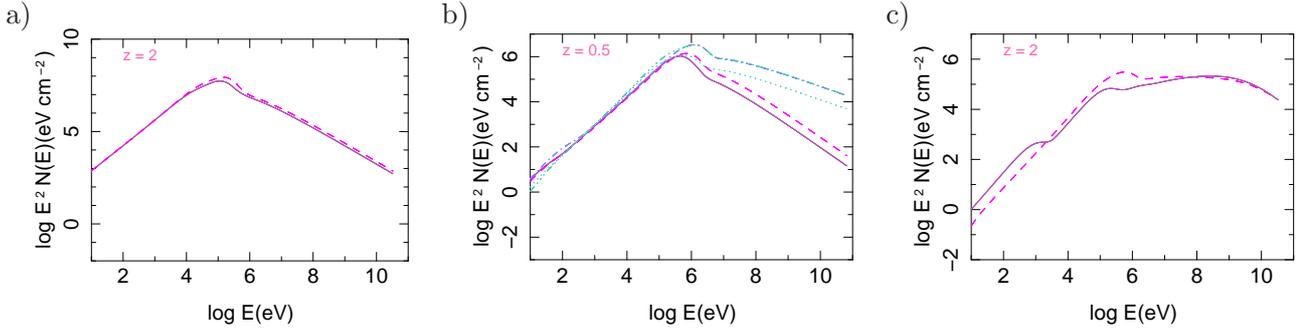

\begin{center}
\begin{tabular}{lll}
\hspace{-1cm}a)\includegraphics[width=4cm,angle=-90]{\plotdir/m1p-122-plcut-3-mag600-tot.ps} &
b)\includegraphics[width=4cm,angle=-90]{\plotdir/m3-322-plcut-tot.ps} &
c)\includegraphics[width=4cm,angle=-90]{\plotdir/spec-m5p-10222-1.5-100mag200-02-spect-tot.eps}
\end{tabular}
\caption {a) Simulation No. 7 with a broken slope at $\omega_{cut}/\omega_m = 3$, $p_1 = 2.1$, $p_2 = 4$ 
(full line); $p = 2.5$ (dash line); 
b) Simulation No. 10 with a broken slope at $\omega_{cut}/\omega_m$ a) and $p_1 = 2.5$, $p_2 = 4$, 
$|B_{ext}| = 17$ kGauss (full line); $p_1 = 2.1$, $p_2 = 4$ and same $|B_{ext}|$ as previous (dash); 
$p_1 = 2.1$, $p_2 = 3$, $|B_{ext}| = 26$ kGauss (dot-dash); same slope and $|B_{ext}| = 35$ kGauss 
(dot); same slope and $|B_{ext}| = 70$ kGauss (dash-3 dots).
c) Simulation No. 5 with same parameters: 4 regimes (full line), 7 regimes (dash). With more gradual 
change of parameters in the latter case the spectrum at low energies becomes smooth and the low 
energy feature disappears.
\label{fig:powercut}}
\end{center}
\end{figure*}
Finally we notice some features in the low energy wings of the total spectrum of some simulated 
bursts - see Fig.\ref{fig:diffp}. Our tests show that these features are artifacts of the abrupt 
change of parameters from one regime to another. Fig. \ref{fig:powercut}-c shows two simulations 
with same parameters but one of them (dash line) has a larger number of regimes and variation of 
parameters from one regime to the next is smaller. The spectrum of this simulation does not show any 
low-energy feature.

We should also remind that in all the simulations presented here the selection of electron index 
and cutoff is completely arbitrary and no connection between these parameters and other quantities 
that determine the physics of the jet and shock are considered, simply because we ignore any 
relation. Some PIC simulations confirm a power-law with exponential cutoff distribution for 
electrons~\cite{fermiaccspec}, but do not give any hint about the relation between the value of 
index and cutoff energy and other quantities. Considering the random nature of instabilities, it is 
plausible that feedback between various quantities and the randomness of the latter smear any 
feature and create the observed featureless spectra - see also below for more explanation. 

\subsubsection{spectrum slope at low energies} \label{sec:lowener}
As we mentioned above, the spectrum of simulated bursts at energies much smaller than peak energy - 
at least two or three orders of magnitude - is consistent with the theoretical prediction of 
$\sim 4/3$. This is a well known result. From properties of synchrotron 
emission~\cite{emission0,emission1,deathline,piranrev}, in another word the Bessel functions in 
equation (\ref{powerdopcorr}), this slope is obtained for $\omega' \ll \omega_c$. However, 
the average low energy slope of observed long bursts is about 1~\cite{batcat}, although brighter 
bursts are somehow harder and have steeper spectra at low energies~\cite{slopebright}. Such a flat 
slope can be explained if electrons are in what is called {\it slow cooling} 
regime~\cite{emission1} (see also Appendix \ref{app:d} for more details), but the efficiency of 
synchrotron emission in this regime is low. For this reason the general believe is that in GRBs 
electrons do not cool slowly and are in fast cooling regime. Thus only steep slopes close to the 
theoretical value of $4/3$ are expected if the origin of emission is synchrotron. A number of 
solutions such as the effect of Klein-Nishina and adiabatic expansion~\cite{simulother1} as well as 
thermal emission have been suggested to explain the flatness of low energy spectrum. These works 
conclude that synchrotron self-Compton processes can explain observations although for GRBs with 
very flat low energy spectra another origin, e.g. thermal emission may be necessary. 

In this section we consider this issue in the framework of the formulation of synchrotron discussed 
in~\cite{hourigrb} and simulated in this work. On the theoretical side we should remind 
that the strict classification of accelerated electrons in a shock to slow or fast cooling by 
comparing $\gamma_m$ with cooling Lorentz factor $\gamma_c$ defined in (\ref{gammac}) is 
meaningful only for a power-law distribution with an abrupt break at $\gamma_m$. In this case most 
of electrons have a Lorentz factor close to $\gamma_m$, thus comparison of $\gamma_m$ with $\gamma_c$ 
indicates whether majority of photons are emitted with energies below the peak or above it. If the 
distribution of $\gamma_e$ is such that there is no abrupt break but it has a tail of low energy 
electrons that preferentially emit synchrotron photons at low energies, a flat spectrum in 
both sides of the peak energy can be obtained. Because our theoretical calculation is performed 
only for simple power-law or power-law with exponential break, here we try to simulate 
more complex simulations by considering a variable distribution for electrons in different regimes 
of a burst.

\begin{figure*}
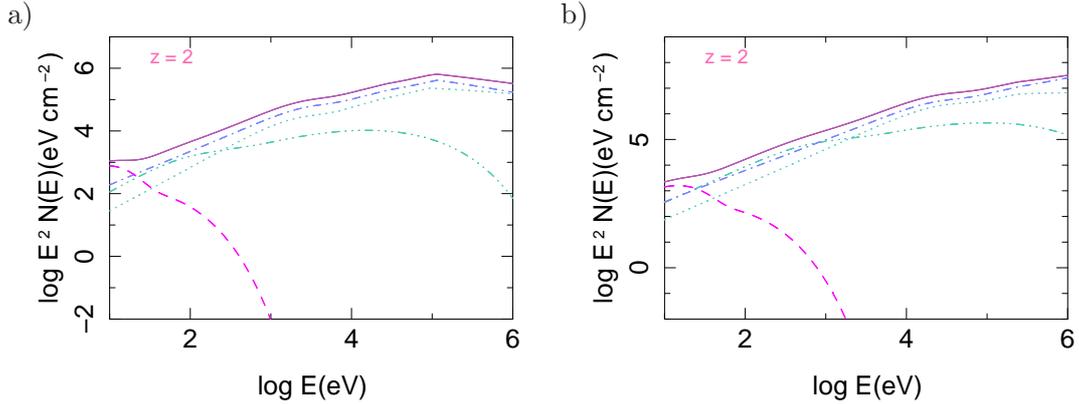

\begin{center}
\begin{tabular}{ll}
a)\includegraphics[width=5cm,angle=-90]{\plotdir/m6p-Band05-10-1.5-2.5-10-2.5-3-10-1-10mag60spect.eps} &
b)\includegraphics[width=5cm,angle=-90]{\plotdir/m6p-Band05-10-1.5-2.2-5-2.2-5-10-1-1mag600spect.eps}
\end{tabular}
\caption {Spectra of simulation No. 1 at $E \lesssim E_{peak}$ in 4 regimes/time intervals with 
power-law-exponential cutoff distribution for electrons in the first and last regimes and broken 
power-law in two middle regimes. The full line is the total spectrum. a) Electron distribution 4 
regimes: 1- power-law with exponential cutoff $p = 0.5$, $\gamma_{cut}/\gamma_m = 10$; 2- broken 
power-law $p_1 = 1.5,~p_2 = 2.5,~\gamma_{cut}/\gamma_m = 10$; 3- broken power-law 
$p_1 = 2.5,~p_2 = 3,~\gamma_{cut}/\gamma_m = 10$; 4- power-law with exponential cutoff 
$p = 1$, $\gamma_{cut}/\gamma_m = 10$; $|B_{ext}| = 10$ kG. b) 1- power-law with exponential cutoff 
$p = 0.5$, $\gamma_{cut}/\gamma_m = 10$; 2- broken power-law $p_1 = 1.5,~p_2 = 2.2,~
\gamma_{cut}/\gamma_m = 5$; 3- broken power-law $p_1 = 2.2,~p_2 = 5,~\gamma_{cut}/\gamma_m = 10$; 
4- power-law with exponential cutoff $p = 1$, $\gamma_{cut}/\gamma_m = 1$; $|B_{ext}| = 100$. 
Note that in both cases the slope of the spectra up to $\sim 2$ orders of magnitude from the 
peak are flatter than $4/3$ and break to this value at very low energies. \label{fig:lowenerslope}}
\end{center}
\end{figure*}

Spectra presented in Figs. \ref{fig:tail}, \ref{fig:sepectmonep}, \ref{fig:diffp}, 
\ref{fig:powercut} show that with evolution of a burst, the peak energy gradually moves to lower 
energies. The spectrum at high energies can have variety of behaviour, notably it can be very flat 
or even rising if electrons spectrum is shallow. Fig. \ref{fig:lowenerslope} shows the spectra of 
the simulation No. 1 with various spectral behaviour for electrons. The slope of the fluence in these 
simulations is $\lesssim 1$ from the peak up to $\sim 1$ keV for Fig.\ref{fig:lowenerslope}-a and 
up to $\sim 10$ keV for Fig.\ref{fig:lowenerslope}-b which has much larger external magnetic field 
and is harder. Luminosity of the simulation in Fig.\ref{fig:lowenerslope}-a is not very strong. 
Therefore such a model can not represent a typical GRB, but the emission from a sub-population 
of mildly accelerated electrons which are present at the same time as highly accelerated electron 
making the high energy component. Simulation in Fig.\ref{fig:lowenerslope}-b is hard and presents 
the spectrum of a typical GRB.

In exceptional cases such as in GRB 061121~\cite{grb061121} it was possible to observe 
simultaneously the burst in gamma-ray and x-ray. Similar to these simulated bursts, the slope at 
low energy wing of GRB 061121 is shallow and approaches $4/3$ only in soft x-ray and 
optical/IR~\cite{grb061121spect}.

\subsubsection{Electrons distribution}
In the previous sub-section we showed that the presence of mildly accelerated electrons can 
produce a flat low energy spectrum. Does such a population of low energy electrons exist ? 
Both PIC~\cite{fermiaccspec} and Monte 
Carlo simulations~\cite{montecarlo0} of charged particle acceleration in relativistic shocks 
show that only about $10\%$ of electrons are accelerated to very high energies. The rest of 
electrons keep their original thermal distribution. But no signature of a thermal emission is 
found in the observations. In presence of the magnetic field induced by the shock they emit 
synchrotron radiation which can contribute to low energy emission and notably flatten the 
spectrum. On the other hand, it seems unlikely that such a large fraction of electrons stay 
unmodified. Moreover, the absence of any thermal signature means that either even before the 
shock they had a nonthermal distribution, or their distribution somehow has been modified 
during the shock. In fact spectra of GRB 080916C which was enough bright to allow the 
determination of the broad band spectra in different time intervals~\cite{spect080916c0} show 
that the low energy slope of the first interval between trigger until about 7 sec is much 
steeper than later intervals~\cite{spect080916c1}. We interpret this as the presence of an 
abrupt break, similar to simulations here, in the electron distribution during the first time 
interval. By contrast, in the second interval the break is smoother meaning that a larger 
number of electrons are moderately accelerated and their distribution at low energies is 
flatter. Considering the results of PIC simulations of~\cite{fermiaccspec1}, the microphysics 
of these observations can be qualitatively explained as the following:\\
During the first interval the amplitude of the electromagnetic wave is not yet very high and 
free path length of electrons scattered by their local environment i.e. protons/ions and other 
electrons $\lambda < r_g = m\gamma_e/eB$ gyro-radius (Lamb radius). Therefore, before these 
electrons can diffuse and accelerate, they are scattered by other particles. It is why 
simulations find them in the same distribution as protons which are practically unperturbed by 
the collision~\cite{fermiaccspec1,montecarlo0}. Gradually the amplitude of electromagnetic wave 
increases, thus more electrons are accelerated and lose their energy by synchrotron emission 
before being scattered by their environment. These mildly accelerated electrons which necessarily 
emit low energy photons play an important role in smoothing electrons Lorentz factor distribution, 
and thereby flattening the photon distribution at low energies. This should also make smaller lag 
between high energy and low energy band, consistent with observations. In conclusion, we should 
use smoother distributions at low energies for electrons. 

\subsection{Compton scattering} \label{sec:compton}
Inverse Compton (IC) emission from GRBs is studied by various authors see e.g.~\cite{simulother0}. 
In this section we briefly describe Compton scattering in the framework of model and simulations 
presented here. As the interaction between photons and accelerated electrons does not significantly 
affect dynamics of the shock we can determine the flux of IC and its spectrum separately. We should 
also remind that for far observers both electrons and synchrotron photons are boosted in their 
direction. Therefore, the Center of Mass (CM) energy of their interactions is small and very little 
energy is exchanged between them. This qualitative argument shows that inverse Compton scattering 
cannot be a significant source of high energy photons. By contrast, due to delay between 
production of photons by synchrotron emission and their scattering by electrons, their light curves 
are more extended in time. Simulations discussed below are consistent with this description.

\begin{figure*}
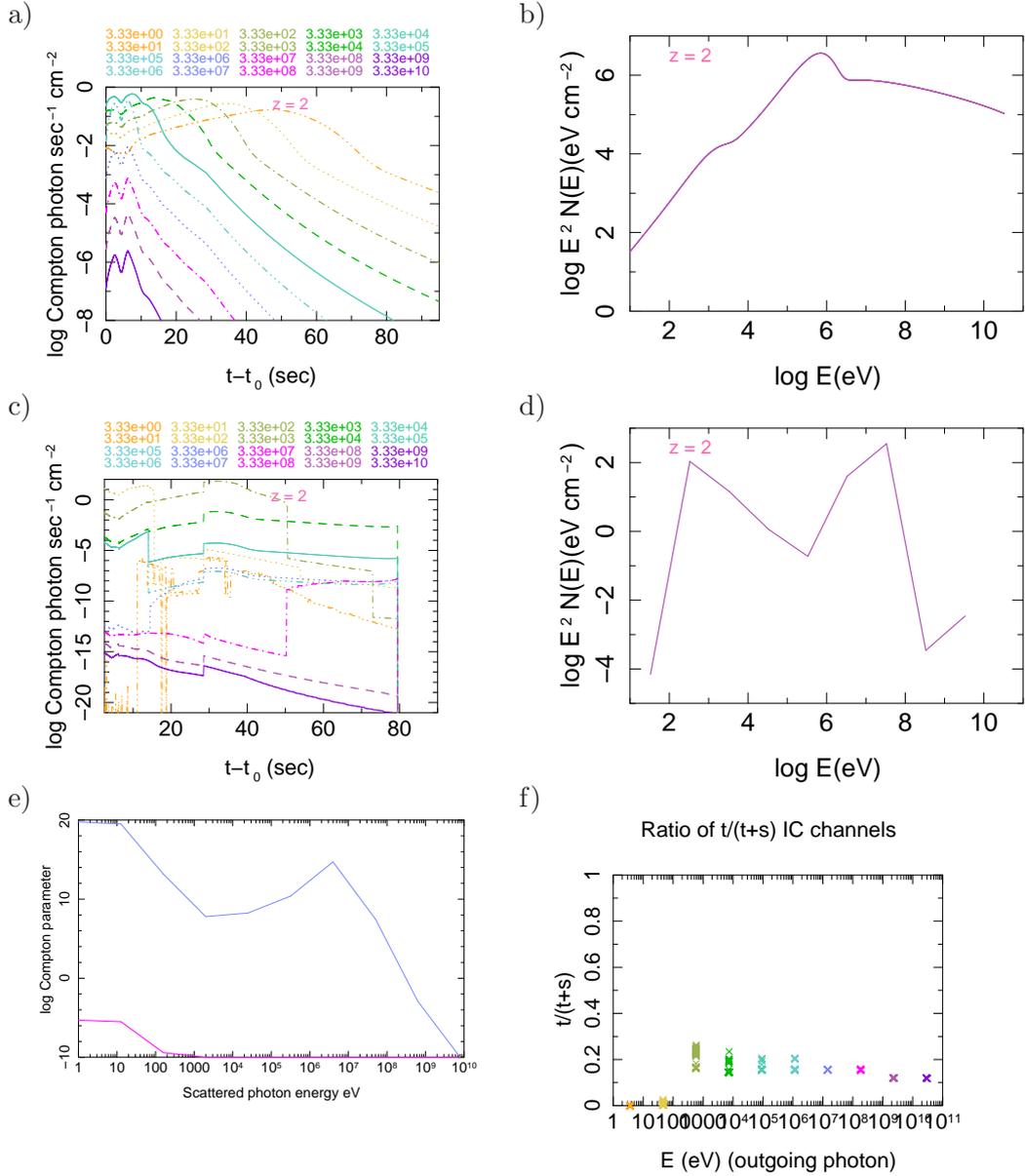

\begin{center}
\begin{tabular}{ll}
a) \includegraphics[width=5cm,angle=-90]{\plotdir/m6p-2.5-1000-mag60-lc-log.eps} & 
b) \includegraphics[width=5cm,angle=-90]{\plotdir/m6p-2.5-1000-mag60-compton-spect-synch.ps} \\
c) \includegraphics[width=5cm,angle=-90]{\plotdir/m6p-2.5-1000-mag60-compton-lc.eps} & 
d) \includegraphics[width=5cm,angle=-90]{\plotdir/m6p-2.5-1000-mag60-compton-spect.eps} \\
e) \includegraphics[width=4cm,angle=-90]{\plotdir/m6p-2.5-1000-mag60-50-comp-cparam.ps} &
f) \includegraphics[width=5cm,angle=-90]{\plotdir/m6p-2.5-1000-mag60-50-comp-t.ps}
\end{tabular}
\caption {Synchrotron light curves (a) and spectrum (b), inverse Compton light curves (c) and 
spectrum (d) for simulation No. 1 with $p = 2.5$ and $\gamma_{cut}/\gamma_m = 1000$. Compton parameter 
(e): assuming the same density of accelerated electron everywhere in the slow shell 
(blue), for the layer containing accelerated electrons according to PIC simulations (magenta). 
We have neglected scattering/absorption of photons by cold plasma (see Sec. \ref {sec:proc}) in 
calculation of Compton parameter. Compton scattering light curves and spectrum correspond to the 
latter case. Contribution of t-channel (f) with respect to energy band of the scattered photon. 
Each $\times$ symbol indicates the ratio of t-channel to total in the right hand side of equation 
(\ref{boltz}) at a given time. Sudden break at low energies is due to truncated synchrotron spectrum 
in our simulations. \label{fig:synchcompton}}
\end{center}
\end{figure*}

Using the Boltzmann equation, the rate of Compton scattering can be written as~\cite{houricr}:
\bea
&&\frac{\partial {F'}_C^{(\gamma)}(E')}{\partial t'} = 4 \pi^3 c \int_{\gamma_m}^\infty 
\frac{d\gamma_e}{\gamma_e} \int \frac{dE'_1}{E'_1}{F'}_{sy}^{(\gamma)}(E_1) n'_e (\gamma_e) 
\int d(cos \theta') \theta') A(\gamma_e, E'_1) \frac{d^2 \sigma (\gamma_e, E'_1, \theta', E')}
{E'dE'd\theta'}, \nonumber \\
&& A (\gamma_e, E'_1) \equiv P'_e.P'_{\gamma} \label{boltz}
\eea
where ${F'}_C^{(\gamma)}$ and ${F'}_{sy}^{(\gamma)} = dP_{sy} / \omega' d\omega'$ are respectively the 
number density of photons 
produced by Compton scattering and photons produced by synchrotron emission, $\theta'$ is the 
angle between the scattered photon and its original direction, $P'_e$ and $P'_{\gamma}$ are 
respectively 4-momentum vectors of incoming electron and photon. We have assumed head-to-head 
collision to obtain an upper limit on the Compton scattered photons and integrated over all 
possible directions in the rest frame of the shock front/active region. Both Thompson and 
Klein-Nishina terms ($s$ and $t$ channels) are included in the Compton cross-section $\sigma$. 
To determine ${F'}_C^{(\gamma)}$, one has to integrate the right hand side of (\ref{boltz}) with 
respect to time. However, as the final aim is to calculate the flux seen by a far observer, before 
integrating with respect to $\theta'$ one has to multiply the right hand side by $\Gamma_f (1 + 
\beta_f \cos (\theta')$~\cite{radproc} to take into account angular transformation with respect to 
observer. Another important fact to be considered is the extent of distribution of accelerated 
electrons inside the slow shell. As explained in Sec. \ref{app:C}, synchrotron photons are emitted 
by accelerated electrons in the Energetic Electromagnetic Structure (EES) very close to shock front. 
PIC simulations show that the width of EES is only $\sim 10$ times of $\lambda_{ep}$ the plasma 
wavelength of electrons, and accelerated electrons diffuse inside slow shell 
$\sim 1000 \lambda_{ep}$~\cite{fermiaccspec,fermiaccspec1}. Therefore synchrotron photons that are 
not scattered in this thin layer, do not meet any accelerated electron coming toward them (in the 
slow shell frame). For this reason the integration of (\ref{boltz}) must be limited to the time 
that photons pass through this region. If one naively integrates this equation for the full 
passage time through slow shell - assuming that its end coincides with the end of the synchrotron 
emission - one obtains, in contradiction with observations~\cite{spect080916c0}, a significant 
deformation of spectrum from synchrotron. Compton parameter for both cases are shown in 
Fig. \ref{fig:synchcompton}-e. Evidently, the absence of concrete information about micro-physics 
of the shock create a large uncertainty on the determination of Compton parameter and thereby, 
the contribution of Compton scattering in shaping the observed spectrum. Therefore simulations 
reported here must be considered as a rough estimation.

For angular cross-section we have used full relativistic quantum expression~\cite{crosssec}. 
Fig. \ref{fig:synchcompton} shows the light curves, spectrum, and Compton parameter of Compton 
scattered photons in simulation No. 1. For comparison the light curves and total spectrum of 
synchrotron alone are also shown. As expected, Compton emission decreases with time much slower 
than synchrotron emission. Nonetheless, it stays - specially at high energies - orders of magnitude 
smaller than synchrotron flux. This would not be the case if we assume a uniform distribution of 
accelerated electrons everywhere along synchrotron outgoing path, see large Compton parameter for 
this case in Fig. \ref{fig:synchcompton}-e. Therefore, according to these simulations inclusion 
of inverse Compton scattering cannot explain the high energy tail of emission observed by the 
Fermi-LAT. These results are consistent with the conclusion of~\cite{spect080916c0} that finds only 
one process responsible for the emission in all energy ranges. The suggestion of~\cite{simulother1} 
that Klein-Nishina inverse Compton cooling can explain the shallow slope of Fermi and Swift 
observations at $E < E_{peak}$ can be true only if the penetration length of accelerated electrons 
inside slow shell is much longer than what PIC simulations obtain. However, this will also change 
the spectral shape at low energies and should produce a bright emission in UV/optical bands unless 
these photons are absorbed. Nonetheless, a small contribution from Compton scattering can somehow 
flatten low energy spectrum and makes it more consistent with observations. As we suggested in 
Sec. \ref{sec:lowener}, a flat electron distribution is most probably necessary to explain the high 
energy slope. In the next section we suggest an alternative explanation for the observed high 
energy tail emission.

\subsection{Origin of the high energy tail} \label{sec:tail}
For having a strong Compton emission, accelerated electrons moving toward the observer must 
collide - in preference head-to-head - with photons emitted by a source at rest or slow moving, 
for instance photons from material surrounding the source. The problem with this suggestion 
is that at late times the shock is much weaker, thus not sufficient high energy electrons are 
around. Moreover, at these late times the jet is well outside the star envelop, thus the density of 
environment photons is relatively low. A number of other explanations are suggested for the tail 
emission. For instance, external forward 
shocks~\cite{forwardsh,forwardsh0,forwardsh1,forwardsh2,forwardsh3,
forwardsh4,forwardsh5,fermipeak,forwardssc0,forwardssc1}, Compton scattering of photospheric 
photons by upstream scattered high energy electrons~\cite{delayphotospher}, etc. Although both 
of these processes can have some contribution in the observed flux, it seems unlikely that they 
provide a satisfactory explanation for all the high energy bursts observed by the Fermi-LAT. 
External shocks are too weak to explain the observed flux~\cite{critisism} and its fast 
variabilities~\cite{delayphotospher}. Interaction of up-scattered electrons is evidently a 
possibility, but it is originally suggested in the context of standard fireball models with 
photospheric emission before shocks. It needs some fine-tuning on the delay between two successive 
shells (or fireballs) and distance between shocks and photosphere. Moreover, the peak of high 
energy emission must be simultaneous with a later peak at lower energies, but the peak of high 
energy light curve does not always correspond to a peak at lower energies, see for instance 
GRB 090902B~\cite{grb090902b} where 2 late peaks in the Fermi-LAT light curve of photons with 
$E > 100$ MeV do not correspond to any peak at lower energies.

Comparing simulated light curves and spectra in Fig. \ref{fig:synchcompton} with observations, 
we find that synchrotron emission can explain the main high energy emission which in all the 
observed bursts coincides with early peaks in lower energies. On the other hand, it seems 
that in bursts with long high energy tails later peaks are significantly harder than earlier 
ones. Either they have occurred at the same time as earlier ones but at shorter distances 
from the engine, or more probably they are due to the collision of late ejected shells with 
the remnant of the earlier collisions which are slowed down. In this case their hardness can 
be due to the accumulation of material in the slowed down earlier shells, what we called 
{\it highway} effect in Sec. \ref{sec:simuloscil}. As we showed in the previous section, 
although inverse Compton scattering last for a much longer time, its flux is not enough to 
explain the tail emission. If there are additional photons from another source rather than 
synchrotron emission, the efficiency of Compton scattering would increase. However, no 
signature of these photons which must also contribute to low energy emission has been found 
but it is possible that they can be absorbed~\cite{thermal6}. Indeed, only very few 
observations of the prompt emission in low energy bands have been performed and the quality of 
data does not allow to estimate the column density of material and the absorption at the time of 
prompt emission. An exceptional case is GRB 061121~\cite{grb061121} which its main peak occurred 
about 70 sec after the precursor. Its average column density during the peak is estimated to be 
$N_H\sim 1.4 \times 10^{22}$ but Swift-XRT data show fast evolution~\cite{grb061121spect}, and 
therefore the initial column density can be much larger. In any case, it does not seem that 
this column density be enough to completely absorb low energy emission and in fact the 
afterglow of this burst was observed in optical and IR~\cite{grb061121}.
 
Here we suggest that high energy tail emission can be due to a delayed synchrotron emission of 
the most energetic electrons. To be accelerated, electrons must oscillate back and forth in the 
collided region - the active region - where the electromagnetic wave is 
concentrated~\cite{fermiaccspec,fermiaccspec1,montecarlo0}. However, due to the strength of normal 
field their movement is oblique to the flow direction and helical, and the minority of them which 
are accelerated to very high energies have also a very large Larmor radius. With the progress of 
the shock gradually the electric and magnetic fields inside the active region would not be enough 
to confine these high energy particles and they escape to the down stream where the magnetic 
field flux is very small, see Fig 1 of ~\cite{fermiaccspec}. A similar process is 
believed to be the origin of cosmic rays in the shocks at much larger spatial scales - see 
e.g.~\cite{cosmicray}. These particles do not lose their energy until either are Compton scattered 
by photons or in presence of an external magnetic field dissipate their energy as synchrotron photon. 
We must also remind that the highly charged region in the shock front~\cite{fermiaccspec} at least 
partially screens the down stream from the magnetic field of the source. This process 
has been studied in the context of accretion from a disk in which the accretion flow partially 
screens the disk from the magnetic field of the central source, see e.g.~\cite{accretionscreen}. 
Such a {\it reservoir} for high energy electrons and their delayed dissipation explains why the 
spectrum of GRBs is consistent with one component emission~\cite{spect080916c0} but their light 
curves seem to show two components. 

To estimate the delay time we assume that the energy of observed high energy photons is close 
to the synchrotron characteristic frequency of the emitting electrons. 
Using (\ref{syncchar}) and the characteristic synchrotron emission time of electrons 
$\tau' = c m_e\gamma_e / eB' = 3\pi\gamma_e^3/\omega'_c$, and a Lorentz factor of $\Gamma \sim 1000$, 
we find $B' \lesssim 300$ Gauss to have a maximum delay of $\sim 100$ sec for photons with 
$E \sim 10^{10}$ eV for the observer. This means that the environment around the shock front must 
be either very unmagnetized or very efficiently screened from the magnetic field. 

Another possibility is a resonant movement of the most energetic electrons with Electromagnetic 
Energy Structure (EES) that forms in the two sides of the shock 
front~\cite{fermiaccspec,fermiaccspec1}. The maximum of electrostatic and electromagnetic fields 
are situated very close to the shock front and electrons are accelerated by moving back and forth 
between down stream and up stream in this field. The electromagnetic component is dominated by 
transverse electric and magnetic fields and is approximately a standing wave expanding more in the 
downstream than upstream, see figure 2 of~\cite{fermiaccspec1} and figure 3 of~\cite{fermiaccspec}. 
Due to the phase difference between electric and magnetic fields, electrons are accelerated where 
the electric field is strong and lose energy where the magnetic field is strong. Although we do 
not yet have a complete simulation of this process, it is perceivable that with the evolution of 
the shock the electromagnetic field gradually becomes non-standing and propagates to down stream. 
Electrons with highest Lorentz factors can follow the propagation of the field for sometime and 
ones which are trapped in the electric field dominated region continue to be accelerated until 
their lag from the electromagnetic field bring them to the magnetic field dominated region where they 
lose their energy by synchrotron emission. This phenomenon also explains why a tail emission in 
low energies is not observed. Low accelerated electrons which emit in lower energies  
cannot follow the propagation of the electromagnetic wave (EES) and their lag with respect to EES 
take them quickly to the the magnetic field dominated region where they dissipate their energy 
quickly. Figure 2 of~\cite{fermiaccspec1} shows that the wavelength of EES $\lambda'_{EES} \sim 5 
\lambda'_{skin}$ where $\lambda'_{skin} \equiv c/\omega'_p$ and 
$\omega'_p \equiv (en'/m_p\epsilon_0)^{1/2}$ are respectively skin depth and plasma characteristic 
frequency of protons in the fast shell. Using the same typical values for these quantities as what 
we used in the previous paragraph, $n' \sim 10^{15}$ cm$^{-3}$, and assuming that EES propagate with 
a speed very close to light propagation in vacuum, we find a time delay of $\sim 800$ sec for 
$E \gtrsim 10$ photons. Evidently, this is just an order of magnitude estimation because PIC 
simulations used here are not realistic for a GRB and we ignore the exact evolution of EES. 
Nonetheless, this exercises shows that in principle this idea is viable, and more realistic PIC 
simulations should be able to verify it.

\subsection {Efficiency} \label {sec:efficien}
The efficiency of synchrotron self-Compton models has been a serious issue in the internal shock 
models~\cite{critisism}. PIC simulations find that only $\sim 10\%$ of the available kinetic 
energy is transferred to accelerated electrons~\cite{fermiaccspec}. Table \ref{tab:param} shows 
that in our phenomenological simulations also $\epsilon_e$ is always small and its maximum value 
is always $\lesssim 10-15\%$. For higher values the duration of the burst becomes very short. The 
second issue is how much of the energy transferred to electrons is effectively emitted as high 
energy synchrotron. In absence of an external magnetic field the efficiency of emission depends on 
$\epsilon_B$. PIC simulations find $\epsilon_B \lesssim 0.001$ for a baryonic plasma. Although 
higher values apriori can increase the efficiency, they also shift the peak of spectrum to very 
high energies inconsistent with observations, see equations (\ref {magener}) and (\ref{syncchar}). 
According to (\ref{synchpower}) increasing the initial relative Lorentz factor $\gamma'_0$ also 
increases the efficiency. However, it also increases the peak energy of the burst. For instance, 
increasing $\gamma'_0$ from 2 to 5 in simulation No. 1 increases the total emission by a factor of 
$\gtrsim 30$ i.e. $E_{iso} \sim 10^{55}$ erg, but the peak of energy is at $\gtrsim 10$ MeV which 
is in contradiction with both Swift-BAT and Fermi-GBM and Fermi-LAT observations. They find the 
peak energies in the range of few hundred keV for long bursts and $\sim 1-2$ MeV in short hard 
bursts. The shape of the light curve also becomes strange. It includes a very short hard $\sim 0.1$ 
sec peak at the beginning, then becomes softer and longer peaks later. This setup is not a usual 
characteristic of GRBs. Although in few apparently long bursts such as 
GRB 060614~\cite{grb060614,swiftgrb060614} very short spikes have been observed, they are rare 
and not a common feature of GRBs. Therefore, the range of parameters that can increase the 
efficiency is quite restricted by other constraints.

\begin{figure*}
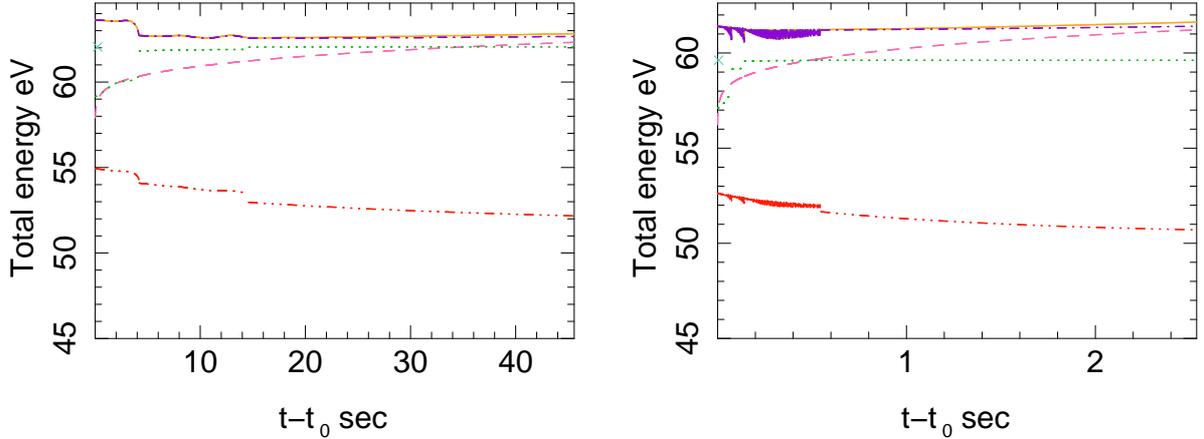

\begin{center}
\begin{tabular}{ll}
\includegraphics[width=7cm,angle=-90]{\plotdir/m1p-2.1-1000-mag300-kinetic.ps} &
\includegraphics[width=7cm,angle=-90]{\plotdir/m4-1.5-200-mag200-kinetic.ps}
\end{tabular}
\caption {Evolution of kinetic energy of colliding shells in the rest frame of a far observer at 
the redshift as source. Left: simulation No. 7 with $B = 50$ kGauss and densities 3 times larger 
than what is reported in Table \ref{tab:param}. This burst is very bright with $E_r \sim 10 
\times 56$ erg i.e. brighter than GRB 080319B and GRB 080607. Right: simulation No. 10, a bright 
short burst. Total kinetic energy (full line), kinetic energy of fast shell (dash-dot), kinetic 
energy of slow shell particle passed through shock front (dash line), magnetic energy inside fast 
shell assuming uniform density (dash-dot), and accumulative emitted energy (dot). The $\times$ 
symbol indicates the value of total emitted synchrotron emission (its position on x-axis is 
arbitrary). Fast variations in magnetic field energy and accumulative energy is due to abrupt 
change in parameters as can be found in Table \ref{tab:param}. \label{fig:kinetic}}
\end{center}
\end{figure*}

The total kinetic energy of the shock for an observer on the central - for whom shells have 
spherical symmetry is the sum of initial kinetic energy of the fast shell and the kinetic 
energy of material from slow shell that are absorbed by the shock:
\be
E_{tot} (r) = 4 \pi m_p c^2 r_0^2 \biggl [n (r_0)\Delta r (r_0) \Gamma (r_0) + 
\frac{r_0 \Gamma_f N_0}{3-\kappa} \biggl ((\frac{r}{r_0})^{3-\kappa} - 1 \biggr) \biggr ] 
\label{totkin}
\ee
and define the efficiency as:
\be
\zeta = \frac{E_{ssc}}{E_{tot}} \label{efficiency}
\ee
where $E_{ssc}$ is the total energy of emitted synchrotron self-Compton and $T$ is shock duration 
until total coalition of shells. We use total duration of each simulation as $T$. But this is 
somehow arbitrary because it does not correspond to total coalition time. Nonetheless, the 
contribution of slow shell in the total energy is small, see Fig. \ref{fig:kinetic} for examples, 
thus its value does not significantly affect the absolute efficiency, see the caption of this 
figure for more details about these examples.
Note that an observer in the center and a far observer do not find the same angular distribution 
for the emission. Due to the relativistic modification of emission distribution and different 
Doppler shift of 
photons emitted at different angles, the emitted and received photons do not have the same angular 
flux distribution, see e.g.~\cite{radproc,pow} and one has to take into account this difference 
in the determination of absolute efficiency (\ref{efficiency}). Equation (\ref{powerdopcorr}) for 
the synchrotron power corresponds to the received power, see equations 50 and 56 in~\cite{hourigrb}. 
For determining $E_{ssc}$ we cannot use equation (\ref{powerdopcorr}), but must use the angular 
power distribution, transfer it to emitted distribution using the relation between emitted and 
received powers~\cite{radproc}:
\be
\frac{\frac{dP^e}{d\omega d\Omega}}{\frac{dP^r}{d\omega d\Omega}} = \frac{1}{\Gamma^2 (1 + 
\beta cos\theta')} \label{emittorecieve}
\ee
where $\theta'$ is the angle between photon direction and line of sight of the observer in the 
slow shell frame. After applying this to equation 56 in~\cite{hourigrb} and integration over 
$theta'$ one finds that at first order of approximation and when terms proportional to 
${\mathcal G} (r)$ in (\ref{powerdopcorr}) are negligible:
\be
\frac{\frac{dP^e}{d\omega}}{\frac{dP^r}{d\omega}} \approx 1 - \beta \approx 
\frac{1}{2\Gamma^2} \label{emrec}
\ee
This relation can be understood if we remind that for far observers emission is beamed in their 
direction. Therefore, the integration for determining total emitted energy must be performed on the 
beam shape profile rather than on a sphere. 

\begin{figure*}
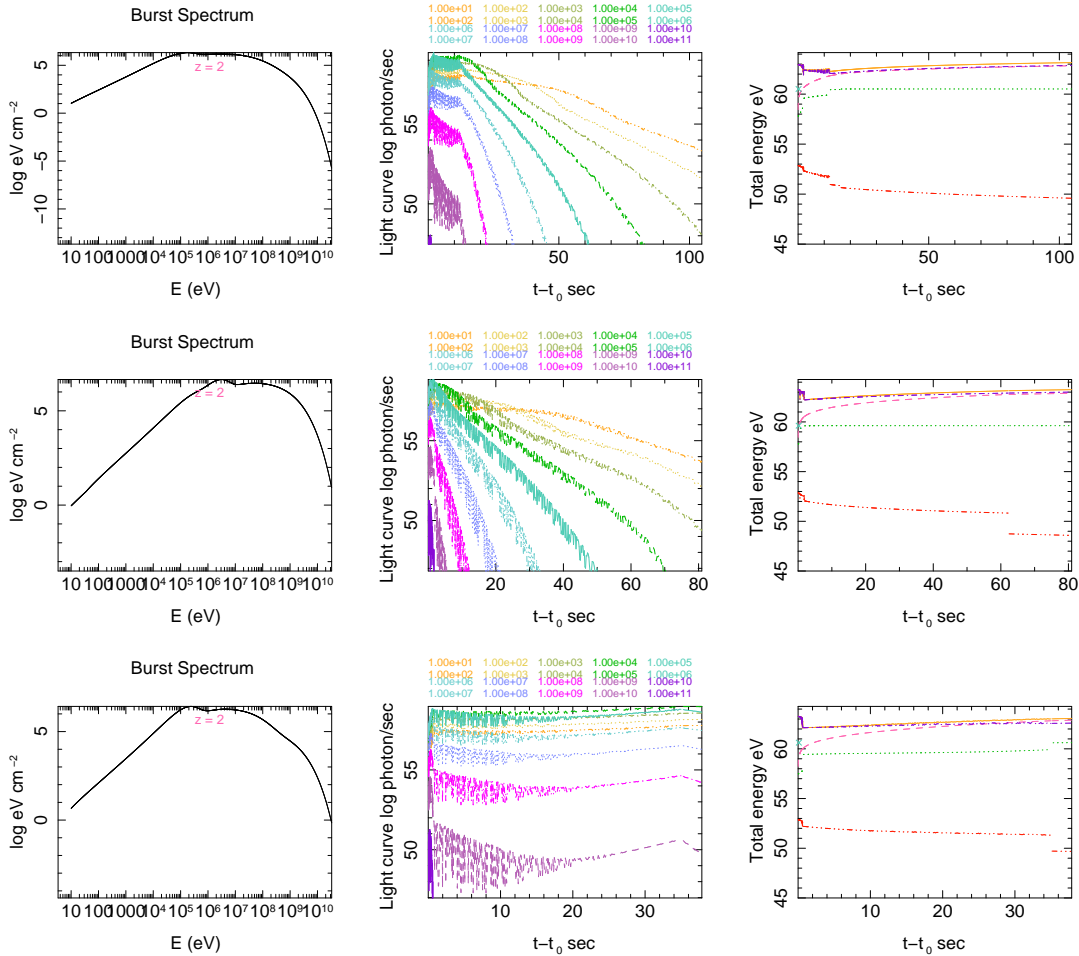

\begin{center}
\begin{tabular}{lll}
\includegraphics[width=4cm,angle=-90]{\plotdir/m2-gammap2-spect.ps} & 
\includegraphics[width=4cm,angle=-90]{\plotdir/m2-gammap2-lc.ps} & 
\includegraphics[width=4cm,angle=-90]{\plotdir/m2-gammap2-kinetic.ps} \\
\includegraphics[width=4cm,angle=-90]{\plotdir/m2-gammap4-spect.ps} &
\includegraphics[width=4cm,angle=-90]{\plotdir/m2-gammap4-lc.ps} &
\includegraphics[width=4cm,angle=-90]{\plotdir/m2-gammap4-kinetic.ps} \\
\includegraphics[width=4cm,angle=-90]{\plotdir/m2-gammap4-rr0-1.5-5-5-5-spect.ps} &
\includegraphics[width=4cm,angle=-90]{\plotdir/m2-gammap4-rr0-1.5-5-5-5-lc.ps} &
\includegraphics[width=4cm,angle=-90]{\plotdir/m2-gammap4-rr0-1.5-5-5-5-kinetic.ps}
\end{tabular}
\caption {Simulation No. 6 with various values for $\gamma'$ and $r/r_0$: $\gamma' = 2$ (top), 
$\gamma' = 4$ (middle), $\gamma' = 4$ and for 4 regimes define in Table \ref{tab:param} 
$r/r_0 = 1.5-5-5-5$. Other parameters are the same as what is reported in Table \ref{tab:param}. 
Note that light curves are presented for an observer at the redshift of the burst. Highest efficiency 
is obtained with $\gamma' = 2$, and highest energy output for $\gamma' = 4$ and $r/r_0 = 1.5-5-5-5$. 
Note that variation of $r/r_0$ without changing indices of $\epsilon_e$ and $\epsilon_B$ varies the 
peak value of the latter and thereby the synchrotron emission. We have also tried $\gamma' = 20$ and 
$\gamma_f = 50$. This leads to the same initial $\Gamma$, but synchrotron emission has a low flux, 
is soft except for a very short peak at the beginning, and efficiency is low $\sim 0.1\%$. For 
these reasons they are not shown here. \label{fig:largegammap}}
\end{center}
\end{figure*}

\begin{figure*}
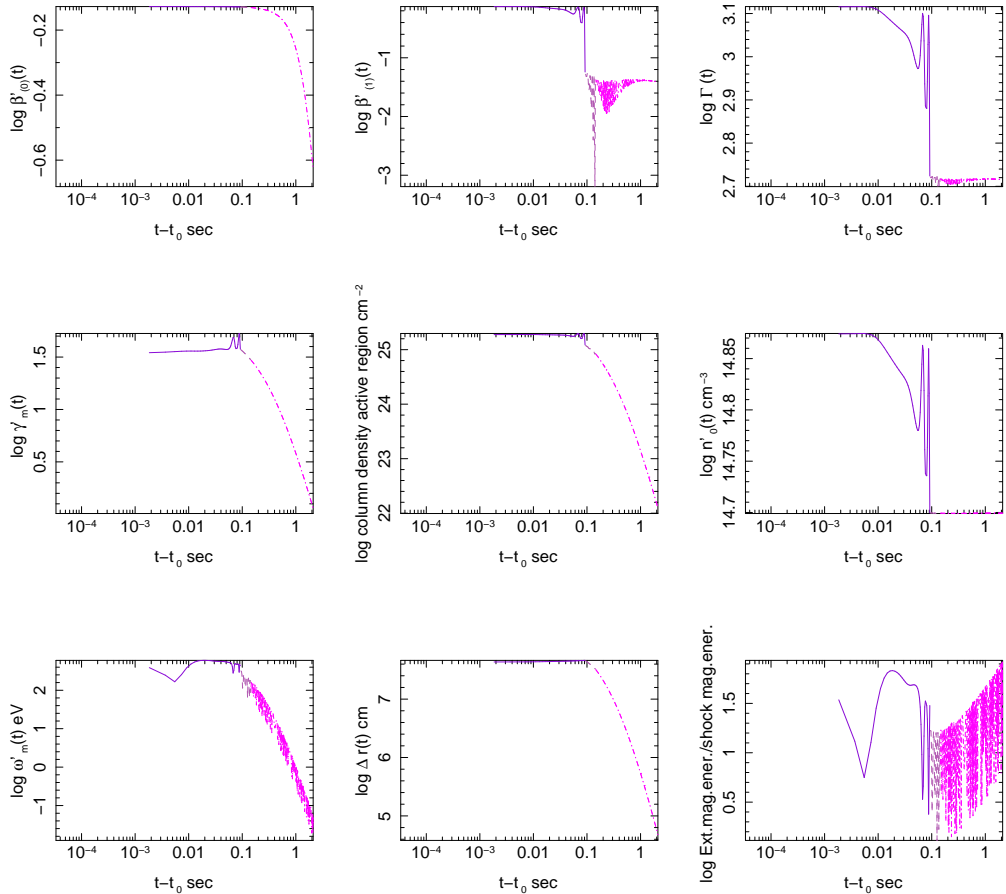

\begin{center}
\begin{tabular}{lll}
\includegraphics[width=4cm,angle=-90]{\plotdir/m3-322-plcut-2.1-3-3-3-3-mag0-highener-param-beta0.ps}& 
\includegraphics[width=4cm,angle=-90]{\plotdir/m3-322-plcut-2.1-3-3-3-3-mag0-highener-param-beta1.ps}& 
\includegraphics[width=4cm,angle=-90]{\plotdir/m3-322-plcut-2.1-3-3-3-3-mag0-highener-param-gamma.ps} \\
\includegraphics[width=4cm,angle=-90]{\plotdir/m3-322-plcut-2.1-3-3-3-3-mag0-highener-param-gammam.ps} & 
\includegraphics[width=4cm,angle=-90]{\plotdir/m3-322-plcut-2.1-3-3-3-3-mag0-highener-param-columndens.ps} & 
\includegraphics[width=4cm,angle=-90]{\plotdir/m3-322-plcut-2.1-3-3-3-3-mag0-highener-param-nprim.ps} \\
\includegraphics[width=4cm,angle=-90]{\plotdir/m3-322-plcut-2.1-3-3-3-3-mag0-highener-param-omegam.ps} & 
\includegraphics[width=4cm,angle=-90]{\plotdir/m3-322-plcut-2.1-3-3-3-3-mag0-highener-param-deltar.ps} &
\includegraphics[width=4cm,angle=-90]{\plotdir/m3-322-plcut-2.1-3-3-3-3-mag0-highener-param-mag.ps} 
\end{tabular}
\caption {Evolution of quantities determining dynamics for simulation No. 10. All quantities except 
$\Gamma$ and $\Delta r$ are in slow shell frame. The time is in the rest frame of an observer at 
redshift of the burst.\label{fig:gammambeta}}
\end{center}
\end{figure*}

The efficiency of some simulations are reported in Table \ref{tab:efficiency}. The absolute 
efficiency is the ratio of total emitted energy calculated above to total kinetic energy from 
equation (\ref{totkin}). The table shows also the weighted efficiency by the maximum values of 
$\epsilon_e$ for each simulation as well as the ratio of emitted energy to the kinetic energy of 
material fallen to the shock front. For presentation in this table, we have chosen one example of 
long bursts and one example of short bursts without and with various values of magnetic field. Two 
other example are presented which show how the variation of other improtant parameters such as 
$\gamma'$ and $\epsilon_e$ affects the total emission and its efficiency. Fig. \ref{fig:largegammap} 
show the spectrum, light curves and energy of various components for variety of choices of $\gamma'$, 
and duration of regimes.
As expected this efficiency is usually small, only few percent. Large value of weighted efficiencies, 
specially when a large external magnetic field is present means that the radiative efficiency of 
shocks is high and most of the energy transferred to electrons is emitted very rapidly. As for the 
absolute efficiency, we should remind that the 
efficiency of synchrotron emission in a shock is always small and this can be understood as the 
following: Considering a neutral plasma, for every accelerated electron that falls to the shock front, 
one or more baryons with a rest mass $\sim 2000$ times larger than electrons also falls - see 
Fig. \ref{fig:gammambeta} for an examlpe of $\beta'$ evolution. Fallen baryons make the fast shell 
heavy, decreases relative Lorentz factor, and thereby fields and synchrotron emission. Only the 
small fraction of kinetic energy which is transferred to fields can be emitted. On the other hand, 
this fraction cannot be high because larger the fields, larger and faster emission, thus faster 
energy dissipation and mixing of two shells which reduces fields and emission. Therefore, this process 
is self-regulatory. Despite low efficiency of emission the energy budget of simulations in 
Table \ref{tab:efficiency} show that even for brightest bursts, the total kinetic energy of the 
outflow is of the order of energy released in normal supernovae, thus plausible.

Apriori the end of the shock is when the two shells are completely mixed and move together with 
the same Lorentz factor. In reality this complete mixing occurs much later than the end of high 
energy emission. This can be seen in Fig. \ref{fig:kinetic}. It shows that during the evolution of 
the burst most of the energy is emitted at early times - consistent with observed FRED-like peaks. 
We also note that in absence of an external magnetic field emission efficiency is usually very low. 
These features can be described as the following: Initially, the discontinuity of density 
and other physical characteristics of the shells is large and turbulence important. Formation 
of the shocked and active regions, and thereby energy dissipation changes the discontinuity 
to a gradual variation that connects high and low density regions. At this point, the lack of 
sufficient turbulence stops high energy emission and only low energy synchrotron emission due to 
a weak remnant magnetic or due to the weakening external field would persist. Evidently our simple 
phenomenological model does not include these details. Nonetheless, parameters of this model are 
meant to present these processes, thus we expect that their evolution grossly reproduce missing 
details. Fig. \ref{fig:gammambeta} shows an example of evolution of various quantities in the 
shocked region. The peak of emission is during the time interval around the maximum of $\omega_m$. 

If the width of the fast shell is much shorter than slow one, it dissipates completely, but this 
can take a very long time. If the flow of fast material continues, accumulation and compression 
inside and in front of the transition region form another shock front and a new spike of high 
energy emission. Therefore, an emission gap is expected from the formation of a gradual transition 
region to accumulation of material and formation of a new shock. During this period the fast shell 
propagates freely and without making any radiation. These qualitative descriptions along with our 
simulations suggest a solution for the efficiency of energy transfer to electrons and thereby the 
synchrotron emission. Because shocks are efficient only for a short time, if anisotropies are 
strong at small spatial scales but are very frequent, there is a large probability that the fast 
shell collides with another shell before significant suppression of discontinuity. In this way 
the high efficiency of the emission can persist for much longer time and a larger fraction of the 
kinetic energy can be transferred to emission. This means that long bursts consist of overlaping 
short bursts. 

\section{Outline}\label{sec:conclude}
In this work we introduced an external magnetic field in a previously developed formulation for 
relativistic shocks and synchrotron self-Compton emission as a model for the prompt emission of 
GRBs. Simulation of a number of bursts with properties similar to observed ones showed that due 
to the fast variation of many quantities in this model, oscillations of an external field do not 
leave significant evidence of their presence in the PDS of light curves. Therefore, the lack of 
a clear oscillatory signature is not the evidence for absence of an external magnetic in GRBs. 
In fact according to the present knowledge of the physics of massive stars - presumed progenitors 
of GRBs - some sort of magnetism must exists in their environment. We also argued that a distance 
of order $10^{10-12}$ cm for the shock from the central engine produces more realistic bursts and 
solves the problem of the undetected photospheric emission of a standard fireball.

The spectra of simulated bursts show that the spectral properties of observed GRBs can be 
explained by internal shock synchrotron self-Compton model. In particular, diversities of the 
high energy wing of the spectrum and flatness of its low energy wing seem to be related to the 
properties of electrons Lorentz factor distribution. The spectrum of very hard and usually 
short bursts can be explained with a power-law distribution which continues to very large energies 
without any cutoff. Softer and usually longer bursts have either broken power-law or power-law with 
exponential cutoff. The shallow slope of observed GRBs is also most probably related to the 
distribution of electrons and its evolution. It can be explained with a distribution that continues 
smoothly to very low accelerations rather than having an abrupt break as usually considered.

Inverse Compton scattering can have an observable contribution at low energies, consistent with 
the results of~\cite{simulother1}, but at high energies its contribution is orders of magnitude 
smaller than synchrotron. Nonetheless, it is not straight forward to estimate the total 
contribution of inverse Compton scattering to GRBs because there can be more low energy photons 
of non-shock origin in their environment. They can increase the flux of inverse Compton scattered 
photons.

As for the reason for much longer emission in high energies we suggested that it should be related 
to a resonant movement of highly accelerated electrons with the electromagnetic standing wave created 
by the shock. This coherent movement delays the time of passage from an acceleration zone to a 
dissipation zone, increases the fly time of high energy electrons and postpones their dissipation.

\begin{table}
\begin{center}
\vspace{-2cm}
\caption{Parameter set of the simulated models. \label{tab:param}}
\end{center}
{\scriptsize
\begin{center}
\begin{tabular}{p{1mm}|p{2mm}p{10mm}p{10mm}p{2mm}p{3mm}p{5mm}p{3mm}p{3mm}p{3mm}p{3mm}p{5mm}p{2mm}p{5mm}p{12mm}p{12mm}p{8mm}p{3mm}p{3mm}p{3mm}p{2mm}p{3mm}}
\hline
No. & mod. & $r_0$ (cm) & $\frac{\Delta r_0}{r_0} $ & $p$ & $\gamma_{cut}$ & $\kappa$ & $ \gamma'_0$ & $ \tau$ & $ \delta$ & $\epsilon_B$ & $\alpha_B$ & $\epsilon_e$ & $\alpha_e$ & $N'_0$ (cm$^{-3}$) & $n'_c$ (cm$^{-2}$) & $(\frac{r'}{r'_0})_{max}$ & $\Gamma_f$ & $|B|$ (kG) & $f$ (Hz) & $\alpha_x$ & phase (rad.)\\
\hline 
\multirow{4}{1mm}{1} & 3 &  $9\times 10^{10}$ & $10^{-3}$ & 2.5 & - & 0 & 1.2 & - & 1 & $10^{-4}$ & -2 & 0.02 & -2 & $1.25\times 10^{15}$ & $5\times10^{28}$ & 1.5 & 500 & 35 & 0.5 & 1 & 0 \\
 & 2 & - & - & - & - & 0 & - & - & 0 & - & -1 & - & -1 & - & - & 1.5 & - & - & 0.5 & 2 & - \\
 & 2 & - & - & - & - & 0 & - & - & 1 & - & 2 & - & 2 & - & - & 3 & - & - & 0.5 & 3 & - \\
 & 2 & - & - & - & - & 0 & - & - & 2 & - & 3 & - & 3 & - & - & 3 & - & - & 0.5 & 3 & - \\
\hline 
\multirow{5}{1mm}{2} & 1 & $7\times 10^{10}$ & $10^{-3}$ & 2.5 & - & 1 & 2 & - & 1 & $10^{-4}$ & -0.1 & 0.02 & -0.1 & $1.25 \times 10^{14}$ & $2.5 \times 10^{26}$ & 2 & 200 & 10 & 50 & 1 & 0 \\
 & 0 & - & - & - & - & 0 & - & -1 & - & - & -2 & - & -2 & - & - & 2 & - & - & 50 & 1 & - \\
 & 2 & - & - & - & - & 2 & - & - & 2 & - & 2 & - & 2 & - & - & 4 & - & - & 50 & 2 & - \\
 & 2 & - & - & - & - & 2 & - & - & 3 & - & 2 & - & 2 & - & - & 15 & - & - & 50 & 3 & - \\
 & 2 & - & - & - & - & 2 & - & - & 4 & - & 4 & - & 4 & - & - & 5 & - & - & 50 & 3 & - \\
\hline 
\multirow{4}{1mm}{3} & 1 & $8\times 10^{10}$ & $10^{-3}$ & 2.1 & - & 0 & 1.2 & 1 & - & $10^{-4}$ & -1 & 0.02 & -1 & $1.25\times 10^{15}$ & $2.5\times 10^{28}$ & 1.5 & 500 & 100 & 0.5 & 1 & 0 \\
 & 2 & - & - & - & - & 0 & - & - & -1 & - & -2 & - & -2 & - & - & 1.5 & - & - & 0.5 & 2 & - \\
 & 2 & - & - & - & - & 0 & - & - & 3 & - & 2 & - & 2. & - & - & 3 & - & - & 0.5 & 3 & - \\
 & 2 & - & - & - & - & 0 & - & - & 2 & - & 3 & - & 3. & - & - & 3 & - & - & 0.5 & 3 & - \\
\hline 
\multirow{4}{1mm}{4} & 1 & $2 \times 10^{10}$ & $5 \times 10^{-3}$ & 1.5 & 200 & 0 & 1.5 & - & 1 & $10^{-4}$ & -1. & 0.02 & -1 & $5\times 10^{15}$ & $1.25\times 10^{27}$ & 1.5 & 400 &35 & $500$ & 1 & 0 \\
 & 0 & - & - & - & 200 & 0 & - & -1 & - & - & -2 & - & -2 & - & - & 1.5 & - & - & $500$ & 1 & - \\
 & 2 & - & - & - & 200 & 0 & - & - & 3 & - & 2 & - & 2 & - & - & 1.5 & - & - & $500$ & 2 & - \\
 & 2 & - & - & - & 200 & 0 & - & - & 5 & - & 4 & - & 4 & - & - & 3 & - & - & $500$ & 3 & - \\
\hline 
\multirow{5}{1mm}{5} & 1 & $7\times 10^{10}$ & $10^{-3}$ & 2.5 & 0.5 & 1 & 2 & - & 1 & $10^{-4}$ & -0.1 & 0.02 & -0.1 & $1.25\times 10^{14}$ & $6.25\times 10^{26}$ & 1.5 & 400 & 100 & 0.2 & 1 & 0 \\
 & 0 & - & - & - & 0.5 & 0 & - & -0.5 & - & - & -2 & - & -2 & - & - & 1.5 & - & - & 0.2 & 1 & - \\
 & 2 & - & - & - & 0.5 & 2 & - & - & 2 & - & 2 & - & 2 & - & - & 3 & - & - & 0.2 & 2 & - \\
 & 2 & - & - & - & 0.5 & 2 & - & - & 3 & - & 3 & - & 3 & - & - & 5 & - & - & 0.2 & 3 & - \\
 & 2 & - & - & - & 0.5 & 2 & - & - & 4 & - & 4 & - & 4 & - & - & 5 & - & - & 0.2 & 3 & - \\
\hline 
\multirow{4}{1mm}{6} & 3 & $4\times 10^{11}$ & 4.e-3 & 1.5 & - & -0.5 & 2 & - & 0.5 & $10^{-3}$ & -2.5 & 0.01 & -2 & $1.25\times 10^{13}$ & $5\times 10^{26}$ & 2 & 500 & 35 & 10 & 2 & 1 \\
 & 2 & - & - & - & - & 0 & - & 0.5 & 0 & - & 0 & - & 1 & - & - & 3 & - & - & 10 & 3 & - \\
 & 2 & - & - & - & - & 1 & - & - & 2 & - & 2 & - & 2 & - & - & 5 & - & - & 10 & 3 & - \\
 & 2 & - & - & - & - & 2 & - & - & 3 & - & 3 & - & 3 & - & - & 5 & - & - & 10 & 3 & - \\
\hline 
\multirow{3}{1mm}{7} & 1 & $8\times 10^{11}$ & $4\times 10^{-3}$ & 2.5 & - & 0 & 2 & - & 1 & $10^{-4}$ & -3 & 0.01 & -3 & $1.25\times 10^{12}$ & $5\times 10^{26}$ & 2 & 500 & 10 & 0.2 & 3 & 1 \\
 & 0 & - & - & - & - & 0 & - & -0.5 & - & - & 0 & - & 0 & - & - & 3 & - & - & 0.2 & 5 & 1 \\
 & 2 & - & - & - & - & 0 & - & - & 2 & - & 2 & - & 2 & - & - & 3 & - & - & 0.2 & 5 & 1 \\
\hline 
\multirow{3}{1mm}{8} & 1 & $2 \times 10^{11}$ & $2\times 10^{-3}$ & 2.5 & - & - & 3 & - & 1 & $2\times 10^{-3}$ & -2 & 0.005 & -2 & $1.25\times 10^{13}$ & $1.25\times 10^{26}$ & 1.5 & 200 & 12 & 0.5 & 1 & 0 \\
 & 2 & - & - & - & - & -1 & - & - & -1 & - & -0.5 & - & -0.5 & - & - & 4 & - & - & 0.5 & 2 & 0 \\
 & 2 & - & - & - & - & 1 & - & - & 2 & - & 3 & - & 3 & - & - & 5 & - & - & 0.5 & 3 & 0 \\
\hline 
\multirow{4}{1mm}{9} & 1 & $2 \times 10^{11}$ & $4\times 10^{-3}$ & 2.5 & - & -0.5 & 2 & - & 0.5 & $10^{-3}$ & -4 & 0.05 & -4 & $1.25\times 10^{13}$ & $5\times 10^{26}$ & 1.2 & 500 & 40 & 0.2 & 2 & 1 \\
 & 2 & - & - & - & - & 0 & - & - & 0 & - & 0 & - & 0 & - & - & 2 & - & - & 0.2 & 3 & 1 \\
 & 2 & - & - & - & - & 1 & - & - & 2 & - & 2 & - & 2 & - & - & 3 & - & - & 0.2 & 3 & 1 \\
 & 2 & - & - & - & - & 2 & - & - & 3 & - & 3 & - & 3 & - & - & 5 & - & - & 0.2 & 3 & 1 \\
\hline 
\multirow{3}{1mm}{10} & 3 & $2 \times 10^{10}$ & $5\times 10^{-3}$ & 2.1 & $10^3$ & 0 & 1.5 & - & 2 & $10^{-4}$ & -2 & 0.02 & -2 & $5\times 10^{14}$ & $5\times 10^{26}$ & 1.3 &  400 & 70 & 500 & 1 & 0 \\
 & 2 & - & - & - & $10^3$ & 0 & - & - & 3 & - & 0 & - & 0 & - & - & 3 & - & - & 500 & 1 & 0 \\
 & 2 & - & - & - & $10^3$ & 0 & - & - & 4 & - & 3 & - & 2 & - & - & 3 & - & - & 500 & 1 & 0 \\
\hline 
\multirow{3}{1mm}{11} & 1 & $5\times 10^{12}$ & $10^{-3}$ & 2.5 & - & 1 & 2 & - & 1 & $10^{-3}$ & -0.1 & 0.3 & -0.1 & $1.25\times 10^{12}$ & $2.5\times 10^{25}$ & 1.005 & 400 & 0.1 & 0.2 & 2 & 0 \\
 & 0 & - & - & - & - & 0 & - & - & -1 & - & -2 & -1 & -2 & - & - & 1.005 & - & - & - & 2 & 0 \\
 & 2 & - & - & - & - & 2 & - & - & 2 & - & 2 & - & 2 & - & - & 1.1 & - & - & - & 3 & 0 \\
 & 2 & - & - & - & - & 2 & - & - & 3 & - & 3 & - & 3 & - & - & 1.1 & - & - & - & 5 & 0 \\
 & 2 & - & - & - & - & 2 & - & - & 4 & - & 4 & - & 4 & - & - & 1.2 & - & - & - & 5 & 0 \\
\hline 
\multirow{3}{1mm}{12} & 1 & $10^{13}$ & $10^{-4}$ & 1.5 & 1 & 1 & 1.05 & - & 1 & $10^{-4}$ & -1 & 0.02 & -1 & $1.25\times 10^{13}$ & $2.5\times 10^{25}$ & 1.005 & 400 & 0.1 & 0.2 & 2 & 0 \\
 & 0 & - & - & - & - & 0 & - & - & -1 & - & -2 & -1 & -2 & - & - & 1.005 & - & - & - & 2 & 0 \\
 & 2 & - & - & - & - & 2 & - & - & 2 & - & 2 & - & 2 & - & - & 1.1 & - & - & - & 3 & 0 \\
 & 2 & - & - & - & - & 2 & - & - & 3 & - & 3 & - & 3 & - & - & 1.1 & - & - & - & 5 & 0 \\
 & 2 & - & - & - & - & 2 & - & - & 4 & - & 4 & - & 4 & - & - & 1.2 & - & - & - & 5 & 0 \\
\hline
\end{tabular}
\end{center}
\footnote*{phase $= \omega t_0$ in (\ref{ac})} \\
\footnote*{Some plots shown in the text differs only in one variable. In this case the model} \\
\footnote*{parameters is shown in this table only once.}\\
\footnote*{Each line corresponds to one simulated regime. Horizontal lines separate independent 
simulations, dashes mean either the quantity is evolved from its initial value according to 
evolution equations described in the text or kept constant because its variation induces 
discontinuity to light curves and the model is not adapted for their variation. Parameters 
$\tau$ and $\delta$ are used in different $\Delta r'$ evolution models and are mutually exclusive.}
}
\end{table}

\begin{table}
\caption{Efficiency of synchrotron emission \label{tab:efficiency}}
\begin{center}
\begin{tabular}{p{1cm}p{5cm}p{1.5cm}p{1.5cm}p{1.5cm}p{1cm}p{1cm}p{0.5cm}p{0.5cm}}
Simul. & Parameters & Total kinetic (erg) & Infall (erg) & $E_e$ (erg) & $E_r\sim E_{iso}$ (erg) & $\zeta$ (\%) & $\zeta / \epsilon_e$ (\%) & $E_e/E_{infall}$ (\%)\\
\hline
No. 7 (long) & $|B| = 0$, expo.cutoff, $\gamma_{cut}/\gamma_m= 1000$, $p = 2.1$ & $4 \times 10^{51}$ & $0.6 \times 10^{47}$
& $2.6 \times 10^{50}$ & $4 \times 10^{53}$ & $1.3 \times 10^{-4}$ & $1.6 \times 10^{-3}$ & $2 \times 10^{-2}$ \\
\hline
No. 7 (long) & $|B| = 0$, PL.cutoff, $\gamma_{cut}= 1000/\gamma_m$, $p_1 = 2.1,~p_2 = 3$ & $4 \times 10^{51}$ & $3 \times 10^{49}$ & $2 \times 10^{45}$ & $1.3 \times 10^{52}$ & $10^{-5}$ & $1.2 \times 10^{-4}$ & $7 \times 10^{-3}$ \\
\hline
No. 7 (long) & $|B| = 35$ kG, expo.cutoff, $\gamma_{cut}/\gamma_m= 1000$, $p = 2.1$ & $4 \times 10^{51}$ & $1.2 \times 10^{50}$ & $7 \times 10^{49}$ & $6 \times 10^{55}$ & 1.6 & 20 & 59 \\
\hline
No. 7 (long) & $|B| = 100$ kG, expo.cutoff, $\gamma_{cut}/\gamma_m= 1000$, $p = 2.1$ & $4 \times 10^{51}$ & $5.6 \times 10^{49}$ & $1.2 \times 10^{50}$ & $10^{56}$ & 2.7 & 34 & 210 \\
\hline
No. 7 (long) & $|B| = 17$ kG, PL.cutoff, $\gamma_{cut}/\gamma_m= 1000$ & $4 \times 10^{51}$ & $5 \times 10^{47}$ & $7 \times 10^{47}$ & $6 \times 10^{53}$  & 0.03 & 0.4 & 135 \\
\hline
No. 1 (long){\scriptsize $^{14}$} & $|B| = 100$ kG, expo.cutoff, $\gamma_{cut}/\gamma_m= 1000$, $p = 2.5$, $\gamma' = 5$, $\Gamma_f = 50$, $\epsilon_e^{max} = 0.17$ & $3.8 \times 10^{51}$ & $2.6 \times 10^{51}$ & $1.6 \times 10^{50}$ & $0.9 \times 10^{55}$ & 4.2 & 37 & 6.2 \\
\hline
No. 4 (short) & $|B| = 0$, expo.cutoff, $\gamma_{cut}/\gamma_m= 200$ & $4 \times 10^{49}$ & $4.4 \times 10^{47}$ & $2.3 \times 10^{47}$ & $10^{53}$ & 0.6 & 9 & 53 \\
\hline
No. 4 (short) & $|B| = 17$ kG, expo.cutoff, $\gamma_{cut}/\gamma_m= 200$ & $4 \times 10^{49}$ & $4.4 \times 10^{49}$ & $4.3 \times 10^{47}$ & $4 \times 10^{53}$ & 1.1 & 16.3 & 1 \\
\hline
No. 4 (short) & $|B| = 35$ kG, expo.cutoff, $\gamma_{cut}/\gamma_m= 200$ & $4 \times 10^{49}$ & $2.6 \times 10^{49}$ & $7 \times 10^{47}$ & $10^{54}$ & 1.7 & 25.2 & 2.5 \\
\hline
No. 9 (long) & $|B| = 0$ kG, expo.cutoff, $\gamma_{cut}/\gamma_m= 1$, $\gamma' = 2$ & $3.5 \times 10^{50}$ & $4.5 \times 10^{50}$ & $1.1 \times 10^{47}$ & $3.5 \times 10^{53}$ & 0.03 & 0.3 & 0.03 \\
\hline
No. 9 (long){\scriptsize $^{15}$} & $|B| = 0$ kG, expo.cutoff, $\gamma_{cut}/\gamma_m= 1$, $p = 1.1$, $\gamma' = 5$, $\epsilon_e^{max} = 0.86$ & $10^{51}$ & $3.6 \times 10^{50}$ & $1.3 \times 10^{47}$ & $3 \times 10^{53}$ & 0.014 & 0.14 & 0.036 \\
\hline
No. 6 (long) & $|B| = 10$ kG, expo.cutoff $\gamma_{cut}/\gamma_m= 10$, $\gamma' = 2$ & $1.4 \times 10^{51}$ & $10^{51}$ & $5.3 \times 10^{48}$ & $4.4 \times 10^{54}$ & 0.4 & 10 & 0.5 \\
\hline
No. 6 (long) & $|B| = 10$ kG, expo.cutoff $\gamma_{cut}/\gamma_m= 10$, $\gamma' = 4$ & $3 \times 10^{51}$ & $1.2 \times 10^{51}$ & $7 \times 10^{47}$ & $8.5 \times 10^{54}$ & 0.02 & 0.5 & 0.06 \\
\hline
No. 6 (long) & $|B| = 10$ kG, expo.cutoff $\gamma_{cut}/\gamma_m= 10$, $\gamma' = 4$, $(r/r_0)_{max}= 1.5, 5, 5, 5$ & $3 \times 10^{51}$ & $1.2 \times 10^{51}$ & $7 \times 10^{48}$ & $5 \times 10^{54}$ & 0.23 & 6 & 0.6 \\
\hline
\end{tabular}
\end{center}
{\small
\footnote*{This simulation has a $\gamma'$ larger than others and $\Gamma_f$ much smaller. Apriory 
it must have higher efficiency and indeed it has, but its efficiency yet only few percent and only 
$56\%$ largeer than simulation No. 7 with same magnetic field.}
\footnote*{The idea behind this simulation is to see whether large $\epsilon_e$ and $\gamma'$ but 
flat electron distribution which reduces $\gamma_m$, see (\ref{gammam}) and (\ref{gammamcut}) produces 
a higher efficiency. It seems that this is not the case. The peak energy spectral shape are also 
inconsistent with observations. \label{tab:note1}}
}
\end{table}
\acknowledgments{BG was supported by Nuffield Trust under the grant No. URB/36645. HZ thanks 
Mullard space Science Laboratory (MSSL) where part of this work has been performed for their 
hospitality. We also thank the GRB group in MPE and the Swift science team members for helpful 
discussions. Special thanks to Kim Page for providing us with spectral properties of GRB 061121.}

\appendix
\section{Analytical approximation of ${\mathcal N}_{(0)}(r')$}\label{app:a}
When $\eta_1 /(3-\kappa)$ is large we can formally expand the integrand of the integral in 
(\ref{nmath}) and integrate term by term:
\bea
{\mathcal N}_{(0)}(r') &=& \frac {\Delta r'_0}{(3-\kappa)} \biggl (\frac{\gamma'_0}{\beta'_0} 
\biggr )^{\tau} ((3-\kappa) {\mathcal D})^{1-\frac{\eta_1}{3-\kappa}} \biggl \{ \frac {1}
{\tau - 2 + \frac{\eta_1}{3-\kappa}} \nonumber \\
& & \bigg [{\beta'_0}^{\tau - 2 + \frac{\eta_1}{3-\kappa}}~_2F_1 
[\frac {\eta_1}{3-\kappa}, \tau - 2 + \frac{\eta_1}{3-\kappa}; \tau - 1 + \frac{\eta_1}{3-\kappa}; 
1 - \frac {\beta'_0}{(3-\kappa) {\mathcal D}}] - \nonumber \\
& & {\beta'_{(0)}}^{\tau - 2 + 
\frac{\eta_1}{3-\kappa}}~_2F_1 [\frac {\eta_1}{3-\kappa}, \tau - 2 + \frac{\eta_1}{3-\kappa};
\tau - 1 + \frac{\eta_1}{3-\kappa}; \frac{\beta'_{(0)}(r')}{\beta'_0}(1 - \frac {\beta'_0}
{(3-\kappa) {\mathcal D}})]\biggr ] + \nonumber \\
& & \frac {(\frac{5}{2} -\frac{\tau}{2})}{\tau + \frac{\eta_1}{3-\kappa}} 
\biggl [{\beta'_0}^{\tau + \frac{\eta_1}{3-\kappa}}~_2F_1 [\frac {\eta_1}{3-\kappa}, \tau + 
\frac{\eta_1}{3-\kappa}; \tau + 1 + \frac{\eta_1}{3-\kappa}; 1 - \frac {\beta'_0}{(3-\kappa) 
{\mathcal D}}] - \nonumber \\
& & {\beta'_{(0)}}^{\tau + \frac{\eta_1}{3-\kappa}}~_2F_1 [\frac {\eta_1}{3-\kappa}, 
\tau + \frac{\eta_1}{3 - \kappa}; \tau + 1 + \frac{\eta_1}{3-\kappa}; 
\frac {\beta'_{(0)}(r')}{\beta'_0}(1 - \frac {\beta'_0}{(3-\kappa) {\mathcal D}})]\biggr ] + 
\ldots \biggr \} \nonumber \\
\label {dynmzeromag}
\eea
where $\beta'_0$ in the initial $\beta \equiv v/c$ of the fast shell with rest to the slow shell and 
$\beta'_{(0)}(r')$ is the zero order solution i.e. when the energy loss due to synchrotron 
emission is ignored:
\be
\beta'_{(0)} (r') = 
\begin{cases}
\frac {(3-\kappa){\mathcal D}}{(\frac {r'}
{{r'}_0})^{3-\kappa} - 1 + \frac {(3-\kappa){\mathcal D}}{\beta'_0}} & 
\kappa \neq 3 \\
\frac {{\mathcal D}}{\ln \frac {r'}{{r'}_0} + \frac {{\mathcal D}}{\beta_0}} 
& \kappa = 3 
\end{cases} \label {betasolzero}
\ee
$~_2F_1$ is the hypergeometric function and can be expanded as a polynomial of its last 
argument. When the latter is less than $1$ and the power of the terms in the polynomial are 
positive, they converge rapidly to zero. When the last argument in$~_2F_1$ is larger than one an 
analytical extension of this function with negative power in the polynomial expansion exists. 
Therefore, the dominant term is always $\beta'_{(0)}$ in front of each $_2F_1$ term. For small 
$\eta_1$ this approximation is not valid. In this case we expand $(1 + (\frac {1}{(3-\kappa) 
{\mathcal D}} - \frac {1}{\beta'_0}) y)^{-\frac{\eta_1}{3-\kappa}}$ and obtain:
\bea
{\mathcal N}_{(0)}(r') &=& \frac {\Delta r'_0}{(3-\kappa)} \biggl (\frac{\gamma'_0}{\beta'_0} 
\biggr )^{\tau} [(3-\kappa)\mathcal D]^{1-\frac{\eta_1}{3-\kappa}} \biggl \{\frac{1}{\tau - 2 + 
\frac{\eta_1}{3-\kappa}} \biggl [{\beta'_0}^{\tau - 2 + \frac{\eta_1}{3-\kappa}}~_2F_1 
(\frac{5}{2} - \frac{\tau}{2}, \frac{\tau}{2} - 1 + \nonumber \\
& & \frac{\eta_1}{2(3-\kappa)}, \frac{\tau}{2} + 
\frac{\eta_1}{2(3-\kappa)}; {\beta'_0}^2) - {\beta'_{(0)}}^{\tau - 2 + \frac{\eta_1}{3 - \kappa}}~
_2F_1 (\frac{5}{2} - \frac{\tau}{2}, \frac{\tau}{2}-1 + \frac{\eta_1}{2(3-\kappa)}, \nonumber \\
& & \frac{\tau}{2} + \frac{\eta_1}{2(3-\kappa)}; {\beta'_{(0)}}^2(r')) \biggr ] - 
\frac {\frac{\eta_1}{3-\kappa} [\frac{1}{(3-\kappa){\mathcal D}} - \frac{1}{\beta'_0}]}
{\tau - 1 + \frac{\eta_1}{3-\kappa}} \biggl [{\beta'_0}^{\tau - 1 + \frac{\eta_1}
{3-\kappa}}~_2F_1 (\frac{5}{2} - \frac{\tau}{2}, \nonumber \\
& & \frac {\tau}{2} - \frac{1}{2} + \frac{\eta_1}
{2(3-\kappa)}; \frac {\tau}{2} + \frac{1}{2} + \frac{\eta_1}{2(3-\kappa)}; {\beta'_0}^2) - 
{\beta'_{(0)}}^{\tau - 1 + \frac{\eta_1}{3-\kappa}}~\nonumber \\
& & _2F_1 (\frac{5}{2} -\frac{\tau}{2}, \frac{\tau}{2} - \frac {1}{2} + \frac{\eta_1}
{2(3-\kappa)}; \frac{\tau}{2} + \frac{1}{2} + \frac{\eta_1}{2(3-\kappa)}; {\beta'_{(0)}}^2(r')) 
\biggr ] + \ldots \biggr \} \label {dynmzeroetazeromag}
\eea

\section {Projection of precessing field on the shock front surface} 
\label{app:b}
\begin{figure*}
\begin{center}
\begin{tabular}{ll}
a) & b) \\
\includegraphics[width=7cm]{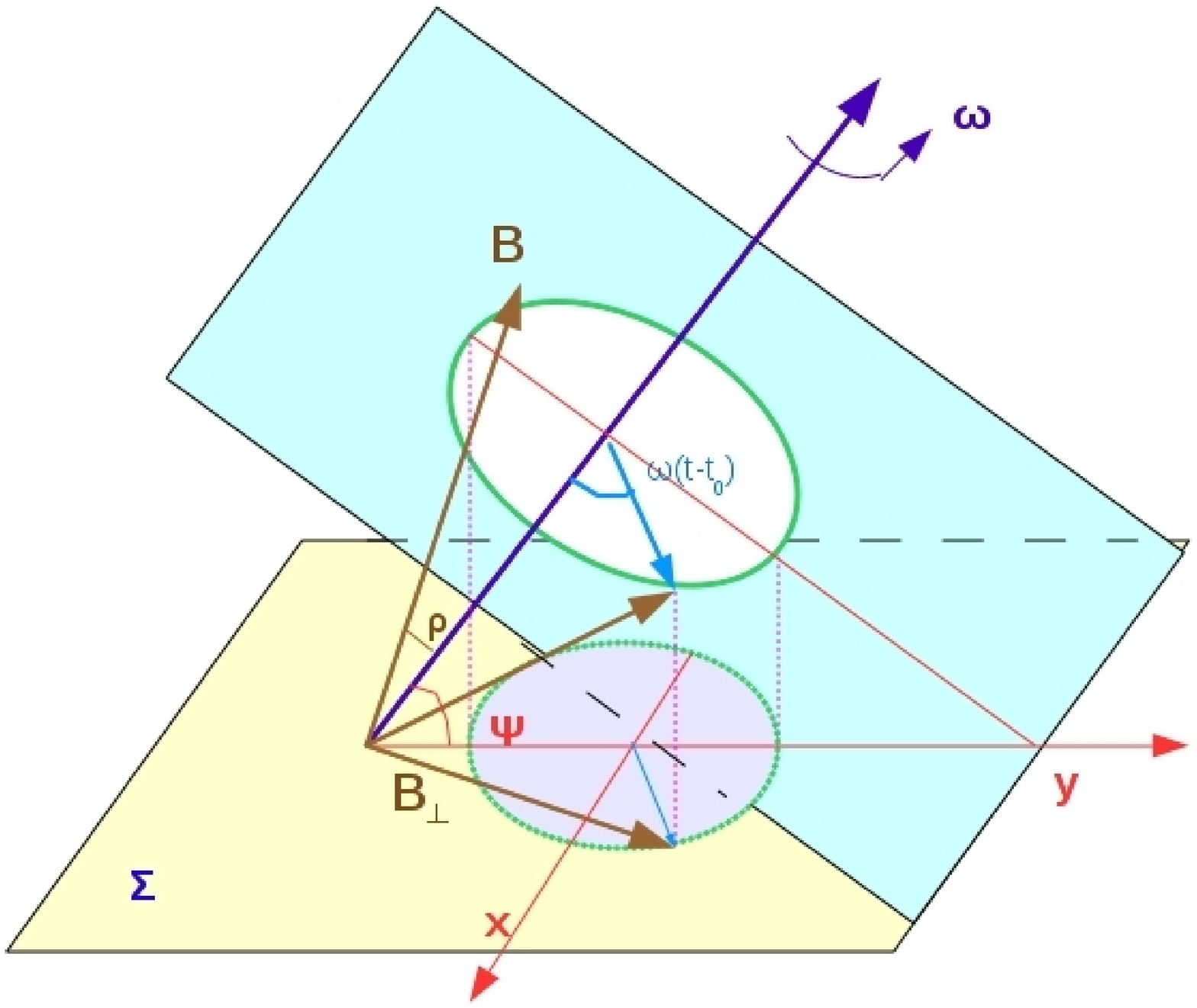} & 
\includegraphics[width=7cm]{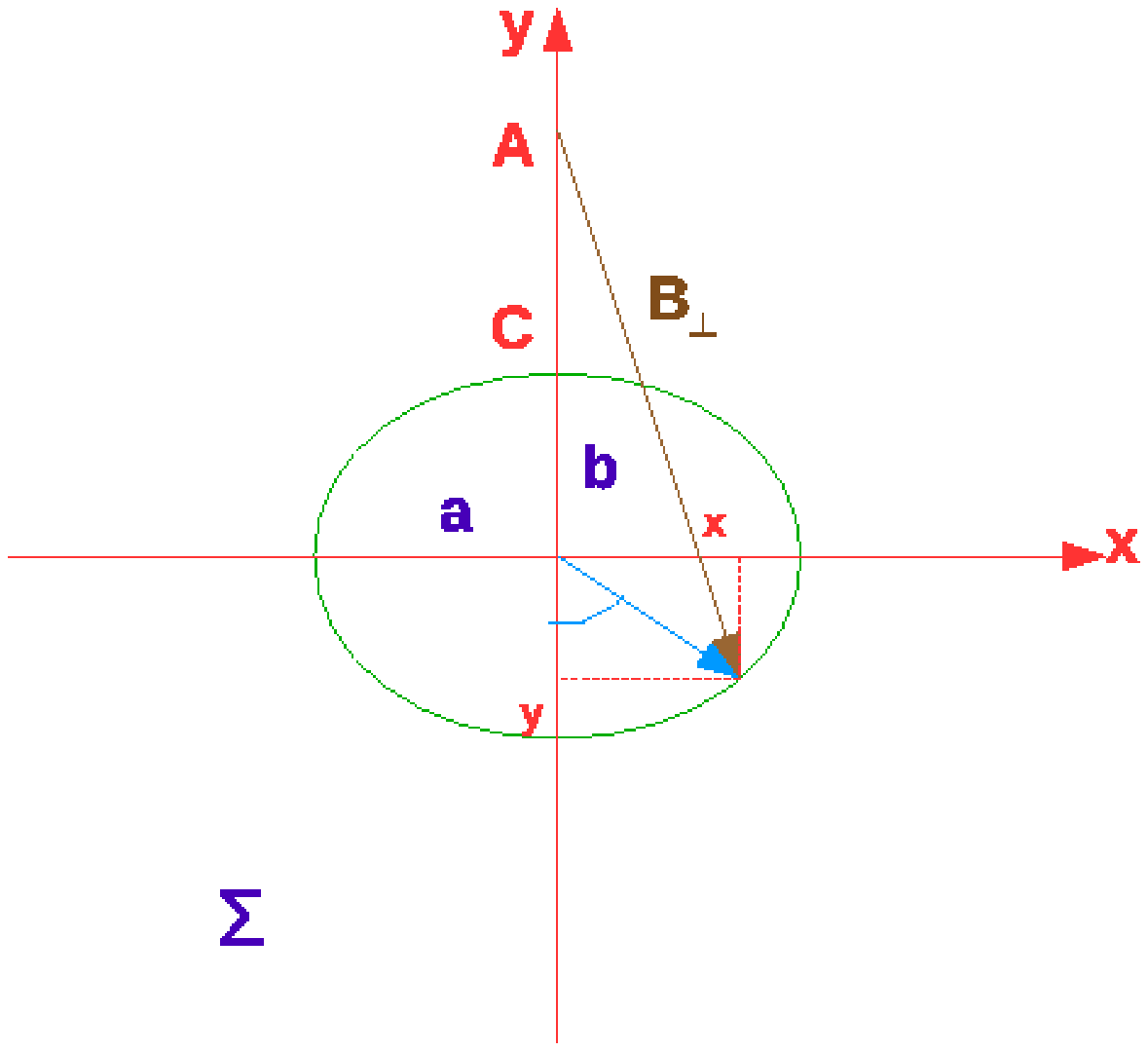}
\end{tabular}
\caption{a) Precessing external magnetic field and its projection on $\Sigma$ the shock front 
source - surface orthogonal to the propagation direction. The magnetic field $\vec{B}$ precesses 
with period $\omega$ around a constant axis having angle $\Psi$ with $\Sigma$. $\rho$ is the 
precession angle and $\vec{B}_\bot$ is the projection of $\vec{B}$ on $Sigma$. b) The precession 
loci of $\vec{B}_\bot$ on $Sigma$. \label{fig:proj}}
\end{center}
\end{figure*}

Figure \ref{fig:proj}-a presents the geometry of the magnetic field with respect to the shock front. 
$\psi$ is the angle between the precession axis of the external magnetic field and the shock 
front surface $\Sigma$. The opening angle of precession is $\rho$. For the sake of simplicity we 
first determine the projection of precession on a surface at rest with respect to the central 
engine and parallel to the shock front\footnote{For simplicity here we identify the active region 
with the shock front because the latter is effectively very close to a flat plane and assume that 
the active region is thin. The case of a thick active region can be considered as superposition of 
multiple thin emitting layers.}. Then, we transform projected quantities to the slow shell frame. 

Projection of the circular precession orbit of the field is an ellipse with half diagonal $a$ 
and $b$ where $a$ is parallel to the intersection between shock front and rotation surface:
\be
a = B\sin \rho \quad \quad b = B\sin\psi\sin\rho \label{ellipse}
\ee
Figure \ref{fig:proj}-b shows the projection of the magnetic field and its component parallel to 
the surface of rotation on $\Sigma$. Using the ellipse equation we find following expressions for 
$x$ and $y$ components of the field parallel to rotation surface and the distance 
between points $A$ and $C$ (see Fig. \ref{fig:proj}):
\be
x = |\vec{B}| \sin\rho\cos (\omega (t-t_0)) \quad \quad y = |\vec{B}| \sin\rho\sin\psi
\sin (\omega (t-t_0)) \quad \quad \bar{AC} = |\vec{B}| \cos (\psi + \rho) \label{ac}
\ee
Finally:
\be
B_\bot^2 = (\bar{AC}^2 + b + y)^2 + x^2 = |\vec{B}|\biggl\{\biggl [\cos (\psi+\rho) + 
\sin\rho\sin\psi \biggl (1+ \sin(\omega (t-t_0))\biggr )\biggr ] \biggr \} \label{bproj}
\ee

It is assumed that no large scale electric field is generated in the engine, thus the 
transformation of magnetic field to a frame moving in the direction of z-axis with a Lorentz 
factor of $\Gamma$ is very simple~\cite{electrody}:
\be
B'_z = B_z \quad \quad B'_\bot = \Gamma B_\bot \label{bboost}
\ee
In contrast to kinematic quantities, the magnetic field along the boost direction is 
preserved, but the transversed field is boosted. Schematically, the observer in the boosted frame 
see that charge particles in the engine are moving with a Lorentz factor of $\Gamma$ creating a 
current which in its turn creates a magnetic field in the surface transverse to current direction.

\section {Plasma instabilities in presence of a background magnetic field} 
\label{app:c}
In the simple simulations presented in Sec. \ref{sec:simul} it was only possible to consider 
the effect of an external magnetic field on the synchrotron emission. However, its influence on 
the development of instabilities in the plasma, the properties of instabilities and their evolution 
are much more complex. Unfortunately, it is very difficult to formulate these processes and find 
analytical or even numerical solutions for various physical quantities. For this reason in this 
section we briefly and qualitatively review the properties of a plasma in presence of a magnetic 
field according to latest PIC simulations. As mentioned before, due to computational restrictions, 
none of these simulations fulfills all the requirements necessary for the production of a GRB. 
But each of them targets some aspects of the physics of ultra-relativistic shocks. Conclusions of 
these studies are important for understanding the reasons for absent or very weak signature of a 
magnetic field in GRB data and are complementary to the conclusions obtained from simulations 
presented in this work.

In absence of an external magnetic field in a plasma environment, Weibel 
instabilities~\cite{weibel} produce a coherent field only at scales much shorter than the size 
of a shell. In the GRB context, the shock between relativistic shells can be considered as a 
{\it beam-background plasma} process~\cite{weibelrate,bgrmagfield}. Anisotropies in the 
distribution of oppositely charged particles across the shock front induces a magnetic field that 
grows for some of fluctuation modes $\vec{k}$ until particles get trapped in this field and their 
growth stops. The growth rate $\Gamma_w$ of the Weibel instabilities is defined as the imaginary 
part of the wave number $\Gamma_w \equiv \Im m \omega (\vec{k})$. The amplitude of the induced 
magnetic field depends on $\Gamma_w$ which in its turn depends on the distribution of both ions 
and electrons in the two sides of the shock front~\cite{weibelrate0}. For thermal distributions 
an explicit solution for $\Gamma_w$ exists~\cite{weibelrate1,weibelrate2}. For relativistic shocks 
in absence of a large scale magnetic field $\Gamma_w^{(max)} \approx \omega'_{pe}/{\gamma'}^{1/2}$ 
where $\gamma'$ is the relative Lorentz factor of the shells and 
$\omega'_{pe} = (4\pi e^2 n'_e/m_e)^{1/2}$ is the plasma frequency of electrons~\cite{bgrmagfield}.
For typical densities used in our simulations, the time scale for the growth of Weibel 
instabilities is very short $\sim 10^{-9}$ sec for a far observer.

In presence of an external magnetic field the dispersion relation of the plasma is more complex and 
depends on the angle between the external field and fluctuation modes. In fact due to the magnetic 
force particles of the slow shell penetrate only a short distance into the fast shell region 
before they are deflected~\cite{magshock}. This makes the mixing of the two shells more difficult, 
but increases the efficiency of synchrotron emission. Moreover, the external magnetic field 
induces a phase shift between electric and magnetic potentials that helps to separate the 
acceleration electrons from their dissipation. These results are also confirmed by 
3D-simulations~\cite{fermiaccspec1}.

Startsev, \etal~\cite{weibelrate2} have studied a special case in which the external field is 
parallel to the direction of the movement of the plasma, both analytically and numerically.
They show that depending on $\theta$ the angle between plasma wavenumber $\vec{k}$ and the 
external magnetic field as well as the relative importance of parallel and transverse modes, 
instabilities can be either Weibel-like, electrostatics lower-hybrid or higher-hybrid. Detailed 
definitions and properties of these instabilities can be found in~\cite{weibelrate2}. 
Weibel instabilities occur when $\theta$ is very small and only for long parallel modes 
(parallel to the field and to the bulk velocity of plasma) and $k_\parallel\ll k_\perp$. For larger 
$k_\parallel$ and the same $\theta$ other type of instabilities are formed. The main physical 
difference between these instabilities is whether ions (protons) of fast shell or slow shell 
stream with respect to electrons and create the induced fields. Notably, although in 
presence of an external magnetic field instabilities continue to grow, their rate saturates - 
see Figure 4 in~\cite{weibelrate2}. The saturated growth rate (width) for 
$\beta' \omega'_{pe}/ \omega'_{ce} \ll 1$: 
\be
\Gamma_{w\infty} = \frac{\beta' \omega'_{pef}\omega'_{pe}}{\sqrt {{\omega'}^2_{pef} + 
{\omega'}^2_{pe}}}  \label{weibelrate}
\ee
Here $\omega'_{ce}\equiv eB_{ext}/m_e$ is the cyclotron (Larmor) frequency of electrons in the slow 
shell, $\omega_{pe} \equiv (4\pi e^2 n_e/m_e)^{1/2}$ and $\omega'_{pef}$ are plasma frequency of 
protons in the slow and fast shells respectively. We assume a plasma composed of protons and 
electrons. For the range of densities and magnetic fields 
we have used in the simulations, the plasma of slow and fast shells are in the saturated regime. 
This means that the shock produce a transverse magnetic field in the active region, and 
Fermi or other similar processes~\cite{accelproc} accelerate electrons and produce synchrotron 
radiation, but this field is limited. For simulations in this work the growth width 
$\Gamma_{w\infty} \gtrsim 10^{-6}$ eV in slow shell rest frame, corresponding to a time scale of 
$\lesssim 10^{-7}$ sec in the rest frame of a far observer. It is longer than unmagnetized plasma 
but very short with respect to time resolution of available observations. We should also mention 
that in a magnetized plasma Weibel instabilities are not always dominant~\cite{weibelrate2}. 

Propagation of shock waves and instabilities and maximum achievable induced field depend also on 
the value of $\omega_{ce}/\omega_{pe}$ and on $v_{sh}/v_A$ where $v_{sh}$ is the velocity of the 
shock front/fast shell and $v_A = B /(4\pi n m_p)^{1/2}$ is the Alfv\'en velocity of the slow 
shell~\cite{magshock}. For mildly relativistic internal shocks and values of $|B|$ and 
density of the plasma considered in our simulations, $v_{sh}/v_A \gg 1$ thus the propagation 
distance of ions inside the shocked region is large. In fact, both simulations and observations 
find jets and shocks in sources with much larger magnetic fields than what is used in these 
simulations, such as in neutron stars and pulsars, see e.g.~\cite{nstarjetsimul,nstarjetobs}. 

The fact that instabilities become 
active at distinct range of plasma fluctuations and external magnetic fields strength is very 
important for GRBs. As we mentioned above plasma modes have a quasi-particle behaviour with a 
non-linear and sophisticated relation between frequency and wavenumber. During propagation of the 
ejecta and collision of shells some modes can get a lag/lead with respect to others. 
Moreover, the amplitude of the induced magnetic field depends in a non-linear manner on the 
external magnetic field which can be time dependent. Here we should remind that the wavelength 
of plasma waves expected to be of order of few centimeters for electrons and few thousands of 
centimeters for protons/ions. Therefore, they can apriori explain very short time scale 
variations - the jitter - of GRBs light curves~\cite{jitter}, but cannot explain a coherent 
oscillation or even random variations of multiple-second scale in long bursts. Therefore, these 
features must be due to large scale variations in the properties of the ejecta, e.g. density, 
Lorentz factor, etc. and/or multiple separate ejection of material by the central engines. 
If the external magnetic field in the direction of the ejecta oscillates - for instance 
due to the precession of the rotation axis of the central engine - the induced magnetic field 
and thereby the synchrotron emission of accelerated electrons vary cyclically. This also induces 
an oscillation into distributions of electrons and ions because dispersion relations depend on 
the external magnetic field.

Due to the nonlinear relation between an external magnetic field and the induced magnetic field 
by instabilities, their oscillation frequencies are not necessarily the same. Moreover, 
during the propagation and collision of shells the nature of dominant instabilities and thereby 
their dependence on the external magnetic field can change. Therefore, one expects that effects 
such as: dominance of higher harmonics, lags due to the differential dispersion relation of modes, 
intrinsic variation of plasma density along the line of sight, etc. smear the coherent behaviour 
in a large extent. Thus, the probability of observation of a significant oscillation 
in the light curve of GRBs is very small. Due to these complexities, a consistent treatment 
of an oscillating external magnetic field on a GRB emission needs a full plasma physics 
treatment of these processes as well as a detailed knowledge of the jet/ejecta properties which 
are not available. Thus, for studying the effect of an oscillating external magnetic field 
on the synchrotron emission we have simply assumed that the total magnetic field consists of a 
magnetic field which depends on the properties of the plasma and an external oscillating component 
which can also varies with the distance of the ejecta from the central engine. In addition, for 
further simplification of our calculation we consider the oscillation as semi-stationary i.e. we 
ignore terms that depend on the derivative of the oscillations.

\section {Electron cooling} \label{app:d}
The usual definition of "slow cooling" and "fact cooling" suggested first in ~\cite{emission1} 
and have been used intensively in GRB literature cannot be applied to this model because 
quantities such as Lorentz factor $\Gamma$ and magnetic $B$, to which the Lorentz factor of 
cooled electrons $\gamma_c$ depends, vary with time. In this case the definition of $\gamma_c$ 
can be changed, for instance to: 
\be
\gamma_c \equiv \int_0^{t_{dyn}} dt P (\gamma_c)/\Gamma m_e c^2 \label{gammac}
\ee
Although the frequency dependence of $\gamma_c$ defined here is apriori the same as what is 
discussed in~\cite{emission1}, due to time-dependence of power $P$, its variation with time 
is not the same as what is obtained in~\cite{emission1}. Slow and fast cooling regimes are 
defined by comparing $\gamma_c$ with $\gamma_m$ which is also time-dependent, 
see Fig. \ref{fig:gammambeta}. 
The presence of an  external magnetic field and electron distributions other than power-law 
make the situation even more different from the simple model used to predict broad band 
spectrum of GRBs. Moreover, both the definition (\ref{gammac}) and the original definition of 
$\gamma_c$ depend on unknown and somehow arbitrarily defined dynamical time of emission. 
Therefore, $\gamma_c$, its comparison with $\gamma_m$, and classification of processes as 
fast and slow cooling are useful for getting an insight into behaviour of the spectrum expected 
from a synchrotron model but not for analyzing a real burst or a more complex model such as 
the one presented in this work. For these reasons we have not used this classification in our 
analysis.

Qualitatively one expects that at the beginning when fields are not yet very strong i.e. 
$\epsilon_e$ and $\epsilon_b$ are small, electrons are in slow cooling according to the 
convention of [96]. When the amplitude of magnetic field, acceleration of electrons, and 
thereby $\gamma_m$ increase, electrons distribution becomes harder and they move to a fast 
cooling regime. Finally when the shock becomes weak, distribution of electrons becomes softer, 
they go back to a slow cooling regime, and emit dominantly at low energies. See Fig. 
\ref{fig:gammambeta} where an example of the evolution of quantities determining characteristics 
of the shock and synchrotron emission are shown. 

In Figs. \ref{fig:sepectmonep} and \ref{fig:diffp} spectra of different time intervals for 
some simulated bursts are plotted separately. They show the gradual movement of the peak to 
lower energies. This is similar to switching from "fast cooling" to "slow cooling". In fact 
the only difference between fast and slow cooling regimes - for simple power-law distribution - 
is the slope of the flux around the peak. For slowly varying afterglows 
which was the subject of study in [96], fast and slow cooling regimes occur in well separated 
time intervals. In the case of fast varying prompt emission both regimes occur during the short 
duration of prompt emission because the distribution of electrons changes with time (similar 
to our simulations). Therefore, the spectrum is formed from superposition of multiple 
fast and slow cooling regimes, and population of accelerated and ones which have kept more or 
less their original distribution - thermal or non-thermal. It is the reason for the absence of 
features corresponding clearly to one or the other regime in observed spectra.


\begin{thebibliography}{99}
\bibitem {helicalmag} K. Asada, M. Inoue, Y. Uchida, S. Kameno, K. Fujisawa, S. Iguchi, \& M. Mutoh, {\it A Helical Magnetic Field in the Jet of 3C 273} \Journal {\PAS}{54}{2002}{39} [astro-ph/0205497].
\bibitem {relfluid} A.M. Anile, {\it Relativistic fluids and magneto-fluids}, Cambrige University Press, (1989).
\bibitem {cosmicray} F.A. Aharonian, {\it Very high energy cosmic gamma radiation}, World Scientific, (2004).
\bibitem {precessionaccre} M.A. Abramowicz, G. Bj\"ornsson, J.E. Pringle, (Edit.), {\it Theory of black hole discs}, Cambridg University Press, (1998).
\bibitem {weibelrate} A.L. Akhiezer, L.A. Akhiezer, R.V. Polovin, A.G. Sitenko, \& K.N. Stepanov, {\it Plasma Electrodynamics}, Pergamon Press (1975). 
\bibitem {protonscatter} K. Asano \& S. Inoue,{\it Prompt GeV-TeV Emission of Gamma-Ray Bursts Due to High-Energy Protons, Muons and Electron-Positron Pairs}, \Journal{\APJ}{671}{2007}{645} [arXiv:0705.2910].
\bibitem {fermi} W.B. Atwood and Fermi collaboration, {\it The Large Area Telescope on the Fermi Gamma-ray Space Telescope Mission}, \Journal {\APJ}{697}{2009}{1071} [arXiv:0902.1089].
\bibitem {montecarlo1} M.G. Baring, {\it Probes of Diffusive Shock Acceleration using Gamma-Ray Burst Prompt Emission}, \Journal {\AIP}{1133}{2009}{294} [arXiv:0901.2535].
\bibitem {snbipolar} A. Burrows, R. Walder, C.D. Ott, E. Livne, {\it Rotating Core Collapse and Bipolar Supernova Explosions}, \Journal{\NPA}{752}{2005}{570} [astro-ph/0409035].
\bibitem {bat} S.D. Barthelmy, J.R. Cummings, E. Fenimore, N. Gehrels, D. Hullinger, H. Krimm, C. Markwardt, \etal, \Journal{\SSR}{120}{2005}{143} [astro-ph/0507410].
\bibitem {fermiacc} J. Bednarz \& M. Ostrowski, {\it Acceleration time scale for the first-order Fermi acceleration in relativistic shock waves}, \Journal {\MRA}{283}{1996}{447} [astro-ph/9608078].
\bibitem {bhpoyting} R.D. Blandford \& R.L. Znajek, {\it Electromagnetic extraction of energy from Kerr black holes}, \Journal{\MRA}{179}{1977}{433}.
\bibitem {simulother0} Z. Bosnjak, F. Daigne \& G. Dubus, {\it Prompt high-energy emission from gamma-ray bursts in the internal shock model}, \Journal{\AA}{498}{2009}{677} [arXiv:0811.2956].
\bibitem {snjet} N. Bucciantini, E. Quataert, J. Arons, B.D. Metzger, T.A. Thompson, {\it Relativistic Jets and Long-Duration Gamma-ray Bursts from the Birth of Magnetars}, \Journal {\MRA}{380}{2007}{1541} [arXiv:0707.2100].
\bibitem {agnprecess} A. Caproni, M. Livio, Z. Abraham, H.J. Mosquera Cuesta, in {\it Co-Evolution of Central Black Holes and Galaxies}, Proceedings of the International Astronomical Union, IAU Symposium, Volume 267 (2010) 325.
\bibitem {grb090709a4} S.B. Cenko, N.R. Buttler, E.O. Ofek, D.A. Perley, A.N. Morgan, D.A. Frail, J. Gorosabel, J.S. Bloom, \etal, {\it Unveiling the Origin of GRB 090709A: Lack of Periodicity in a Reddened Cosmological Long-Duration Gamma-Ray Burst}, \Journal{\AST}{140}{2010}{224} [arXiv:0911.3150].
\bibitem {windexpo} R.A. Chevalier \& Z.-Y. Li, {\it Wind Interaction Models for Gamma-Ray Burst Afterglows: The Case for Two Types of Progenitors}, \Journal{\APJ}{536}{2000}{195} [astro-ph/9908272].
\bibitem {highener080916c2} A. Corsi, D. Guetta, L. Piro, {\it GeV emission from short Gamma-Ray Bursts: the case of GRB 081024B}, \Journal {\AA}{524}{2010}{92} [arXiv:0905.1513].
\bibitem {shocksynch} F. Daigne \& R. Mochkovitch, {\it Gamma-ray bursts from internal shocks in a relativistic wind: temporal and spectral properties}, \Journal{\MRA}{296}{1998}{275} [astro-ph/9801245].
\bibitem {rphoto} F. Daigne \& R. Mochkovitch, {\it The expected thermal precursors of gamma-ray bursts in the internal shock model}, \Journal{\MRA}{336}{2002}{127} [astro-ph/0207456].
\bibitem {shocksynch1} F. Daigne \& R. Mochkovitch, {\it The physics of pulses in gamma-ray bursts: emission processes, temporal profiles and time lags},\Journal{\MRA}{342}{2003}{587} [astro-ph/0303287].
\bibitem {simulother1} F. Daigne, Z. Bosnjak, \& G. Dubus, {\it Reconciling observed GRB prompt spectra with synchrotron radiation ?}, [arXiv:1009.2636], accepted for publication in \AA.
\bibitem {poytingmodel} G. Drenkhahn \& H.C. Spruit, {\it Efficient acceleration and radiation in Poynting flux powered GRB outflows}, \Journal{\AA}{391}{2002}{1141} [astro-ph/0202387].
\bibitem {montecarlo0} D.C. Ellison \& G.P. Double, {\it Diffusive Shock Acceleration in Unmodified Relativistic, Oblique Shocks}, \Journal {\APP}{22}{2004}{323} [astro-ph/0408527]. 
\bibitem {forwardssc0} Y.Z. Fan, T. Piran, R. Narayan, D.M. Wei, {\it High Energy Afterglow from Gamma-ray Bursts}, \Journal{\MRA}{384}{2008}{1483} [arXiv:0704.2063].
\bibitem {poytingprobl0} Y.Z. Fan, {\it The spectrum of Gamma-ray Burst: a clue}, \Journal{\MRA}{403}{2010}{483} [arXiv:0912.1884].
\bibitem {helicalmag1} C. Fendt \& E. memola, {\it Collimating, relativistic, magnetic jets from rotating disks}, \Journal{\AA}{365}{2001}{631} [astro-ph/0010479].
\bibitem {forwardsh1} S.Y. Feng \& Z.G. Dai, {\it Multiband Fitting to Three Long GRBs with Fermi/LAT Data: Structured Ejecta Sweeping up a Density-Jump Medium}, (2010) [arXiv:1011.3103]. 
\bibitem {shelldecel} E.E. Fenimore \& E. Ramirez-Ruiz, {\it Gamma-Ray Bursts as Internal Shocks Caused By Deceleration}, (1999) [astro-ph/9909299].
\bibitem {grb080916c} Fermi GBM/LAT Collaborations {\it Fermi Observations of High-Energy Gamma-Ray Emission from GRB 080916C}, \Journal {\SCI}{323}{2009}{1688}.
\bibitem {grb090510} Fermi GBM/LAT Collaborations, {\it Testing Einstein's special relativity with Fermi's short hard gamma-ray burst GRB090510}, \Journal {\NAT}{462}{2009}{331} [arXiv:0908.1832].
\bibitem {grb090902b} The Fermi/GBM collaboration, The Fermi/LAT Collaborations, The Swift Team {\it Fermi Observations of GRB 090902B: A Distinct Spectral Component in the Prompt and Delayed Emission}\Journal {\APJ}{706}{2009}{L138} [arXiv:0909.2470].
\bibitem {grb090926a} Fermi GBM/LAT Collaborations, submitted, (2011) [arXiv:1101.2082].
\bibitem {swift} N. Gehrels, G. Chincarini, P. Giommi, K.O  Mason, J.A. Nousek, A.A. Wells, N.E. White, S.D. Barthelmy, \etal, {\it The Swift Gamma-Ray Burst Mission}, \Journal{\APJ}{611}{2004}{1005} [astro-ph/0405233].
\bibitem {nstarjetobs} B.M. Gaensler, J. Arons, V.M. Kaspi, M.J. Pivovaroff, N. Kawai, K. Tamura, {\it Chandra imaging of the X-ray nebula powered by pulsar B1509-58}, \Journal{\APJ}{569}{2002}{878} [astro-ph/0110454].
\bibitem {swiftgrb060614} N. Gehrels, J.P. Norris, V. Mangano, S.D. Barthelmy, D.N. Burrows, J. Granot, Y. Kaneko, C. Kouveliotou, \etal, {\it Swift detects a remarkable gamma-ray burst, GRB 060614, that introduces a new classification scheme}, \Journal {\NAT}{444}{2006}{1044} [astro-ph/0610635].
\bibitem {forwardsh5} G. Ghirlanda, G. Ghisellini, \& L. Nava, {\it The onset of the GeV afterglow of GRB 090510}, \Journal{\AA}{510}{2010}{L7} [arXiv:0909.0016].
\bibitem {fermipeak} G. Ghisellini, G. Ghirlanda, L. Nava, \& A. Celotti, {\it GeV emission from Gamma Ray Bursts: a radiative fireball?}, \Journal{\MRA}{403}{2010}{926} [arXiv:0910.2459].
\bibitem {poytingprobl1} G. Ghisellini \& G. Ghirlanda, {\it Fermi/LAT Gamma Ray Burst emission models and jet properties}, (2010) [arXiv:1002.3377].
\bibitem {grb090709a1} S. Golenetskii, R. Aptekar, E. Mazets, V. Pal'shin, D. Frederiks, P. Oleynik, M. Ulanov, D. Svinkin, \& Konus-Wind and Konus-RF Collabs., Cline, T. \& Konus-Wind Collab. \cir{9647}, (2009).
\bibitem {fireball1} J. Goodman, {\it Are gamma-ray bursts optically thick?}, \Journal{\APJL}{308}{1986}{L47}.
\bibitem {grb090709a2} D. Gotz, S. Mereghetti, A. von Kienlin, M. Beck (INTEGRAL-SPI-ACS), \cir{9649}, (2009).
\bibitem {spect080916c0} J. Granot \& Fermi collaboration, {\it GRB Theory in the Fermi Era}, in "Proceedings of 44th Recontres de Moriond - "Very High Energy Phenomena in the Universe", La Thuile (Val d'Aosta, Italy) February 1 - 8, 2009", (2009) [arXiv:0905.2206].
\bibitem {highenerextracompo} H.N. He, X.F. Wu, X.F., K. Toma, X.Y. Wang, P. M\'esz\'aros, {\it On the High Energy Emission of the Short GRB 090510}, (2009) [arXiv:1009.1432].
\bibitem {massstarmag} S. Hubrig, M. Schoeller, M. Briquet, M.A. Pogodin, R.V. Yudin, J.F. Gonzalez, T. Morel, P. De Cat, \etal, {\it Magnetic fields in massive stars}, the CP/AP Workshop, Vienna, Austria, September (2007) [arXiv:0712.0191].
\bibitem {thermal3} K. Ioka, K. Murase, K. Toma, S. Nagataki, T. Nakamura, {\it Unstable GRB photospheres and electron-positron annihilation lines}, \Journal{\APJ}{670}{2007}{L77} [arXiv:0708.1249].
\bibitem {electrody} J.D. Jackson, {\it Classical electrodynamics}, John Wiley \& Sons INC. (2001).
\bibitem {slopebright} Y. Kaneko, R.D. Preece, M.S. Briggs, \etal, \Journal {\APJS}{166}{2006}{298}. 
\bibitem {nstarjetobs} Y. Kato, M.R. Hayashi, R. Matsumoto, {\it Formation of Semirelativistic Jets from Magnetospheres of Accreting Neutron Stars: Injection of Hot Bubbles into a Magnetic Tower}, \Journal{\APJ}{600}{2004}{338} [astro-ph/0308437].
\bibitem {diskmag1} Y. Kato, {\it Formation of large-scale magnetic-towers in quasars}, \Journal{\ANA}{327}{2006}{450}.
\bibitem {magshock} T. Kawashima, S. Myahara, Y. Ohsawa, {\it Amplitude oscillation of shockwaves in a magnetized plasma}, \Journal{\JPJ}{72}{2003}{1664}.
\bibitem {snjet1} S. Komissarov, N. Vlahakis, A. Konigl, M. Barkov, {\it Magnetic acceleration of ultra-relativistic jets in gamma-ray burst sources} \Journal{\MRA}{394}{2009}{1182} [astro0811.1467]
\bibitem {jetcollim} J.H. Krolik \& J.F. Hawley, {\it General Relativistic MHD Jets}, in The Jet Paradigm, Lecture Notes in Physics, Volume 794. Springer-Verlag (2010) 265 [arXiv:0909.2580].
\bibitem {forwardsh} P. Kumar \& R. Barniol-Duran, {\it On the generation of high energy photons detected by the Fermi Satellite from gamma-ray bursts}, \Journal{\MRA}{400}{2009}{L75} [arXiv:0905.2417].
\bibitem {forwardsh0} P. Kumar \& R. Barniol-Duran, {\it External forward shock origin of high energy emission for three GRBs detected by Fermi}, \Journal{\MRA}{409}{2010}{226} [arXiv:0910.5726].
\bibitem {crosssec} L.D. Landau, E.M. Lifshitz, \& B. Berestetskii, {\it Relativistic Quantum Theory}, Pergmon Press (1971).
\bibitem {accretionscreen} R.V.E. Lovelace, M.M. Romanova, G.S. Bisnovatyi-Kogan, {\it Screening of the magnetic field of disk accreting stars}, \Journal{\APJ}{625}{2005}{957} [astro-ph/0508168].
\bibitem {radialvelo} R.V.E. Lovelace, D.M. Rothstein, G.S. Bisnovatyi-Kogan, {\it Advection/diffusion of large scale field in accretion disks} \Journal{\APJ}{701}{2009}{885} [arXiv:0906.0345].
\bibitem {poytingflow} M. Lyutikov \& R.D. Blandford, {\it Gamma Ray Bursts as Electromagnetic Outflows}, (2003) [astro-ph/0312347].
\bibitem {binaryprecess} P.R. Maloney \& M.C. Begelman, {\it The Origin of Warped, Precessing Accretions Disks in X-Ray Binaries}, \Journal{\APJ}{491}{1997}{L43} [astro-ph/9710060].
\bibitem {grb060614} V. Mangano, S.T. Holland, D. Malesani, E. Troja, G. Chincarini, \etal \Journal {\AA}{470}{2007}{105}, [arXiv:0704.2235].
\bibitem {grb090709a0} C.B. Markwardt, F.P. Gavriil, D.M. Palmer, W.H. Baumgartner, S.D. Barthelmy, \cir{9645}, (2009).
\bibitem {jitter} M. Medvedev, {\it Theory of "Jitter" Radiation from Small-Scale Random Magnetic Fields and Prompt Emission from Gamma-Ray Burst Shocks}, \Journal {\APJ}{540}{2000}{704} [astro-ph/0001314].
\bibitem {bgrmagfield} M. Medvedev \& A. Loeb, {\it Generation of Magnetic Fields in the Relativistic Shock of Gamma-Ray-Burst Sources}, \Journal{\APJ}{526}{1999}{697} [astro-ph/9904363].
\bibitem {neutronrich} P. M\'esz\'aros \& M.J. Rees, {\it Multi-GeV Neutrinos from Internal Dissipation in GRB Fireballs}, \Journal{\APJ}{541}{2000}{L5} [astro-ph/0007102].
\bibitem {maghydoshock} P. Mimica \& M.A. Aloy, \Journal {\MRA}{401}{2010}{525} [arXiv:0909.1328].
\bibitem {maghydoshocksyn} P. Mimica, D. Giannios, \& M.A. Aloy, (2010) [arXiv:1004:2720].
\bibitem {grb090709axrt} N. Mirabal \& E.V. Gotthelf, \cir{9696}, (2009).
\bibitem {snbipolar1} S.G. Moiseenko \& G.S. Bisnovatyi-Kogan, {\it Magneto-rotational supernovae. Magneto-rotational instability. Jet formation}, \Journal {\ASS}{311}{2007}{191}.
\bibitem {forwardsh2} K. Murase, K. Toma, R. Yamazaki, R., P. M\'esz\'aros, {\it On the Implications of Late Internal Dissipation for Shallow-decay Afterglow Emission and Associated High-energy Gamma-ray Signals} \Journal{\APJ}{732}{2011}{77} [arXiv:1011.0988].
\bibitem {forwardsh4} K. Murase, K. Toma, R. Yamazaki, S. Nagataki, K. Ioka, {\it High-energy emission as a test of the prior emission model for gamma-ray burst afterglows}, \Journal{\MRA}{402}{2010}{L54} [arXiv:0910.0232].
\bibitem {fermiaccspec1} G.C. Murphy, M.E. Dieckmann, L. O'C Drury, {\it Multidimensional simulations of magnetic field amplification and electron acceleration to near-energy equipartition with ions by a mildly relativistic quasi-parallel plasma collision}, IEEE Transactions on Plasma science, 38, 2985 (2010) [arXiv:1011.4406]
\bibitem {particledata} K. Nakamura, \etal (Particle Data Group), \Journal {\JPg}{37}{2010}{075021}.
\bibitem {reverseshock} E. Nakar \& T. Piran, {\it Early Afterglow Emission from a Reverse Shock as a Diagnostic Tool for GRB Outflows}, \Journal {\MRA}{353}{2004}{647} [astro-ph/0403461].
\bibitem {poytingsimul1} R. Narayan \& P. Kumar, {\it A turbulent model of gamma-ray burst variability}, \Journal{\MRA}{394}{2009}{L117} [arXiv:0812.0018].
\bibitem {grbcompare0} L. Nava, G. Ghirlanda, G. Ghisellini, A. Celotti, {\it Spectral properties of long and short Gamma-Ray Bursts: comparison between BATSE and Fermi bursts}, (2010) [arXiv:1004.1410].
\bibitem {grbcompare1} L. Nava, G. Ghirlanda, G. Ghisellini, A. Celotti, {\it Spectral properties of 438 GRBs detected by Fermi/GBM}, (2010) [arXiv:1012.2863].
\bibitem {warpeddisk} S.H. Lubow, G.I. Ogilvie, J.E. Pringle, {\it The evolution of a warped disc around a Kerr black hole} [astro-ph/0208206]. 
\bibitem {sampaper} S.R. Oates, M.J. Page, P. Schady, M. de Pasquale, T.S. Koch, \etal, \Journal {\MRA}{395}{2009}{490} [arXiv:0901.3597].
\bibitem {binaryprecess1} G.I. Ogilvie \& G. Dubus, {\it Precessing warped accretion discs in X-ray binaries}, \Journal{\MRA}{320}{2001}{485} [astro-ph/0009264].
\bibitem {alfvenradius} G.I. Ogilvie \& M. Livio, {\it Launching of Jets and the Vertical Structure of Accretion Disks}, \Journal{\APJ}{553}{2001}{158}, [astro-ph/0007474].
\bibitem {grb090709a3} M. Ohno, W. Iwakiri, M. Suzuki, M. Kokubun, T. Takahashi, M. Tashiro, Y. Terada, A. Endo, \etal, (Suzaku-WAM collab.), \cir{9653}, (2009).
\bibitem {grb100316b} F. Olivares, J. Greiner, \etal (2011), submitted.
\bibitem {notforward} A. Panaitescu, {\it GRB 090510: a short burst from a massive star ?}, \Journal{\MRA}{414}{2011}{1379} [arXiv:1005.1051].
\bibitem {fireball2} B. Paczy\'nski, {\it Gamma-ray bursters at cosmological distances}, \Journal{\APJL}{308}{1986}{L43}. 
\bibitem {thermal0} B. Paczy\'nski, {\it Super-Eddington winds from neutron stars}, \Journal{\APJ}{363}{1990}{218}.
\bibitem {grb061121} K.L. Page , R. Willingale, J.P. Osborne, B. Zhang, O. Godet, F.E. Marshall, A. Melandri, J.P. Norris, \etal, {\it GRB 061121: Broadband spectral evolution through the prompt and afterglow phases of a bright burst}, \Journal {\APJ}{663}{2007}{1125}.
\bibitem {grb061121spect} K.L. Page, (private communication) (2011).
\bibitem {grb050820a} M. Page, D. Burrows, A. Beardmore, D. Palmer, J. Kennea, \etal, \cir{3830}, (2005).
\bibitem {grb090709ahost} D.A. Perley, S.B. Cenko, \& J.S. Bloom, \cir{10903}, (2010).
\bibitem {epplasma} A. Pe'er \& E. Waxman, {\it Time dependent numerical model for the emission of radiation from relativistic plasma}, \Journal {\APJ}{628}{2005}{857} [astro-ph/0409539].
\bibitem {thermalmod0} A. Pe'er, P. M\'esz\'aros, M.J. Rees, {\it The observable effects of a photospheric component on GRB's and XRF's prompt emission spectrum}, \Journal{\APJ}{642}{2006}{995} [astro-ph/0510114].
\bibitem {thermalmod1} A. Pe'er, P. M\'esz\'aros, M.J. Rees, {\it Radiation from an expanding cocoon as an explanation of the steep decay observed in GRB early afterglow light curves}, \Journal{\APJ}{652}{2006}{482} [astro-ph/0603343].
\bibitem {thermal4} A. Pe'er, in the proceedings of "2008 Nanjing GRB conference", Nanjing, China, June (2008) [arXiv:0809.0903].
\bibitem {thermal6} A. Pe'er, F. Felix Ryde, {\it Observations, theory and implications of thermal emission from gamma-ray bursts}, (2010) [arXiv:1003.2582].
\bibitem {grb090510obs} V. Pelassa, M. Ohno, \& Fermi LAT and GBM collaborations, {\it Fermi and Swift observations of the bright short GRB 090510}, in Proceedings of "The 2009 Fermi Symposium", eConf Proceedings C091122, (2010) [arXiv:1002.2863].
\bibitem {piranrev} T. Piran, {\it Gamma-Ray Bursts and the Fireball Model}, \Journal{\PRE}{314}{1999}{575} [astro-ph/9810256].
\bibitem {helicalsynch} O. Porth, C. Fendt, Z. Meliani, B. Vaidya, {\it Synchrotron radiation of self-collimating relativistic MHD jets}, (submitted) [arXiv:1105.4258].
\bibitem {deathline} R.D. Preece, M.S. Briggs, R.S. Mallozzi, G.N. Pendleton, W.S. Paciesas, D.L. Band, {\it The Synchrotron Shock Model Confronts a `Line of Death' in the BATSE Gamma-Ray Burst Data}, \Journal{\APJ}{506}{1998}{L23} [astro-ph/9808184].
\bibitem {intext} M.J. Rees \& P. M\'esz\'aros, {\it Unsteady Outflow Models for Cosmological Gamma-Ray Bursts}, \Journal{\APJ}{430}{1994}{L93} [astro-ph/9404038].
\bibitem {intext1} M.J. Rees \& P. M\'esz\'aros, {\it Refreshed Shocks and Afterglow Longevity in GRB}, \Journal{\APJL}{496}{1998}{L1} [astro-ph/9712252].
\bibitem {thermal2} M.J. Rees \& P. M\'esz\'aros, {\it Dissipative Photosphere Models of Gamma-ray Bursts and X-ray Flashes}, \Journal{\APJ}{628}{2005}{847} [astro-ph/0412702].
\bibitem {radproc} G.B. Rybicki \& A.P. Lightman, ``Radiative Processes in Astrophysics'', Wiley-VCH (2004).
\bibitem {batcat} T. Sakamoto, S.D. Barthelmy, L. Barbier, J. Cummings, E. Fenimore, \etal, {\it The First Swift BAT Gamma-Ray Burst Catalog}, \Journal{\APJS}{175}{2008}{179} [arXiv:0707.4626].
\bibitem {emission0} R. Sari, R. Narayan, \& T. Piran, {\it Cooling Time Scales and Temporal Structure of Gamma-Ray Bursts}, \Journal{\APJ}{473}{1996}{204} [astro-ph/9605005].
\bibitem {emission1} R. Sari, T. Piran, \& R. Narayan, {\it Spectra and Light Curves of Gamma-Ray Burst Afterglows}, \Journal{\APJ}{497}{1998}{17} [astro-ph/9712005].
\bibitem {revshockoptflash} R. Sari \& T. Piran, {\it Predictions for The Very Early Afterglow and The Optical Flash}, \Journal{\APJ}{520}{1999}{641} [astro-ph/9901338].
\bibitem {grb061007} P. Schady, J. Cummings, C. Guidorzi, C. Pagani, D. Palmer, D., \etal \cir{5707}, \rep{7.1}, (2006b).
\bibitem {grb061007-1} P. Schady, M. De Pasquale, J. Cummings, M.J. Page, S.B. Pandey, \etal, {\it Extreme Properties Of GRB061007: A Highly Energetic OR Highly Collimated Burst?}, \Journal{\MRA}{380}{2007}{1041} [astro-ph/0611089].
\bibitem {grb060218} A.M. Soderberg, S.R. Kulkarni, E. Nakar, E. Berger, D.B. Fox, \etal, {\it Relativistic ejecta from XRF 060218 and the rate of cosmic explosions}, \Journal{\NAT}{442}{2006}{1014} [astro-ph/0604389].
\bibitem {fermiaccspec} A. Spitkovsky, {\it Particle acceleration in relativistic collisionless shocks: Fermi process at last?}, \Journal{\APJ}{682}{2008}{5} [arXiv:0802.3216].
\bibitem {weibelrate2} E.A. Startsev, R.C. Davidson, \& M. Dorf, {\it Streaming instabilities of intense charged particle beams propagating along a solenoidal magnetic field in a background plasma}, \Journal{\PPL}{15}{2008}{062107}.
\bibitem {pow} Sadun, A.C. \& Sadun, L.A., \Journal{\ASS}{185}{1991}{21}.
\bibitem {spect080916c1} H. Tajima \& Fermi collaboration, {\it Fermi Observations of high-energy gamma-ray emissions from GRB 080916C}, in proceedings of "31st International Cosmic-Ray Conference", (2009) [arXiv:0907.0714].
\bibitem {accelproc} S. Takeuchi, {\it new particles accelerations by magnetized plasma shock waves}, \Journal {\PPL}{12}{2005}{102901}.
\bibitem {deathline1} M. Tavani, D. Band, G. Ghirlanda, {\it Time resolved GRB spectroscopy}, \Journal {\AIP}{526}{2000}{185} [astro-ph/0002207].
\bibitem {poytingsimul0} A. Tchekhovskoy, J.C. Mckinney, \& R. Narayan, {\it Force-free Simulations of Ultra-Relativistic Jets}, \Journal{\MRA}{388}{2008}{551}.
\bibitem {bhhelicalorb} E. Teo, {Spherical Photon Orbits Around a Kerr Black Hole}, \Journal{\GRG}{35}{2003}{1909}
\bibitem {thermal1} C. Thompson, {\it A Model of Gamma-Ray Bursts}, \Journal{\MRA}{270}{1994}{480}.
\bibitem {snrot} T.A. Thompson, E. Quataert, A. Burrows, {Viscosity and Rotation in Core-Collapse Supernovae}, \Journal{\APJ}{620}{2005}{861} [astro-ph/0403224].
\bibitem {upscatter} K. Toma, X.F. Wu, P. M\'esz\'aros, {\it An up-scattered cocoon emission model of Gamma-Ray Burst high-energy lags}, \Journal {\APJ}{707}{2009}{1404} [arXiv:0905.1697].
\bibitem {forwardsh3} K. Toma, X.F. Wu, P. M\'esz\'aros, {\it An External Inverse Compton Emission Model of Gamma-Ray Burst High-Energy Lags}, in eConf Proceedings C091122, ``2009 Fermi Symposium'', (2009), [arXiv:0912.3277].
\bibitem {delayphotospher} K. Toma, X.F. Wu, P. M\'esz\'aros, {\it A Photosphere-Internal Shock Model of Gamma-Ray Bursts: Case Studies of Fermi/LAT Bursts}, (2010) [arXiv:1002.2634].
\bibitem {precursdecel} H. Umeda, N. Tominaga, K. Maeda, \& K. Nomoto, {\it Precursors and Main-bursts of Gamma Ray Bursts in a Hypernova Scenario}, \Journal {\APJ}{633}{2005}{L17} [astro-ph/0509750].
\bibitem {diskmag} G.V. Ustyugova, R.V.E. Lovelace, M.M. Romanova, H. Li, S.A. Colgate, {\it Poynting Jets from Accretion Disks: Magneto-hydrodynamic Simulations}, \Journal {\APJ}{541}{2000}{L21}.
\bibitem {grb050820a1} W.T. Vestrand, J.A. Wren, P.R. Wozniak, R. Aptekar, S. Golentskii, \etal, \Journal {\NAT}{442}{2006}{172} [astro-ph/0605472].
\bibitem {massstarmag1} R. Walder, D. Folini, G. Meynet, {\it Magnetic fields in massive stars, their winds, and their nebulae}, to be published in \Journal {\SSR}{}{}{} [arXiv:1103.3777].
\bibitem {highener080916c0} X.Y. Wang, Z. Li, Z.G. Dai, \& P. M\'esz\'aros, {\it GRB 080916C: on the radiation origin of the prompt emission from KeV/MeV to GeV}, \Journal {\APJ}{698}{2009}{L98} [arXiv:0903.2086].
\bibitem {highener080916c1} X.Y. Wang, H.N. He, Li, Z., Wu, X.F. Dai, Z.G., {\it Klein-Nishina effects on the high-energy afterglow emission of gamma-ray bursts}, \Journal {\APJ}{712}{2010}{1232} [arXiv:0911.4189].
\bibitem {weibel} E.S. Weibel, {\it Spontaneously growing transverse waves in a plasma due to an anisotropic velocity distribution}, \Journal{\PRL}{2}{1959}{83}.
\bibitem {xrtafterglow} R. Willingale, P. O'Brien, J.P. Osborne, O. Godet, K.L. Page, \etal, {\it Testing the standard fireball model of GRBs using late X-ray afterglows measured by Swift}, \Journal{\APJ}{662}{2007}{1093} [astro-ph/0612031].
\bibitem {weibelrate0} P.H. Yoon \& R.C. Davidson, {\it Exact analytical model of the classical Weibel instability in a relativistic anisotropic plasma}, \Journal {\PRA}{35}{1987}{2718}.
\bibitem {revshockform} M. Yokosawa, {\it Reverse shock wave in relativistic explosions}, \Journal{\ASS}{107}{1984}{109}.
\bibitem {weibelrate1} P.H. Yoon, {\it Electromagnetic Weibel instability in a fully relativistic bi-Maxwellian plasma}, \Journal {\PFB}{1}{1989}{1336} [correction: \Journal {\PPL}{14}{2007}{024504}].
\bibitem {grbmagform} B. Zhang, \& P. M\'esz\'aros, {\it Gamma-Ray Burst Afterglow with Continuous Energy Injection: Signature of a Highly Magnetized Millisecond Pulsar}, \Journal {\APJ}{552}{2001}{35} [astro-ph/0011133].
\bibitem {thermal5} B. Zhang \& A. Pe'er, {\it Evidence of an Initially Magnetically Dominated Outflow in GRB 080916C}, \Journal{\APJ}{700}{2009}{L65}.
\bibitem {critisism} B. Zhang, H. Yan, {\it The Internal-Collision-Induced Magnetic Reconnection and Turbulence (ICMART) Model of Gamma-Ray Bursts}, \Journal{\APJ}{726}{2011}{90} [arXiv:1011.1197].
\bibitem {fermispect} B.B. Zhang, B. Zhang, E.W. Liang, Y.Z. Fan, X.F. Wu, A. Pe'er, A. Maxham, H. Gao, Y.M. Dong, {\it A Comprehensive Analysis of Fermi Gamma-Ray Burst Data. I. Spectral Components and Their Possible Physical Origins of LAT/GBM GRBs}, \Journal {\APJ}{730}{2011}{141} [arXiv:1009.3338].
\bibitem {grb060105} H. Ziaeepour, A. Blustin, D. Burrows, J. Cummings, \etal \cir{4429}, (2006a).
\bibitem {grb060607a} H. Ziaeepour, S.T. Holland, P.T. Boyd, K. Page, S. Oates, \etal, {\it GRB 060607A: A GRB with Bright Asynchronous Early $X$-ray and Optical Emissions}, \Journal{\MRA}{385}{2008a}{453} [arXiv:0712.3269].
\bibitem {houricr} H. Ziaeepour, {\it Searching the Footprint of WIMPZILLAs}, \Journal {\APP}{16}{2001}{101} [astro-ph/0001137].
\bibitem {grb070721b} H. Ziaeepour, S.D. Barthelmy, A.P. Beardmore, D. Burrows, P.A. Evans, \etal, \rep{73.2}, (2007).
\bibitem {hourigrb} H. Ziaeepour, {\it Gamma Ray Bursts Cook Book I: Formulation}, \Journal {\MRA}{397}{2009}{361} [arXiv:0812.3277].
\bibitem {hourigrb1} H. Ziaeepour, {\it Gamma Ray Bursts Cook Book II: Simulation}, \Journal {\MRA}{397}{2009}{386} [arXiv:0812.3279].
\bibitem {forwardssc1} Y.C. Zou, Y.Z. Fan, \& T. Piran, {Expected high energy emission from GRB 080319B and origins of the GeV emission of GRBs 080514B, 080916C and 081024B}, \Journal{\MRA}{396}{2009}{1163} [arXiv:0811.2997].
\end{thebibliography}
\end{document}